\documentclass[prc,twocolumn,showpacs,superscriptaddress,nofootinbib,floatfix,showkeys,10pt]{revtex4-2}
\usepackage{bm}
\usepackage{times}
\usepackage{braket}
\usepackage{amsfonts,amssymb,stmaryrd,latexsym,amsmath}
\usepackage[usenames,dvipsnames]{color}
\usepackage{slashed}
\usepackage{hyperref}
\usepackage{subfigure}
\usepackage{graphicx}
\usepackage{slashed}
\usepackage{orcidlink}
\usepackage{multirow}
\usepackage{appendix}
\usepackage{dashrule}
\usepackage{float}
\allowdisplaybreaks[1]
\usepackage{verbatim}
\usepackage{blkarray} 

\definecolor{nicered}{rgb}{0.7,0.1,0.1}
\definecolor{nicegreen}{rgb}{0.1,0.5,0.1}
\definecolor{emph}{rgb}{1,0,0}
\definecolor{doub}{rgb}{0.7,0.2,1.0}
\definecolor{navyblue}{RGB}{0, 110, 184}

\hypersetup{colorlinks,citecolor=nicegreen,linkcolor=nicered,urlcolor=navyblue}

\begin{document}

\title{Charge-dependent nucleon-nucleon interaction at N$^3$LO in nuclear lattice effective field theory}

\author{Chengxin Wu}\affiliation{School of Physics, Sun Yat-sen University, Guangzhou 510275, China}
\author{Teng Wang}\affiliation{School of Physics, Peking University, Beijing 100871, China}
\author{Bing-Nan Lu}\email{bnlv@gscaep.ac.cn}\affiliation{Graduate School of China Academy of Engineering Physics, Beijing 100193, China}
\author{Ning Li\,\orcidlink{0000-0003-2987-2809}}\email{lining59@mail.sysu.edu.cn}\affiliation{School of Physics, Sun Yat-sen University, Guangzhou 510275, China}

\date{\today}

%%%%%%%%%%%%%%%%%%
\begin{abstract}
The nuclear lattice effective field theory (NLEFT) is an efficient tool for solving nuclear many-body problems, which takes high-fidelity lattice chiral interactions as input and computes nuclear low-energy observables via quantum Monte Carlo techniques.
In this work, we present the first next-to-next-to-next-to-leading order (N$^3$LO) chiral forces on the lattice with the isospin-breaking effects fully taken into account. We focus on both the charge-independence breaking (CIB) and charge-symmetry breaking (CSB)
effects. Specifically, we include the isospin-breaking effect from the mass difference between the charged and neutral pions in the one-pion-exchange potential (OPEP), the Coulomb force for the $pp$ interaction and the contribution of two additional charge-dependent contact operators.
We also explicitly incorporate the two-pion-exchange potentials which was mostly neglected in previous NLEFT calculations.
With these improvements, we are able to accurately reproduce the $np$ and $pp$ scattering phase shifts up to relative momentum $p \approx 200$~MeV as well as the deuteron properties.
The construction of these charge-dependent lattice nuclear forces establishes a solid foundation for future high-precision nuclear \textit{ab initio} calculations within the NLEFT framework.

\end{abstract}

\maketitle

%%%%%%%%%%%%%%%%%%%%%%%%
\section{Introduction}

The nuclear lattice effective field theory (NLEFT) is a versatile tool for addressing nuclear many-body problems from the first principle~\cite{PRC70-014007, PRC72-024006, PPNP63-117}. 
Starting from the bare nuclear forces constructed from the effective field theory (EFT) principles, we realize the Hamiltonian on a spatial cubic lattice and solve the corresponding eigenvalue problems using the auxiliary-field Monte Carlo techniques~\cite{AFQMC}. 
The recent applications of the NLEFT cover a broad range of active topics, including the properties of nuclear ground states and excited
states~\cite{EPJA31-105,PhysRevLett.104.142501,EPJA45-335,PLB732-110,PLB797-134863,PhysRevLett.128.242501,Nature630-59, PhysRevLett.112.102501,Nat.Comm.14-2777,PhysRevLett.132.062501,arxiV2411.14935},
nuclear clustering~\cite{PhysRevLett.106.192501,PhysRevLett.109.252501,PhysRevLett.110.112502,PhysRevLett.119.222505,arxiV2411.17462}, scattering processes~\cite{PhysRevC.86.034003,Nature528-111}, thermal nuclear
matter~\cite{PhysRevLett.117.132501,PhysRevLett.125.192502,PLB850-138463,PhysRevLett.132.232502}
and hypernucleus~\cite{PhysRevLett.115.185301,EPJA56-24,EPJA60-215}. 
Compared with the shell-model based nuclear \textit{ab initio} methods, the NLEFT works directly in the coordinate representation and incorporates the full quantum correlations using stochastic sampling, thus has advantages for describing phenomena connected with strong many-body correlations, \textit{e.g.}, the nuclear clustering and nuclear thermodynamics.

While modern nuclear forces achieve remarkable precision in continuum calculations (\textit{e.g.}, AV18~\cite{PRC51-38}, CD-Bonn~\cite{PRC63-024001}, and chiral potentials~\cite{PRC68-041001, NPA747-362, PRL128-142002}), implementing the interactions on a lattice is not merely a simple numerical discretization problem and still poses great challenges to nuclear physicists.
For instance, the cubic lattice explicitly breaks the rotational invariance as well as the translational invariance, thus the continuous symmetry group is reduced to a crystal space group $O_h^1$.
Moreover, the stationary lattice sets a special inertial frame of reference and breaks the Galilean invariance for non-relativistic particles.
These symmetry-breaking effects vanish in the continuum limit $a\rightarrow0$.
However, in practical NLEFT calculations, we usually take a relatively large lattice spacing $a\approx 1-2$ fm due to the limited resources, which leads to violations of fundamental symmetries.
For a high-precision calculation, we need to restore these symmetries by either averaging over the continuous group space (\textit{e.g.}, SO(3) group)~\cite{PRD90-034507, PRD92-014506} or adding extra counter terms to the Hamiltonian~\cite{PRC99-064001}.  
The latter method is similar to the Symmanzik improvement scheme~\cite{NPB226-187, NPB226-205} widely used in the study of lattice quantum chromodynamics (lattice QCD) and can be implemented order-by-order.

Another challenge in building lattice interactions is to parameterize the potentials using high-quality nucleon-nucleon ($NN$) scattering data.
Note that the results obtained in the continuum are not directly applicable on the lattice, thus we have to perform the fitting from scratch.
For this purpose, one promising tool is the L\"uscher's formula, which maps the energy levels in a periodic box to the $NN$ scattering phase shifts in the free space~\cite{Comm.Math.Phys.105-153, NPB339-222, NPB354-531, NPB364-237}.
For mixing partial waves, the L\"uscher's formula is of less precision, and an alternative approach is to impose a spherical wall on the boundary and extract the phase shifts and mixing angles by inspecting the asymptotic wave functions~\cite{PLB760-309}.
This method has sufficient precision for scattering momentum $Q\ll \Lambda_{\rm lat}$ and has been routinely used in determining the low-energy constants (LECs) in NLEFT, where $\Lambda_{\rm lat} = \pi / a$ is the lattice momentum cutoff.

Lastly, only a few types of lattice interactions among all possible constructions can be solved precisely at the many-body level~\cite{PRL118-202501, EPJA56-113, Few-Body.Syst.58-26}.
More specifically, the quantum Monte Carlo calculations for nuclei are often hampered by the fermionic sign problem.
In NLEFT this usually occurs for repulsive interactions or unbalanced fermions~\cite{PLB797-134863, Nature630-59, EPJA51-92}. 
So far most of the NLEFT calculations employ a simple non-perturbative interaction without the sign problem as the leading order and treat the residual corrections using the first-order perturbation theory~\cite{PhysRevLett.106.192501,PhysRevLett.109.252501,PhysRevLett.110.112502,PhysRevLett.112.102501}.
However, this protocol is incompatible with most of the popular construction of the nuclear forces, which involve large momentum components that induce significant second-order perturbative corrections by mixing the ground states with highly excited states.
We note that similar difficulty is also encountered in shell-model based methods such as the no-core shell model~\cite{PPNP69-131} and coupled cluster method~\cite{Rept.Prog.Phys.77-096302}, where the calculations with hard interactions require remarkably large model space. 
To resolve this issue, we can either perform a phase-shift-equivalent transformation such as the wave function matching (WFM) method to soften the interaction~\cite{Nature630-59} or calculate the second-order perturbative energies directly using the perturbative quantum Monte Carlo (ptQMC) method~\cite{PhysRevLett.128.242501, arXiv2502.13565}.
The WFM method involves local unitary transformations with a maximal raidus $R$, which sets an upper bound for the range of the lattice interactions. 
On the other hand, the ptQMC method only accept local or semi-local interactions that can be decomposed using the auxiliary field transformations.
As a consequence of these limitations, in NLEFT calculations the lattice nuclear force must be adapted to the specifically chosen many-body 
 algorithm. 

In this work, we focus on the high-fidelity chiral forces that faithfully reproduce the $NN$ scattering phase shifts. 
The early developments involve the first two-body N$^2$LO~\cite{EPJA53-83} and N$^3$LO~\cite{PhysRevC.98.044002} lattice chiral forces with spatial derivatives implemented with numerical finite differences.
In Ref.~\cite{PhysRevLett.128.242501} we built a N$^2$LO lattice chiral force with the Fast Fourier Transform (FFT) algorithm to illustrate the capability of the ptQMC method.
The FFT method computes spatial derivatives using the entire lattice sites, enabling us to employ a local operator basis that can be easily decomposed in the momentum space. 
Recently, for simulations of thermal nuclear matter, a N$^3$LO chiral force with rank-one-operator basis was proposed to facilitate perturbative calculations~\cite{PhysRevLett.132.232502}.
In Ref.~\cite{Nature630-59} the WFM method was combined with a N$^3$LO chiral force to reproduce the binding energies and charge radii of the light nuclei up to $A=56$, where a series of three-body forces and Galilean invariance restoration terms were introduced to restore the symmetries and improve the quality of describing the experiments.
We note that all these interactions reproduce the low-momentum $NN$ phase shifts nearly equally well.
The differences can be related to the realization of spatial derivatives, the handling of broken symmetries, and the implementation of three-body forces. 
As with the case of lattice QCD, we expect that the different lattice actions can give consistent many-body predictions as long as these issues are treated correctly and systematically.

So far most of the lattice chiral forces assume an exact isospin symmetry.
However, in nature the isospin is slightly violated due to the quark mass difference and electromagnetic interactions.
The degree of this violation is quantified by the mass difference between proton and neutron $M_P - M_N \approx 1.293$~MeV, as well as mass splitting of three pion species $M_\pi^\pm - M_\pi^0 \approx 4.593$~MeV,
which are two to three orders smaller than the corresponding particle masses.
Such a seemingly minor symmetry-breaking effect are found to play essential roles in explaining certain experimental phenomena, \textit{e.g.}, energy differences between yrast states of isobaric multiplets~\cite{PRL89-142502,Suzuki:1998jgw,Suzuki:1993zz}, isospin-breaking effects in superallowed Fermi $\beta$-decay~\cite{PLB773-521,Sagawa:1996zz,Sagawa:1995qpt,Kaneko:2017lvv} and energy splitting between mirror nuclei~\cite{PRL92-132502, PRL97-132501, PRL97-152501, NPR41-233,Naito:2025qub,Sarma:2024bow,Li:2023hky}.
For high-precision \textit{ab initio} calculations, it is desirable to include these isospin-breaking interactions as corrections to the nuclear forces.
To achieve this goal, in this work we revisit the nuclear force up to N$^3$LO within the NLEFT framework. 
We fully take into account the isospin-breaking effects from the one-pion-exchange potential (OPEP) due to the mass difference between the charged and neutral pions, and the electromagnetic interaction for $pp$, \textit{i.e.}, the Coulomb force.
In Refs.~\cite{PRC51-38,Stoks:1988vn,Stoks:1993tb_Nij,Bergervoet:1988zz_Nij} it is pointed out that even after taking into account the electromagnetic effects, subtle but important differences remain between the $np$ and $pp$ scattering data. One should also take into account the charge-independence breaking~(CIB) and charge-symmetry breaking~(CSB) effects ~\cite{Machleidt:1989tm_csb1, Li:1998xh_csb2, Miller:1990iz_csb3, Li:1998ya_csb4,Epelbaum:1999zn,Walzl:2000cx, Epelbaum:2005fd,Gardestig:2009ya_csb5}. As in the continuum case, we include the contributions of two additional contact operators, $\mathcal{O}_{\rm CIB}$ and $\mathcal{O}_{\rm CSB}$, which account for the charge-independence breaking and charge-symmetry breaking effects, respectively. 
Our construction of the interactions and fitting procedure is similar to that recently used in Ref.~\cite{arXiv2502.13565}, where a N$^3$LO lattice chiral force with leading-order CSB contact terms was built for ptQMC calculations.
Compared with Ref.~\cite{arXiv2502.13565}, in this work we additionally include the isospin-breaking corrections to OPEP due to the pion mass splitting. 
Moreover, we also discuss the contribution of the lattice two-pion exchange potential (TPEP), which was previously claimed to be absorbed into the contact terms for low-momentum processes~\cite{Epelbaum:2009rkz,Epelbaum:2009zsa,EPJA45-335,PhysRevLett.104.142501}. 
Summarizingly, the aim of this work is to build high-fidelity lattice interactions at the two-body level which can be fed into the NLEFT machinery in the future.

The paper is organized as follows. In Section~\ref{NN-interaction} we present the details of the nucleon-nucleon interaction up to 
N$^3$LO in Chiral effective field theory (ChEFT). Then, we discuss the charge-independence breaking and charge-symmetry breaking of the nuclear interaction in
Section~\ref{charge-dependence}. In Section~\ref{fitting}, we present the method how to extract the scattering phase shifts on a cubic lattice, and the fitting procedure. In Section~\ref{numerical}, we present the numerical 
results and make some discussions. 
Finally, we summarize our results and present an outlook in Section~\ref{conclusion}.

%%%%%%%%%%%%%%%%%%%%%%%%%%%%%%%%%%%%%%%%%%%%%%%%%%%%%%%%%%%%%%%%%%%%%%%%%%
\section{Nucleon-nucleon interaction up to N$^3$LO} \label{NN-interaction}

ChEFT has been widely applied to the low-energy nuclear physics, and gained great success, see ~\cite{Epelbaum:2008ga, Machleidt:2011zz} for reviews. 
In ChEFT, the nuclear interaction is built systematically in powers of the ratio $Q/\Lambda_\chi$ where $Q$ is the 
soft scale associated with the pion mass or external nucleon momenta while $\Lambda_\chi \approx 1$~GeV is the chiral symmetry breaking scale~\cite{NPA747-362, Epelbaum:2008ga}. Starting from the effective Lagrangians which consist of an infinite series of terms, one derives the effective potential order by order according to a power counting rule 
~\cite{Weinberg:1991um_nu}. The most important contribution appears at leading order (LO) or order $\mathcal{O}((Q/\Lambda_\chi)^0)$, the second most important contribution appears at next-to-leading order (NLO) or order $\mathcal{O}((Q/\Lambda_\chi)^2)$, and then next-to-next-to-leading order (N$^2$LO) or order $\mathcal{O}((Q/\Lambda_\chi)^3)$ and 
the next-to-next-to-next-to-leading order (N$^3$LO) or order $\mathcal{O}((Q/\Lambda_\chi)^4)$, refer to ~\cite{Ordonez:1995rz_Q,Epelbaum:1998ka,Epelbaum:1999dj,NPA747-362} for more details about ChEFT. 
From a different perspective, the nuclear force can be decomposed into the short-range interaction accounted by a series of contact terms, and long-range interaction from exchanging pions, i.e., the OPEP and multipion-exchange potential. Generally, up to N$^3$LO the nucleon-nucleon interaction can be written in the following form~\cite{NPA747-362,Machleidt:2011zz}, 
\begin{eqnarray}
    V^{\mathrm{N3LO}} &=& V^{(0)} + V^{(2)} + V^{(3)} + V^{(4)} \nonumber \\
    &=& V_{\mathrm{short-range}} + V_{\mathrm{long-range}}, 
\end{eqnarray}
where 
\begin{equation}
V_{\mathrm{short-range}} = V_{\mathrm{cont}}^{(0)} + V_{\mathrm{cont}}^{(2)} + V_{\mathrm{cont}}^{(4)}
\label{eq-contact}
\end{equation}
is the short-range interaction, and 
\begin{eqnarray}
    V_{\mathrm{long-range}} = V_{1\pi} + V_{2\pi} 
\end{eqnarray}
is the long-range interaction. Notice that there are no contact terms  at N$^2$LO or order $\mathcal{O}((Q/\Lambda_\chi)^3$). 

For the contact terms, the bases are not unique due to the antisymmetry of the wave function 
under the exchange of two nucleons~\cite{Epelbaum:2000kv,Gezerlis:2013ipa}. We take the basis which have the following form in the center-of-mass (COM) frame~\cite{Lu:2023jyz,arXiv2502.13565}, 
%%%%%%%%%%%%%%%%
\begin{widetext}
\begin{eqnarray}
V_{\mathrm{cont}}^{(0)}(\boldsymbol{p^\prime}, \boldsymbol{p}) &=& B_1 + B_2(\boldsymbol{\sigma_1} \cdot \boldsymbol{\sigma_2}), \\
V_{\mathrm{cont}}^{(2)}(\boldsymbol{p^\prime}, \boldsymbol{p})&=& C_1 q^2 
+ C_2 q^2 (\boldsymbol{\tau_1} \cdot \boldsymbol{\tau_2}) 
+ C_3 q^2 (\boldsymbol{\sigma_1} \cdot \boldsymbol{\sigma_2}) 
+ C_4 q^2 (\boldsymbol{\sigma_1} \cdot \boldsymbol{\sigma_2}) (\boldsymbol{\tau_1} \cdot \boldsymbol{\tau_2}) \nonumber \\
&& + C_5\frac{i}{2}(\boldsymbol{q}\times \boldsymbol{k})\cdot (\boldsymbol{\sigma_1} + \boldsymbol{\sigma_2}) 
+ C_6(\boldsymbol{\sigma_1} \cdot \boldsymbol{q})(\boldsymbol{\sigma_2} \cdot \boldsymbol{q}) 
+ C_7(\boldsymbol{\sigma_1} \cdot \boldsymbol{q})(\boldsymbol{\sigma_2} \cdot \boldsymbol{q})(\boldsymbol{\tau_1} \cdot \boldsymbol{\tau_2}),\\
V_{\mathrm{cont}}^{(4)}(\boldsymbol{p^\prime}, \boldsymbol{p}) &=& D_1 q^4 
+ D_2 q^4 (\boldsymbol{\tau_1} \cdot \boldsymbol{\tau_2}) 
+ D_3 q^4 (\boldsymbol{\sigma_1} \cdot \boldsymbol{\sigma_2}) 
+ D_4 q^4 (\boldsymbol{\sigma_1} \cdot \boldsymbol{\sigma_2}) (\boldsymbol{\tau_1} \cdot \boldsymbol{\tau_2}) 
+ D_5 q^2 (\boldsymbol{\sigma_1} \cdot \boldsymbol{q})(\boldsymbol{\sigma_2} \cdot \boldsymbol{q}) \nonumber \\
&& + D_6 q^2(\boldsymbol{\sigma_1} \cdot \boldsymbol{q})(\boldsymbol{\sigma_2} \cdot \boldsymbol{q})(\boldsymbol{\tau_1} \cdot \boldsymbol{\tau_2}) 
+ D_7 q^2k^2 
+ D_8 q^2k^2(\boldsymbol{\sigma_1} \cdot \boldsymbol{\sigma_2}) 
+ D_9 (\boldsymbol{q} \cdot \boldsymbol{k})^2 \nonumber \\
&&+ D_{10} (\boldsymbol{q} \cdot \boldsymbol{k})^2 (\boldsymbol{\sigma_1} \cdot \boldsymbol{\sigma_2})
+ D_{11} \frac{i}{2}q^2 (\boldsymbol{q} \times \boldsymbol{k}) (\boldsymbol{\sigma_1} + \boldsymbol{\sigma_2})
+ D_{12} \frac{i}{2}q^2 (\boldsymbol{q} \times \boldsymbol{k}) (\boldsymbol{\sigma_1} + \boldsymbol{\sigma_2})(\boldsymbol{\tau_1} \cdot \boldsymbol{\tau_2}) \nonumber \\
&&+ D_{13} k^2 (\boldsymbol{\sigma_1} \cdot \boldsymbol{q}) (\boldsymbol{\sigma_2} \cdot \boldsymbol{q})
+ D_{14} k^2 (\boldsymbol{\sigma_1} \cdot \boldsymbol{q}) (\boldsymbol{\sigma_2} \cdot \boldsymbol{q})
+ D_{15}[ (\boldsymbol{q} \times \boldsymbol{k})\cdot \boldsymbol{\sigma_1} ] [ (\boldsymbol{q} \times \boldsymbol{k})\cdot \boldsymbol{\sigma_2} ].
\label{eq-contact-detail}
\end{eqnarray}
\end{widetext}
In above, ${\bm p}$ and ${\bm p}^\prime$ are the momenta of the incoming and outgoing nucleons in COM frame, respectively. ${\bm q} = {\bm p}^\prime - {\bm p}$ is the transfer momentun in the \textit{t} channel while ${\bm k} = ({\bm p} + {\bm p}^\prime )/2$ is the transfer momentum in the \textit{u} channel. $B_{1-2}$, $C_{1-7}$ and $D_{1-15}$ are the low-energy constants (LECs), associated with the LO, NLO and N$^3$LO contact terms, whose values need to be determined by fitting the nucleon-nucleon scattering data. ${\bm \sigma}_i$ and ${\bm \tau}_i$ are the respective spin and isospin Pauli matrices of nucleon $i$. The superscripts denote the power of $Q/\Lambda_\chi$. As in Ref.~\cite{Lu:2023jyz,arXiv2502.13565} we regulate the contact terms nonlocally by 
introducing a regulator, $\exp(-p^6/(2\Lambda^6))$, for each nucleon in the momentum space.

However, it is quite challenging to determine all of the LECs by fitting simultaneously the $S$-, $P$- $D$-wave phase shifts and mixing angles, particularly for the lattice calculation. Alternatively, we decompose the interaction for individual partial waves. After projection and multiplied by the regulator, the specific expressions are~\cite{PhysRevC.98.044002, Epelbaum:2014efa_2pe, Machleidt:2011zz, PhysRevLett.117.132501},

\begin{widetext}
\begin{eqnarray}
\label{eq:partialwaves}
\langle ^1S_0 |V_{\rm short-range}^\Lambda | ^1S_0 \rangle &=& \left[C_{0,{}^1S_0} + C_{2,{}^1S_0} (p^2 + p^{\prime 2}) + C_{4,{}^1S_0}^{1} (p^2  p^{\prime 2}) + C_{4,{}^1S_0}^{2} (p^4 + p^{\prime 4})\right]\exp(-{p^\prime}^6/\Lambda^6 - p^6/\Lambda^6), \nonumber \\
\langle ^3S_1 |V_{\rm short-range}^\Lambda | ^3S_1 \rangle &=& \left[C_{0,{}^3S_1} + C_{2,{}^3S_1} (p^2 + p^{\prime 2}) + C_{4,{}^3S_1}^{1} (p^2  p^{\prime 2}) + C_{4,{}^3S_1}^{2} (p^4 + p^{\prime 4})\right]\exp(-{p^\prime}^6/\Lambda^6 - p^6/\Lambda^6), \nonumber \\
\langle ^1P_1 |V_{\rm short-range}^\Lambda | ^1P_1 \rangle &=& \left[C_{2,{}^1P_1}p p^{\prime} + C_{4,{}1P_1} p p^{\prime} (p^2 +  p^{\prime 2})\right]\exp(-{p^\prime}^6/\Lambda^6 - p^6/\Lambda^6), \nonumber \\
\langle ^3P_1 |V_{\rm short-range}^\Lambda | ^3P_1 \rangle &=& \left[C_{2,{}^3P_1}p p^{\prime} + C_{4,{}3P_1} p p^{\prime} (p^2 +  p^{\prime 2})\right]\exp(-{p^\prime}^6/\Lambda^6 - p^6/\Lambda^6), \nonumber \\
\langle ^3P_0 |V_{\rm short-range}^\Lambda | ^3P_0 \rangle &=& \left[C_{2,{}^3P_0}p p^{\prime} + C_{4,{}3P_0} p p^{\prime} (p^2 +  p^{\prime 2})\right]\exp(-{p^\prime}^6/\Lambda^6 - p^6/\Lambda^6), \nonumber \\
\langle ^3P_2 |V_{\rm short-range}^\Lambda | ^3P_2 \rangle &=& \left[C_{2,{}^3P_2}p p^{\prime} + C_{4,{}3P_2} p p^{\prime} (p^2 +  p^{\prime 2})\right]\exp(-{p^\prime}^6/\Lambda^6 - p^6/\Lambda^6), \nonumber \\
\langle ^1D_2 |V_{\rm short-range}^\Lambda | ^1D_2 \rangle &=& C_{4,{}^1D_2} p^2  p^{\prime 2}\exp(-{p^\prime}^6/\Lambda^6 - p^6/\Lambda^6), \nonumber \\
\langle ^3D_2 |V_{\rm short-range}^\Lambda | ^3D_2 \rangle &=& C_{4,{}^3D_2} p^2  p^{\prime 2}\exp(-{p^\prime}^6/\Lambda^6 - p^6/\Lambda^6), \nonumber \\
\langle ^3D_1 |V_{\rm short-range}^\Lambda | ^3D_1 \rangle &=& C_{4,{}^3D_1} p^2  p^{\prime 2}\exp(-{p^\prime}^6/\Lambda^6 - p^6/\Lambda^6), \nonumber \\
\langle ^3D_3 |V_{\rm short-range}^\Lambda| ^3D_3 \rangle &=& C_{4,{}^3D_3} p^2  p^{\prime 2}\exp(-{p^\prime}^6/\Lambda^6 - p^6/\Lambda^6), \nonumber \\
\langle ^3S_1 |V_{\rm short-range}^\Lambda | ^3D_1 \rangle &=& \left[C_{2,{}SD} p^2  + C_{4,{}SD}^1 p^2  p^{\prime 2} +  C_{4,{}SD}^2 p^4\right]\exp(-{p^\prime}^6/\Lambda^6 - p^6/\Lambda^6), \nonumber \\
\langle ^3D_1 |V_{\rm short-range}^\Lambda | ^3S_1 \rangle &=& \left[C_{2,{}SD} p^{\prime 2}  + C_{4,{}SD}^1 p^2  p^{\prime 2} +  C_{4,{}SD}^2 p^{\prime 4}\right]\exp(-{p^\prime}^6/\Lambda^6 - p^6/\Lambda^6), \nonumber \\
\langle ^3P_2 |V_{\rm short-range}^\Lambda | ^3F_2 \rangle &=& C_{4,{}PF} p^3 p^{\prime}\exp(-{p^\prime}^6/\Lambda^6 - p^6/\Lambda^6), \nonumber \\
\langle ^3F_2 |V_{\rm short-range}^\Lambda | ^3P_2 \rangle &=& C_{4,{}PF} p p^{\prime 3}\exp(-{p^\prime}^6/\Lambda^6 - p^6/\Lambda^6).
\label{eq-basis}
\end{eqnarray}
\end{widetext}
After the decomposition, the fitting process is simplified by performing the calculations for individual channels only, instead of for all the channels together. Notice that the spectroscopic coefficients $C_{0, i}, C_{2, i}$ and $C_{4, i}$ are linear combinations of the original LECs $B_i, C_i$, and $D_i$. Once $C_{0, i}, C_{2, i}$ and $C_{4, i}$ are determined, the original LECs $B_i, C_i, D_i$ can be easily solved through a linear transformation.    

Up to N$^3$LO, there are $2 + 7 + 15 = 24$ LECs in total. However, after taking into acccount the on-shell condition, i.e., $p^2 = p^{\prime2}$, three of the 15 LECs at N$^3$LO are redundant~\cite{Reinert:2017usi_4order}. One is from the $^1S_0$ channel, one is from the $^3S_1$ channel, and the third is from the $^3S_1$-$^3D_1$ channel. After removing the three redundant operators, $C_{4,{}^1S_0}^{2} (p^4 + p^{\prime 4})$, $C_{4,{}^3S_1}^{2} (p^4 + p^{\prime 4})$, and $C_{4,{}SD}^2 p^{\prime 4}$, there are totally 21 contact terms up to N$^3$ LO, i.e., 2 for LO, 7 for NLO and 12 for N$^3$LO. 

The long-range nucleon-nucleon interaction arises from the pion exchanges. The
well-established OPEP appears at LO according to the power counting. In  
present work, we take the following form~\cite{Reinert:2017usi_4order,PhysRevLett.128.242501},
\begin{eqnarray}
V_{1\pi}^{\Lambda_\pi}(M_\pi) &=& -\frac{g^{2}_{A} f^{\Lambda_\pi}_{\pi}(q^2)} {4F^{2}_{\pi}}
\bigg[\frac{({\bm \sigma}_1 \cdot {\bm q}) ({\bm \sigma}_2 \cdot {\bm q})}{q^2 + M_{\pi}^2} \nonumber \\
&+& C(M_\pi) {\bm \sigma}_1 \cdot {\bm \sigma}_2 \bigg],\label{eq-1pe}
\end{eqnarray}
where $g_A = 1.287$ is the axial-vector coupling constant and $F_{\pi} = 92.2$ MeV the pion decay constant. $M_{\pi}$ is the pion mass, for which we take $M_{\pi^0} = 134.98$~MeV and $M_{\pi^{\pm}} = 139.57$~MeV for the neutral and charged pions, respectively. The mass difference between the neutral and charged pions is one source of isospin breaking which will be discussed in the next section. The second term is a LO contact interaction which is adopted to minimize the amount of short-range contributions in the regularized OPEP~\cite{Reinert:2017usi_4order}. The constant $C(M_\pi)$ takes the following form
\begin{eqnarray}
 C(M_\pi) = -\frac{
 \Lambda_{\pi}(\Lambda_{\pi}^2 - 2M_{\pi}^2) 
  + 2\sqrt{\pi} M_{\pi}^3 \exp \left(\frac{M_{\pi}^2}{\Lambda_{\pi}^2}\right) 
 {\rm erf}\left(\frac{M_{\pi}}{\Lambda_{\pi}}\right)}{3\Lambda_{\pi}^3}, \nonumber \\
\end{eqnarray}
with the momentum regulator,  
\begin{equation}
f^{\Lambda_\pi}_{\pi}(q^2) = \exp\left(-\frac{q^2 + M_{\pi}^2}{\Lambda_{\pi}^2}\right).
\end{equation}
Note that $C(M_\pi)$ not only regularizes the OPEP but also ensures that the residue of the static pion propagator at the pion position is unchanged~\cite{Reinert:2017usi_4order}. As in Ref.~\cite{Lu:2023jyz}, we choose the momentum cutoff $\Lambda_{\pi} = 300$ MeV in our calculation. Notice that the isospin factor ${\bm \tau}_1 \cdot {\bm \tau}_2$ which reflects the isospin effect is not included in Eq.~(\ref{eq-1pe}). We will take into account the isospin-breaking effect specifically, and come back to this issue in the next section.

In addition to OPEP, the long-range nucleon-nucleon interaction also arises from TPEP, which first appears at NLO, and then N$^2$LO and N$^3$LO, etc.~\cite{Kaiser:2001pc_2pe, Epelbaum:2014efa_2pe, Epelbaum:2014sza_2pe, Machleidt:2011zz}. Generally, the TPEP can be decomposed into the following form~\cite{Fettes:1998ud_2pe},
\begin{eqnarray}
V_{2\pi} &=& V_C + \boldsymbol{\tau_1}\cdot\boldsymbol{\tau_2}W_C \nonumber \\
&+&[V_S+\boldsymbol{\tau_1} \cdot \boldsymbol{\tau_2}W_S] \boldsymbol{\sigma_1} \cdot \boldsymbol{\sigma_2} \nonumber \\
&+&[V_T +\boldsymbol{\tau_1} \cdot \boldsymbol{\tau_2}W_T] (\boldsymbol{\sigma_1} \cdot \boldsymbol{q})  (\boldsymbol{\sigma_2}\cdot \boldsymbol{q}) \label{eq-2pe} \\
&+&[V_{LS}+\boldsymbol{\tau_1}\cdot\boldsymbol{\tau_2}W_{LS}] i \frac{(\boldsymbol{\sigma_1} + \boldsymbol{\sigma_2})}{2}\cdot (\boldsymbol{q} \times \boldsymbol{k}).\nonumber  
\end{eqnarray}
In above, $V_i$ and $W_i$ are scalar functions which depend on the nucleon momenta. It should be mentioned that the terms in the last line only provide relativistic effects, which we will not take into account in the present work. Therefore, the TPEP in our calculation is purely local, and we refer to ~\cite{Kaiser:2001pc_2pe, Epelbaum:2014efa_2pe} for the specific expressions of scalar functions $V_i$ and $W_i$.

%%%%%%%%%%%%%%%%%%%%%%%%%%%%%%%%%%%%%%%%%%%%%%%%%%%%%%%%%%%%%%%%%%%
\section{Charge-independence breaking and charge-symmetry breaking} \label{charge-dependence} 

In quantum chromodynamics (QCD), the up and down quarks have the same mass and form the isospin doublet in the isospin symmetry limit. In nuclear physics, the neutron and proton have approximately the same mass and form the isospin doublet, 
which reflects the $\mathrm{SU(2)}$ isospin symmetry at the hadronic level. However, the small nucleon mass difference and the pion mass difference indicate the isospin breaking. Therefore, to obtain high precision nuclear forces, the isospin breaking effect should be taken into account. More precisely, one should take into account the  charge-independence breaking (CIB) and charge-symmetry breaking (CSB) effects. CIB arises from the fact that the proton~($I_z = 1/2$) and neutron~($I_z = -1/2$) have different masses and charges~\cite{Niskanen:2001aj_cib1}, and as a result, under a rotation in isospin space, the interaction between a proton and a neutron does not respect charge invariance. 
CSB, as a special case of charge dependence, occurs under a 180$^\circ$ rotation about the $y$-axis in isospin space~\cite{Epelbaum:1999zn,Machleidt:2011zz}. 

To have a clear view of this issue, in the following we list the empirical values of the scattering length $a$ and effective range $r$ for $pp$, $nn$, and $np$ scattering respectively, in the $^{1}S_{0}$ channel~\cite{Miller:1990iz_csb3, GonzalezTrotter:2006wz_181, Chen:2008zzj_182,PRC63-024001},
\begin{align}
a_{pp} &= -17.3\pm 0.4 ~\mathrm{fm},   &r_{pp} = 2.85 \pm 0.04  ~ \mathrm{fm} , \nonumber \\
a_{nn} &= -18.95 \pm 0.40 ~ \mathrm{fm}, &r_{nn} = 2.75 \pm 0.11 ~\mathrm{fm}, \\ \label{eq-nndata} 
a_{np} &= -23.740 \pm 0.020 ~\mathrm{fm}, &r_{np} = 2.77 \pm 0.05 ~ \mathrm{fm}. \nonumber
\end{align}
As discussed in Ref.~\cite{Machleidt:2011zz}, one can quantify and measure CSB via 
\begin{eqnarray}
\Delta a_{\mathrm{CSB}} &=& a_{pp} - a_{nn} = 1.65 \pm 0.60\ \mathrm{fm},  \nonumber \\
\Delta r_{\mathrm{CSB}} &=& r_{pp} - r_{nn} = 0.10 \pm 0.12\ \mathrm{fm},  \label{eq:8}
\end{eqnarray}
and CIB by 
\begin{eqnarray}
\Delta a_{\mathrm{CIB}} &=&\frac{1}{2} (a_{pp} + a_{nn}) - a_{np} = 5.6\pm 0.6\ \mathrm{fm},  \nonumber \\
\Delta r_{\mathrm{CIB}} &=&\frac{1}{2} (r_{pp} + r_{nn}) - r_{np} = 0.03\pm 0.13\ \mathrm{fm}. \label{eq-data}
\end{eqnarray}
Apparently, to investigate the individual contribution of CIB and CSB, it is essential to examine not only neutron-proton~($np$) scattering but also proton-proton~($pp$) and neutron-neutron~($nn$) scattering.

In the present work, we consider the following term to account for the CIB to the nucleon-nucleon interaction~\cite{NPA747-362,Machleidt:2011zz}
\begin{equation}
\mathcal{O}_{\rm CIB} \propto \frac{e^2}{(4\pi)^2}(\bar{N}\tau_3 N)(\bar{N} \tau_3 N),\label{CIB}
\end{equation}
where $\tau_3$ denotes the third Pauli matrix associated with isospin.
The CIB contact term is characterized by an unknown LEC, $C_{\rm CIB}$, which needs to be determined. Notice that the factor $e^2/(4\pi)^2$ is a constant, as the usual way in the continuum case we incorporate this factor 
into the parameter $C_{\rm CIB}$.  
The leading contribution of CSB to the nucleon-nucleon interaction is accounted by the following term~\cite{vanKolck:1996rm_epsilon}
\begin{equation}
\mathcal{O}_{\mathrm{CSB}} \propto \epsilon M^2_{\pi}(\bar{N}\tau_3 N)(\bar{N} N),\label{CSB}
\end{equation}
where $\epsilon = 1/3$ is a constant. There is another LEC, $C_{\rm CSB}$, associated with the CSB contact term. And, the usual way is to incorporate the constant, $\epsilon M^2_\pi$, into $C_{\rm CSB}$.

Instead of determining directly the two LECs, $C_{\mathrm{CIB}}$ and 
$C_{\mathrm{CSB}}$, we introduce another set of two contact terms for $pp$ and $nn$,  respectively~\cite{arXiv2502.13565}, 
\begin{eqnarray}
V_{CD}^{pp} &=& C_{CD}^{pp}\left(\frac{1 + \tau_1^3}{2}\right)\left(\frac{1 + \tau_2^3}{2}\right)\exp(-{p^\prime}^6/\Lambda^6 - p^6/\Lambda), \nonumber \\
V_{CD}^{nn} &=& C_{CD}^{nn}\left(\frac{1 - \tau_1^3}{2}\right)\left(\frac{1 - \tau_2^3}{2}\right) \exp(-{p^\prime}^6/\Lambda^6 - p^6/\Lambda). \nonumber \\ \label{charged-operators}
\end{eqnarray}
The same regulator as those for the contact terms in Eq.~(\ref{eq:partialwaves}) is applied here. Notice that $C_{CD}^{pp}$ and $C_{CD}^{nn}$ are linear combinations of $C_{\rm CIB}$ and $C_{\rm CIB}$. Thus, once $C_{CD}^{pp}$ and $C_{CD}^{nn}$ are determined, $C_{\rm CIB}$ and $C_{\rm CSB}$ can be solved easily via a linear transformation. 

Another source of isospin breaking is the electromagnetic interaction. For $pp$ we introduce the static Coulomb 
interaction~\cite{Lu:2023jyz,arXiv2502.13565}
\begin{eqnarray}
    V_{\rm cou}^{\Lambda_c}(r)  = \alpha \frac{{\rm erf}(\Lambda_c r/2)}{r}, 
\end{eqnarray}
where $\alpha = 1/137$ is the fine structure constant, and the error function is applied to soft the Coulomb force at short range. $\Lambda_{c} = 300$~MeV is adopted in the calculation. 

Additionally, we also take into account the isospin-breaking effect to  OPEP. 
The mass difference between the charged and neutral pions is
\begin{eqnarray}
    M_{\pi^\pm} - M_{\pi^0} \approx 4.593~\mathrm{MeV},
\end{eqnarray}
which causes the symmetry violation under a rotation in the isospin space. To take into account 
the isospin-breaking effect of the OPEP specifically, instead of applying the isospin factor ${\bm \tau}_1\cdot{\bm \tau}_2$, 
we adopt the following expression for OPEP ~\cite{Walzl:2000cx},
\begin{eqnarray}
V_{1\pi}^{np} &=& -V_{1\pi}(M_{\pi^0})+2(-1)^{I+1}V_{1\pi}(M_{\pi^\pm}), \nonumber \\
V_{1\pi}^{pp} &=& V_{1\pi}^{nn} = V_{1\pi}(M_{\pi^0}),
\end{eqnarray}\label{eq-CD in 1pe}
 where $M_{\pi^0} = 134.977$ MeV is the neutral pion mass while $M_{\pi^\pm} = 139.570$ MeV is the charged pion mass. 
 $I=0/1$ denotes the total isospin of the two-nucleon system, and the function  $V_{1\pi}$ is given by Eq.~(\ref{eq-1pe}). 

To have a clear overview of the potential we are using, we summarize them as follows
\begin{eqnarray}
V^{pp} &=& V_{1\pi}^{\Lambda_\pi}(M_{\pi^0}) + V_{2\pi}^{\Lambda_\pi}(M_\pi) + V_{\mathrm{short-range}}^{\Lambda} + V_{\mathrm{cou}}^{\Lambda_c} \nonumber \\
&+& V_{\mathrm{CD}}^{pp, \Lambda}, \nonumber\\
V^{np} &=& \left(-V_{1\pi}^{\Lambda_\pi}(M_{\pi^0})+2(-1)^{I+1}V_{1\pi}(M_{\pi^\pm}^{\Lambda_\pi})\right)  \nonumber \\
&+& V_{2\pi}^{\Lambda_\pi}(M_\pi) + V_{\rm short-range}^{\Lambda}, \\
V^{nn} &=& V_{1\pi}^{\Lambda_\pi}(M_{\pi^0}) +V_{2\pi}^{\Lambda_\pi}(M_\pi) + V_{\mathrm{short-range}}^{\Lambda} + V_{CD}^{nn, \Lambda}. \nonumber 
\end{eqnarray}

%%%%%%%%%%%%%%%%%%%%%%%%%%%%%%%%%%%%%%%%%%%%%%%%%%%%%%%%%%%%%%%
\section{Scattering Phase Shifts on the Lattice}\label{fitting}
We perform calculations on a cubic lattice with lattice spacing $a$ and $N$ 
lattice sites in each of the three dimensions. $N$ is chosen large enough for the nucleon-nucleon system.
In our calculation, we adopt the periodic boundary condition. 

\subsection{Method to Calculate Phase Shifts}
On the cubic lattice, the symmetry group is the cubic rotational group $O_h^1$, instead of the continuous rotational group ${\rm SO(3)}$.  Generally, the states of the two-nucleon system with different quantum angular momenta are classified according to the 
irreducible representations of the cubic group. The scattering phase shifts can be extracted from the calculated energies  via the famous L\"uscher's formula. However, using the L\"uscher's formula one can only obtain the phase shifts corresponding to the first few energy levels. Alternatively, we follow the methodology outlined in 
Refs.~\cite{PLB760-309, Elhatisari:2016hby_phase,PhysRevC.98.044002, Stapp:1956mz_phase}, in which the scattering phase shifts are extracted from the asymptotic radial wave functions.

We define the basis of the radial wave function on the lattice as 
\begin{eqnarray}
\ket{r, LSJM} = \sum_{L_z(S_z)}\mathcal{C}^{JM}_{LL_z,SS_z}\sum_{\bm{r}} \delta _{|\bm{r}|=r} Y_{LL_z}(\hat{r})\ket{\bm{r}}\chi_{S,S_z}, \nonumber \\
\label{basis}
\end{eqnarray}
with $JM$ the total angular momentum and the third component, respectively. $\mathcal{C}^{JM}_{LL_z,SS_z}$ is the Clebsch-Gordan coefficient while $Y_{LL_z}$ is spherical harmonic function. $\chi_{SS_Z}$ is the internal spin wave function, and ${\bm r}$ is 
the Jacobian coordinate of two nucleons on the lattice. 
Using this basis, Schr\"odinger equation becomes a generalized eigenvalue problem, whose matrix form in each partial wave looks like, 
\begin{eqnarray}
    H R(r) = E N R(r). 
\end{eqnarray}
$N$ is the $n_r\times n_r$ norm matrix and $H$ is the $n_r\times n_r$ Hamiltonian defined in the $|r, LSJM\rangle$ basis, with $n_r$ the number of radial lattice sites.  

For the $np$ or $nn$ system, in the range where two nucleons are well-separated so that they cannot interact with each other, i.e., $V(r) \to 0$,  the Schr\"odinger equation for the radial wave function becomes the spherical Bessel equation,
\begin{equation}
\left [ \frac{d^2}{d\rho^2} + \frac{2}{\rho} \frac{d}{d\rho} + \left ( 1-\frac{L(L+1)}{\rho^2} \right ) \right] R_L(\rho) = 0,
\end{equation}
where $\rho \equiv kr$ and $k \equiv \sqrt{2\mu E}$ with $\mu$ the reduced mass and $E$ the scattering energy of the two-nucleon system. $L$ is the orbital-angular quantum number while $R_L(\rho)$ represents the radial wave function of the $L$-th partial wave. The radial wave function $R_L(\rho)$ has the following asymptotic form, 
\begin{equation}
R_L(\rho) \propto A_L h^-_L(\rho) - B_L h^+_L(\rho), 
\end{equation}
with $h^-_L(\rho)$ and $h^+_L(\rho)$ the respective first and second spherical Bessel functions. We take the general  parametrization for phase shifts, 
\begin{equation}
S_L \equiv e^{2i\delta _L} = \frac{B_L}{A_L},
\end{equation}
with $\delta_L$ the  phase shift of the $L$-th wave.

Different from the $np$ and $nn$ cases, there are long-range electromagnetic interactions between two protons. In the range where the strong 
interaction vanishes but two protons still interact with each other via the long-range Coulomb force, the radial wave function $R_L(\rho)$ has the following asymptotic behavior~\cite{Vincent:1974zz_cou},
\begin{equation}
\label{eq:pp asymptotic}
R_L(\eta, \rho) \propto A_L F^-_L(\eta, \rho) - B_L G^+_L(\eta, \rho),
\end{equation}
where $F^-_L(\eta, \rho)$ and $G^+_L(\eta, \rho)$ are the spherical Coulomb wave functions. Similarly, the phase shifts of $pp$ scattering can be extracted following Eq.~(\ref{eq:pp asymptotic}).

%%%%%%%%%%%%%%%%%%%%%%%%%%%%%%%
\subsection{Fitting Procedure}
As in the continuum case, we determine the LECs by fitting the nucleon-nucleon scattering data, i.e., the phase shifts, $S$-wave scattering lengths and effective ranges.  After decomposing the wave functions into partial waves, we can perform the fit for individual channels separately. For the partial waves except $^1S_0$ we determine directly the LECs by fitting the $np$ scattering phase shifts. However, the $^1S_0$ channel which accounts for the majority of CIB and CSB effects is more complicated. 

For small momentum $p$, the $S$-wave phase shifts of nucleon-nucleon scattering can be expanded as~\cite{Bethe:1949yr_r_a,PPNP63-117},
\begin{equation}
p\cot\delta(p) = -\frac{1}{a} + \frac{1}{2}r p^2 + \mathcal{O}(p^4), 
\label{eq-r-a}
\end{equation}
where the parameters $a$ and $r$ are the scattering length and effective range, respectively. 
For partial wave $^1S_0$, we perform a joint fit. We determine the three LECs, $C_{0,^1S_0}$, $C_{2,^1S_0}$, and $C_{4, ^1S_0}^1$, as well as the two LECs, $C_{CD}^{pp}$ and $C_{CD}^{nn}$ reflecting the CIB and CSB effects, by simultaneously fitting the phase shifts of $np$ and $pp$ in $^1S_0$ channel, and the $S$-wave scattering length $a_{nn}$ for $nn$ scattering.

%%%%%%%%%%%%%%%%%%%%%%%%%%%%%%%%%%%%%%%%%%%%
\section{Numerical Results}\label{numerical}
We first perform calculations without TPEP since it is argued that for low momentum, TPEP is analytical and can be absorbed by redefining the LECs associated with the contact terms. And, this is what it is usually done in the Monte Carlo simulations of the few- and many-body systems within 
NLEFT~\cite{Epelbaum:2009rkz,Epelbaum:2009zsa,EPJA45-335,PhysRevLett.104.142501}. To check the contribution from TPEP, we also perform calculations with specific TPEP included. 

 Given the lattice spacing $a$, the corresponding momentum cutoff of the lattice regularization is $\Lambda_{\mathrm{lat}} = \pi/a$. So for the small lattice spacing $a = (200~\mathrm{MeV})^{-1} = 0.99$ fm, the 
momentum cutoff is $\Lambda_{\mathrm{lat}} = 628$~MeV. To investigate the lattice artifacts arising from the 
non-zero lattice spacing, we choose several lattice spacings, i.e., $a = 1.97,~1.64,~1.32$ and $0.99$~fm. Additionally, we explore the dependence on the momentum cutoff $\Lambda$ introduced for 
the contact terms, by using several values of $\Lambda$, i.e., $\Lambda = 250, 350$ and $450$~MeV.

%%%%%%%%%%%%%%%%%%%%%%%%%%%%%%%%%%%%%%%%%%%%%%%%%%%%
\subsection{$np$ and $pp$ Scattering Phase Shifts}
We present the LECs and corresponding errors associated with the contact terms in Table~\ref{LECs-OPEP}. The errors are extracted from the covariance matrix in the fit. Notice that the LECs are given in lattice units, and classified 
according to partial waves. The original LECs, $B_i, C_i$ and $D_i$, as well as $C_{\rm CIB}$ and $C_{\rm CSB}$, which are not given here, can be obtained 
easily via a linear transformation. 

\begin{figure*}[htp]
\centering
\includegraphics[width=\textwidth]{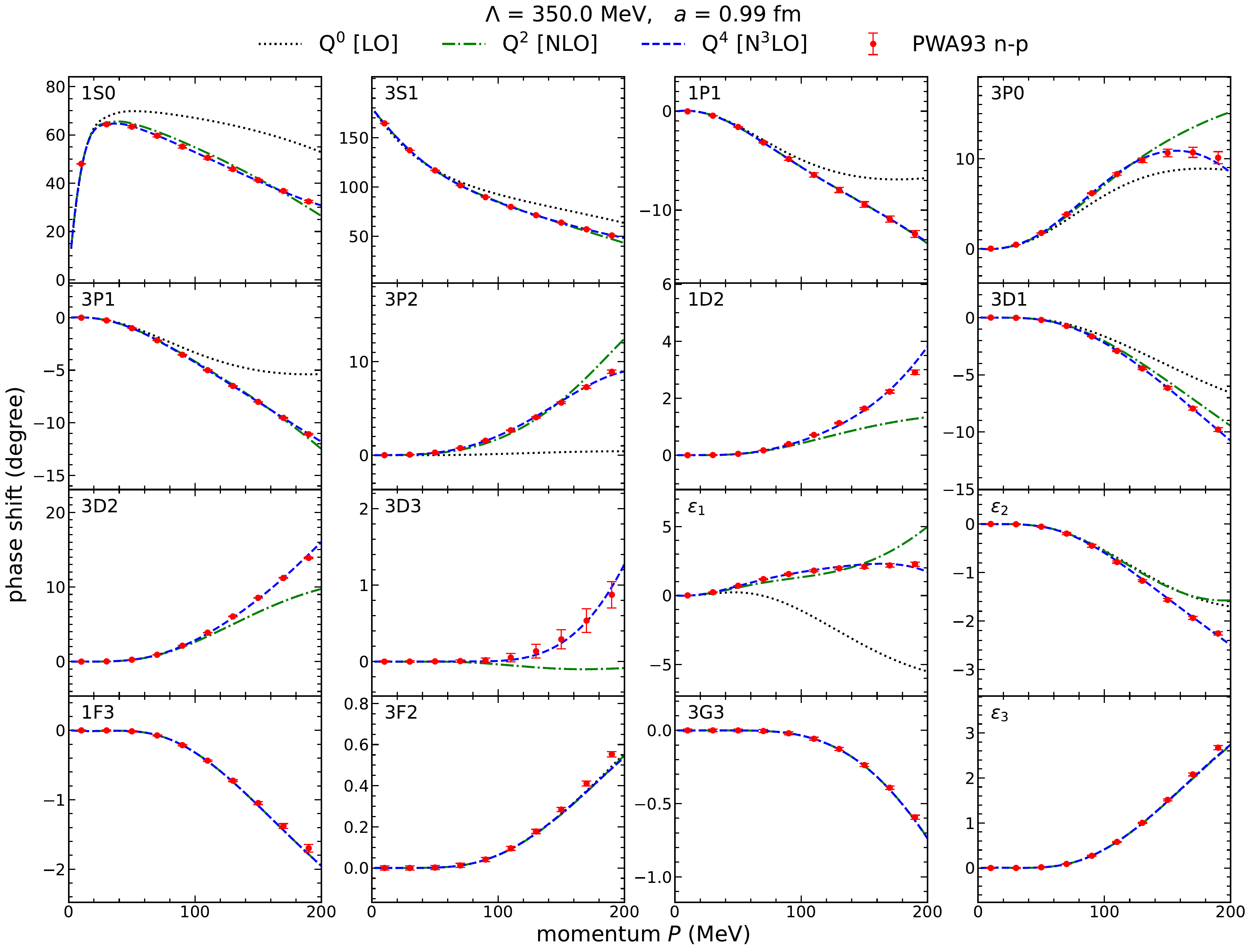} 
\caption{(Color online) Scattering phase shifts and mixing angles \textit{versus} the relative momenta $P$ between two nucleons for  the $np$ systems. The lattice spacing is set to $a = 0.99$~fm, and the momentum cutoff is set to $\Lambda = 350$~MeV. Notice that the TPEP is not included, which is consistent with the Monte Carlo simulations of the few- and many-body systems using NLEFT.}\label{plot-np-0.99}
\end{figure*}

\begin{figure*}[htp]
\centering
\includegraphics[width=0.8\textwidth]{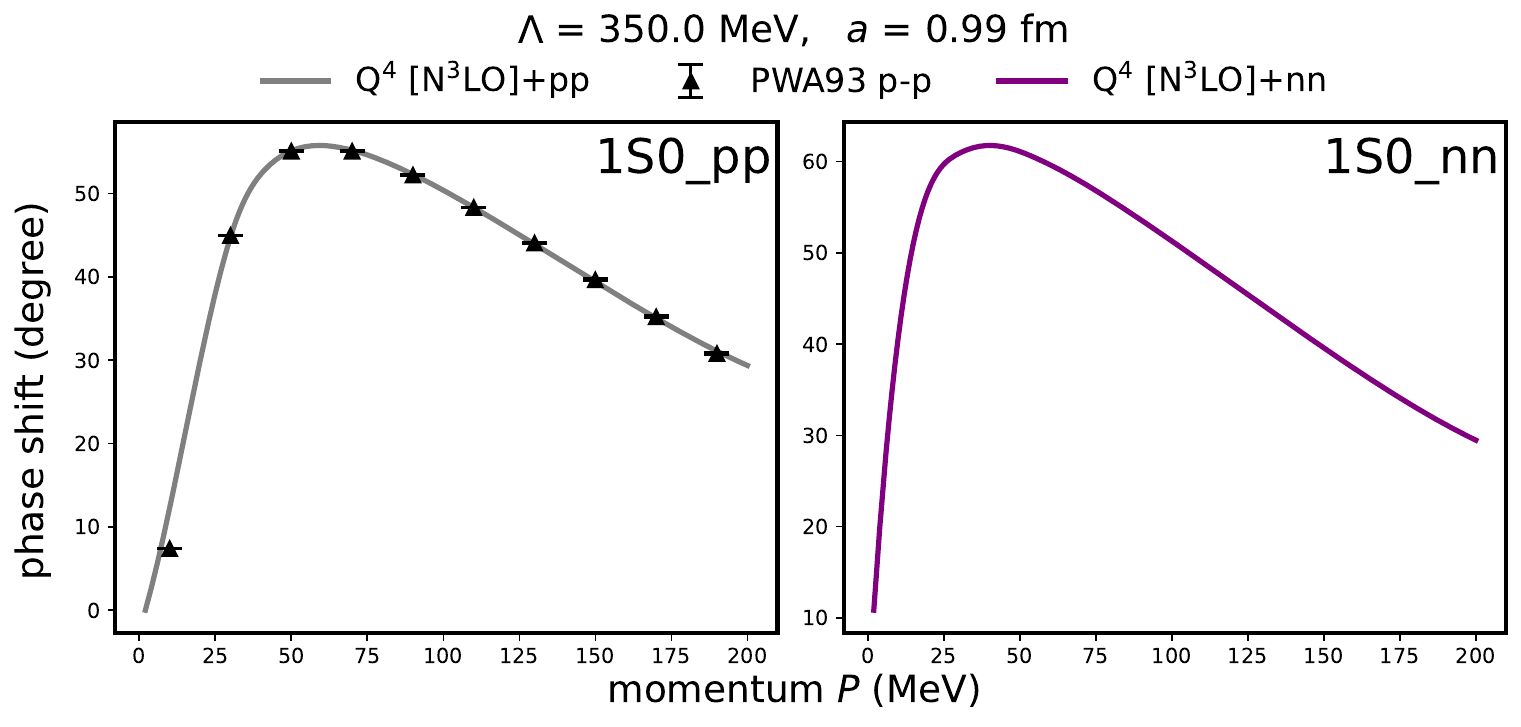} 
\caption{(Color online) $pp$ and $nn$ scattering phase shifts for partial wave $^1S_0$. Notice that the $nn$ scattering phase shifts are not used in the fits, thus the results in right panel should be taken 
as a lattice prediction of the $nn$ phase shifts in partial wave $^1S_0$.}\label{plot-pp-nn-0.99}
\end{figure*}

We show the plots of scattering phase shifts for $np$ in Fig.~\ref{plot-np-0.99}, and those for $pp$ and $nn$ 
in the $^1S_0$ channel in Fig.~\ref{plot-pp-nn-0.99}. Those calculations are performed with a small lattice 
spacing $a = 0.99$~fm and momentum cutoff $\Lambda = 350$~MeV. We emphasize that we only take the $S$-, $P$- and $D$-wave phase shifts, as well as the mixing angles $\epsilon_1$ between $^3S_1$ and $^3D_1$ and $\epsilon_2$ between $^3P_2$ and $^3F_2$ as inputs in the fit. Thus, the results for peripharal $F$ and 
$G$ waves should be viewed as predictions. From the plots, one can clearly see that with the N$^3$LO nuclear force, the $np$ phase shifts in the $S$, $P$, $D$, and even $F$ and $G$ waves can be well described for the relative momentum $p<200$ MeV. Additionally, after taking into account the Coulomb force and charge-dependent contact term $V_{CD}^{pp}$, as well as the isospin breaking term of the OPEP due to the pion mass difference, the phase shifts of $pp$ scattering in the $^1S_0$ channel from Nijmegen partial wave analysis can be accurately reproduced.

\begin{table*}
\renewcommand{\arraystretch}{1.5}
\centering
\caption{LECs and corresponding errors (in lattice units) determined from a N$^3$LO fit to the empirical values of 
$np$ and $pp$ scattering phase shifts, and $nn$ scattering length $a_{nn}$ in the $^1S_0$ partial wave. Four lattice spacings, 
$a = 1.97, 1.64, 1.32$, and $0.99~{\rm fm}$, are used. The momentum cutoff is set to $\Lambda = 350$ MeV. Notice that $C_{CD}^{nn}$ and $C_{CD}^{pp}$ are the same order of $C_{0,^1S_0}$ because we directly use the operators in Eq.~(\ref{charged-operators}) in the fits, which works as LO contact terms.}\label{LECs-OPEP}
\begin{ruledtabular}
\begin{tabular}{c|cccc}
% \hline\hline
 LECs                & $ a = 1.97$ fm  & $a = 1.64$ fm   &  $a = 1.32$ fm & $a = 0.99$ fm   \\   
  \hline
 $C_{0,{}^1S_0}$  &$-1.045\pm0.001     $&$-1.505\pm0.002     $&$-2.351\pm0.003     $&$-4.179\pm0.005     $  \\
 $C_{0,{}^3S_1}$  &$-1.22\pm0.01       $&$-1.61\pm0.02       $&$-2.51\pm0.03       $&$-4.47\pm0.06      $  \\
 \hline
 $C_{2,{}^1S_0}$  &$ 0.1135\pm0.0006   $&$ 0.236\pm0.001     $&$ 0.575\pm0.003     $&$ 1.818\pm0.009     $  \\
 $C_{2,{}^3S_1}$  &$ 0.102\pm0.002     $&$ 0.184\pm0.005     $&$ 0.45\pm0.01       $&$ 1.45\pm0.04      $  \\
 $C_{2,SD}$       &$ 0.084\pm0.002     $&$ 0.182\pm0.004     $&$ 0.444\pm0.009     $&$ 1.40\pm0.03      $  \\
 $C_{2,{}^1P_1}$  &$ 0.027\pm0.003     $&$ 0.057\pm0.006     $&$ 0.14\pm0.01       $&$ 0.45\pm0.04      $  \\
 $C_{2,{}^3P_0}$  &$-0.032\pm0.001     $&$-0.068\pm0.003     $&$-0.167\pm0.008     $&$-0.53\pm0.02      $  \\ 
 $C_{2,{}^3P_1}$  &$ 0.0315\pm0.0005   $&$ 0.0674\pm0.0009   $&$ 0.165\pm0.002     $&$ 0.520\pm0.007     $  \\
 $C_{2,{}^3P_2}$  &$-0.0436\pm0.0007   $&$-0.089\pm0.001     $&$-0.219\pm0.003     $&$-0.69\pm0.01      $  \\
 \hline
 $C_{4,{}^1S_0}$  &$-0.0176\pm0.0002   $&$-0.0529\pm0.0007   $&$-0.202\pm0.002     $&$-1.13\pm0.01      $  \\
 $C_{4,{}^3S_1}$  &$-0.0150\pm0.0005   $&$-0.039\pm0.002     $&$-0.153\pm0.006     $&$-0.87\pm0.03      $  \\ 
 $C_{4,SD}$       &$-0.0084\pm0.0004   $&$-0.029\pm0.001     $&$-0.111\pm0.004     $&$-0.62\pm0.02      $  \\ 
 $C_{4,{}^1P_1}$  &$ 0.0000\pm0.0007   $&$ 0.000\pm0.002     $&$-0.001\pm0.007     $&$-0.01\pm0.04      $  \\
 $C_{4,{}^3P_0}$  &$ 0.0035\pm0.0004   $&$ 0.012\pm0.001     $&$ 0.045\pm0.005     $&$ 0.26\pm0.03      $  \\ 
 $C_{4,{}^3P_1}$  &$-0.0009\pm0.0001   $&$-0.0032\pm0.0004   $&$-0.012\pm0.001     $&$-0.067\pm0.008    $  \\ 
 $C_{4,{}^3P_2}$  &$ 0.0026\pm0.0001   $&$ 0.0070\pm0.0004   $&$ 0.027\pm0.001     $&$ 0.149\pm0.008     $  \\ 
 $C_{4,PF}$       &$-0.00056\pm0.00003 $&$-0.00197\pm0.00008 $&$-0.0074\pm0.0003   $&$-0.042\pm0.002    $  \\ 
 $C_{4,{}^1D_2}$  &$-0.00169\pm0.00004 $&$-0.0051\pm0.0001   $&$-0.0193\pm0.0004   $&$-0.108\pm0.002     $  \\
 $C_{4,{}^3D_1}$  &$ 0.0007\pm0.0002   $&$ 0.0020\pm0.0007   $&$ 0.006\pm0.002     $&$ 0.037\pm0.014    $  \\ 
 $C_{4,{}^3D_2}$  &$-0.00321\pm0.00002 $&$-0.00893\pm0.00007 $&$-0.0338\pm0.0003   $&$-0.190\pm0.001     $  \\ 
 $C_{4,{}^3D_3}$  &$-0.0003\pm0.0001   $&$-0.0028\pm0.0003   $&$-0.011\pm0.001     $&$-0.065\pm0.008    $  \\
 \hline 
 $C_{CD}^{pp}$&$-1.010\pm0.001  $&$-1.452\pm0.002    $&$-2.269\pm0.003     $&$-4.034\pm0.005   $  \\ 
 $C_{CD}^{nn}$&$-1.028\pm0.001  $&$-1.479\pm0.002    $&$-2.311\pm0.003     $&$-4.108 \pm0.005    $  \\
 \end{tabular}  
 \end{ruledtabular}
\end{table*}

%%%%%%%%%%%%%%%%%%%%%%%%%%%%%%%%%%%%%%%%%%
\subsection{Dependence on Lattice Spacing}
\begin{figure*}
\centering
\includegraphics[width=\textwidth]{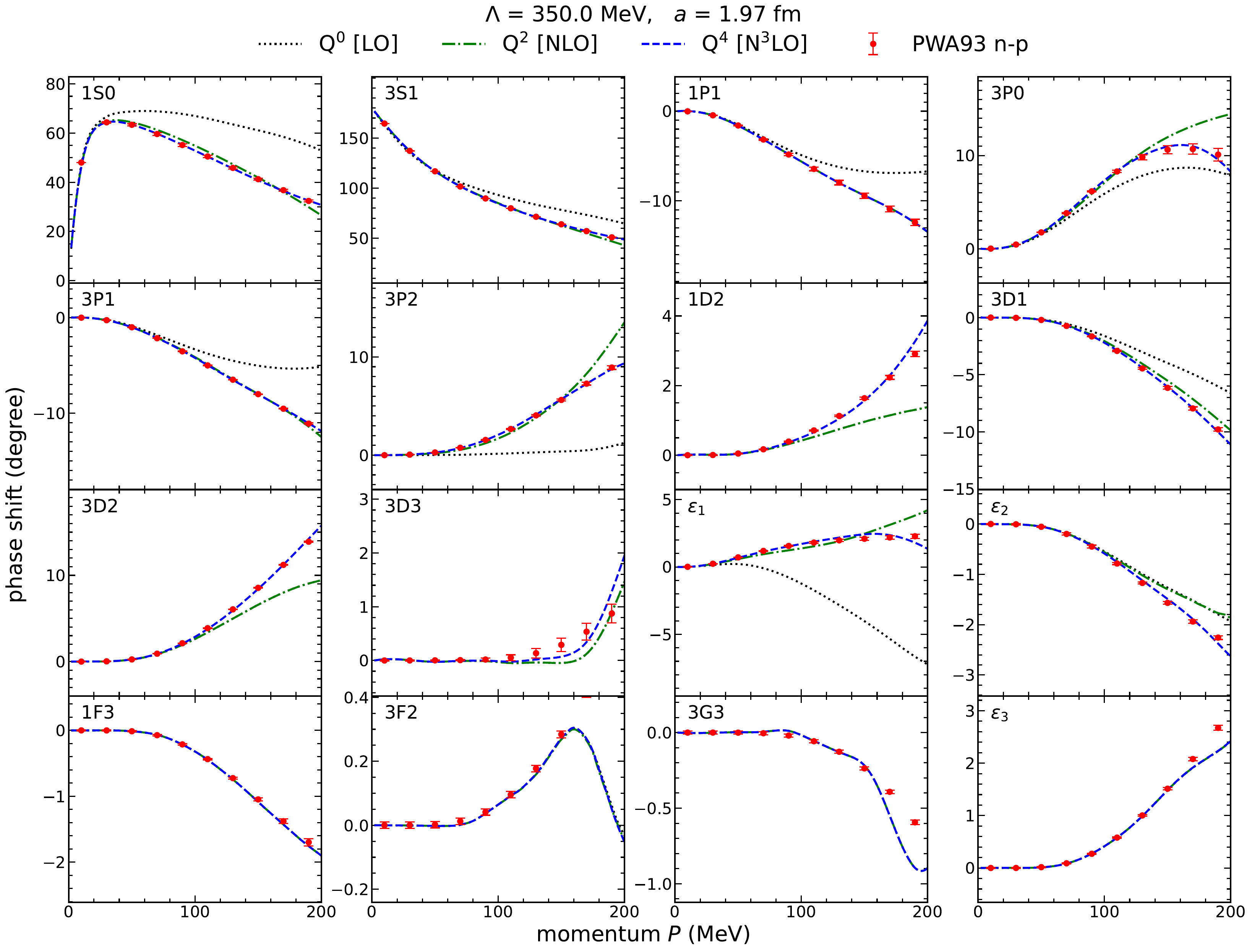} % Here is how to import EPS art
\caption{(Color online) $np$ scattering phase shifts and mixing angles \textit{versus} relative momentum $P$ between neutron and proton, calculated with a lattice spacing of $a = 1.97~{\rm fm}$. The momentum cutoff is fixed as $\Lambda = 350$ MeV.}\label{a1.97}
\end{figure*}

\begin{figure*}
\centering
\includegraphics[width=\textwidth]{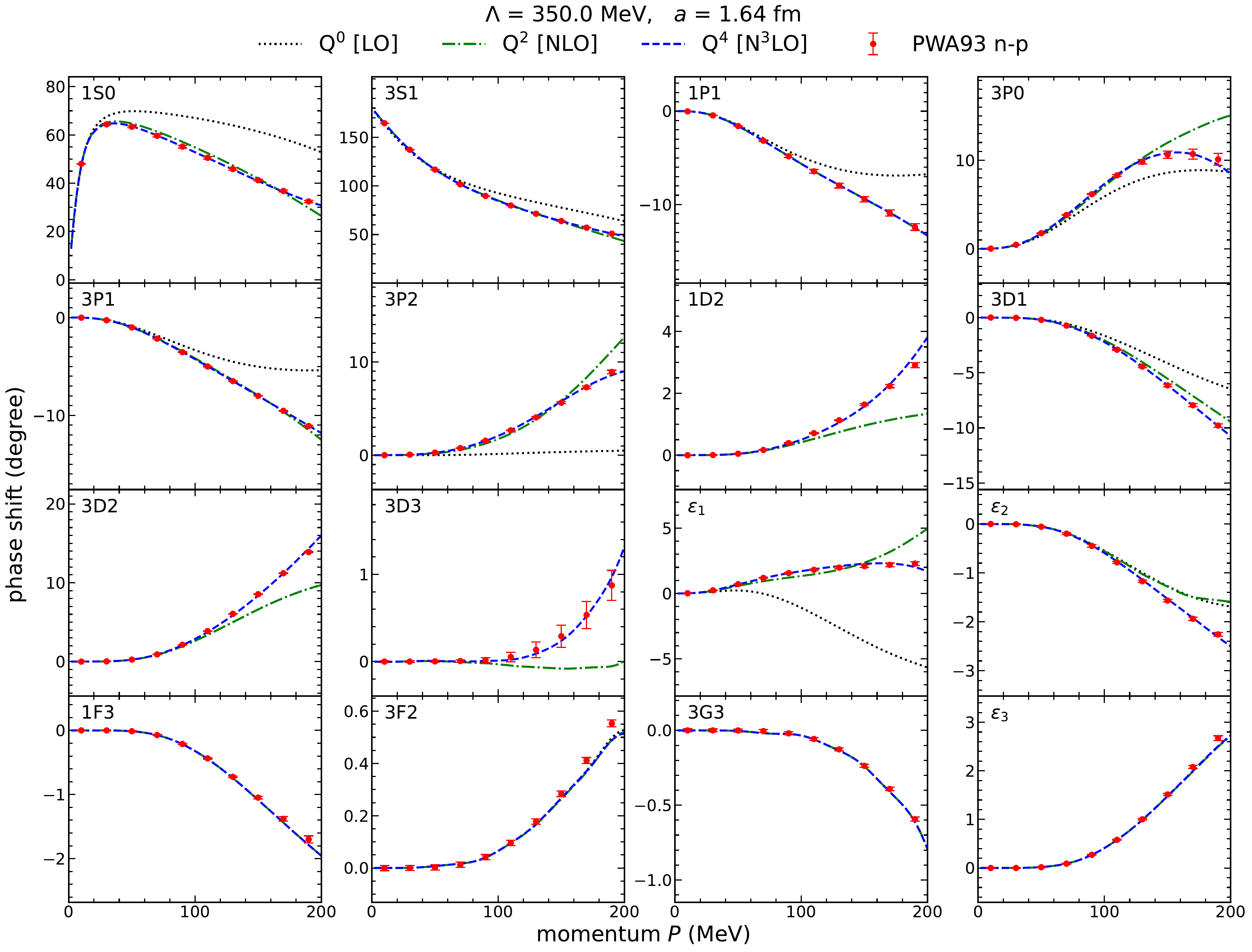} % Here is how to import EPS art
\caption{(Color online) $np$ phase shifts and mixing angles \textit{versus} relative momentum $P$ between neutron and proton, 
calculated with a lattice spacing of $a = 1.64~{\rm fm}$. The momentum cutoff is fixed as $\Lambda = 350$ MeV. }\label{a1.64}
\end{figure*}

\begin{figure*}
\centering
\includegraphics[width=\textwidth]{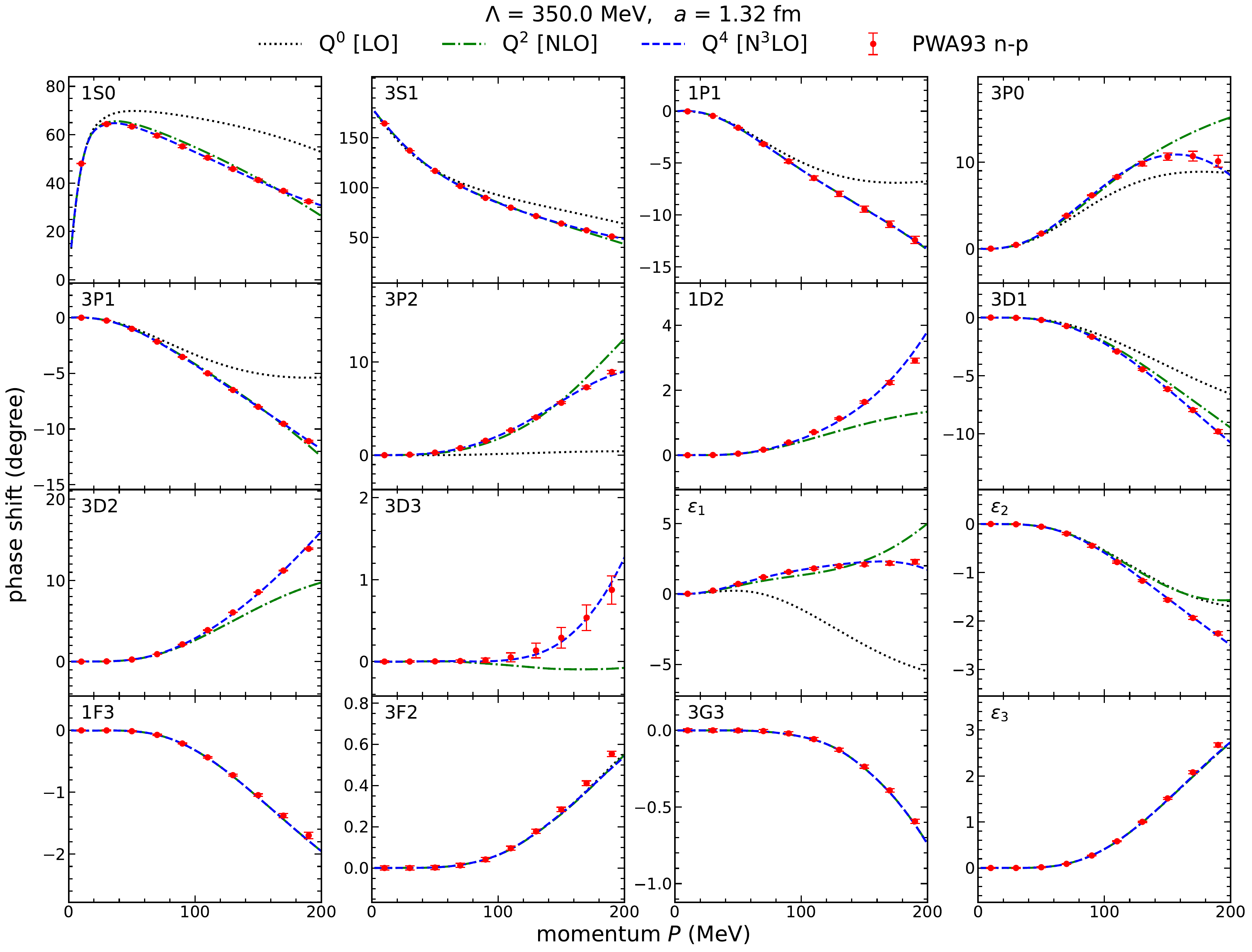} % Here is how to import EPS art
\caption{(Color online) $np$ phase shifts and mixing angles \textit{versus} the relative momentum $P$ between neutron and proton, 
calculated with a lattice spacing of $a = 1.32~{\rm fm}$. The momentum cutoff is fixed as $\Lambda = 350$ MeV. } \label{a1.32}
\end{figure*}

As is known that for a lattice calculation the use of nonzero lattice spacings will induce lattice artifacts~\cite{Wilson:1974sk,Kogut:1974ag}. However, 
it is impossible to let the lattice spacing go to zero due to the huge requirement of computing resources.
Alternatively, the usual way to obtain the results in the continuum limit is to perform calculations with several different lattice spacings, then extrapolate the results to the limit of zero lattice spacing.  
In the present work, we perform calculations with four different lattice spacings, i.e., $a = 0.99, 1.32, 1.64$
and $1.97$~fm, to explore dependence on the lattice spacings. In addition to the results given in Fig.\ref{plot-np-0.99}, we also show the plots of $np$ phase shifts in Figs.~\ref{a1.97}, \ref{a1.64}, and \ref{a1.32} for  $a=1.97, 1.64$, and $1.32$~fm, respectively.  

From the plots, one can clearly see that using the coarsest lattice with $ a= 1.97$~fm, the low partial waves, such as $S$ and $P$ waves, can be described pretty well for relative momentum $p < 200$~MeV. However, for higher partial waves, such as $^3D_3$, $^3F_2$, and $^3G_3$, it is quite difficult to reproduce the empirical phase shifts for relative momentum beyond $150$~MeV, which is mainly because the used lattice spacing is big. After decreasing the lattice spacing to $1.64$~fm, all of the $S$, $P$, $D$ and $F$ waves can be described very well for relative momentum $p<200$~MeV. If we further decrease the lattice spacing to $1.32$~fm and then to $0.99$~fm, the changes are almost invisible from the plots. This indicates that the lattice artifacts for $a=1.32$ fm and $0.99$ fm are small, which is suitable for precise calculations of low-energy nuclear physics with NLEFT.

%%%%%%%%%%%%%%%%%%%%%%%%%%%%%%%%%%%%%%%%%%%%%%%%%%%%%%%%%%
\subsection{Deuteron Properties and $S$-wave Parameters}

To validate the interaction, we also calculate the deuteron properties and S-wave parameters. 
Notice that we have not included any information of the deuteron in the fits. Thus, all the properties of the deuteron calculated in the present work should be taken as predictions. 
 
As a bound state formed by a proton and a neutron, deuteron is a mixing state of $^3S_1$ and $^3D_1$. 
In the range where the proton and neutron are well-separated, the asymptotic behavior of the deuteron radial wave function in the $^3S_1$ channel is given by~\cite{NPA747-362}
\begin{equation}
u(r) = A_S e^{-\gamma r} \label{asymptotic-s},
\end{equation}
with $A_S$ the normalization constant. $\gamma$ is defined as $\sqrt{m |E_d|}$ with $E_d$ the binding energy of the deuteron. In the $^3D_1$ channel, the normalized asymptotic form of the radial wave function
is 
\begin{equation}
w(r) = \zeta A_S \left [ 1+ \frac{3}{\gamma r} +\frac{3}{(\gamma r)^2} \right ] e^{-\gamma r}\label{asymptotic-d}
\end{equation}
with $\zeta$ the ratio of the normalization coefficients between $^3D_1$ and $^3S_1$. 

The root-mean-square radius and quadrupole momentum are also important physical quantities of deuteron, which are defined as 
\begin{equation}
r_d = \frac{1}{2}\left[\int dr r^2 [u^2(r) + w^2(r)]\right]^{1/2},
\end{equation}
and 
\begin{equation}
Q_d = \frac{1}{20} \int dr r^2  w^2(r) \left[\sqrt{8}u(r)-w(r) \right].
\end{equation}
Additionally, we also calculate the S-wave parameters, i.e., the scattering length $a_{^3S_1}$ and $a_{^1S_0}$
and effective range $r_{^3S_1}$ and $r_{^1S_0}$ for the $^3S_1$ and $^1S_0$ channels, respectively. 

In Table~\ref{property:deuteron}, we summarize the numerical results of deuteron properties and S-wave parameters, i.e., $E_d$, $A_S$, $\zeta$, $Q_d$ and $r_d$ for the deuteron, and $a_{^3S_1}$, $a_{^1S_0}$, $r_{^3S_1}$, and $r_{^1S_0}$ for $np$ scattering. 
In the calculation, a small lattice spacing $a = 0.99$~fm and momentum cutoff $\Lambda = 350$~MeV are used. It is clear that all the results for the deuteron and S-wave parameters, except $\zeta$, are consistent with the empirical values. $\zeta$ is slightly smaller than the empirical value. Given that we do not take any information of the deuteron in 
the fits, this is not surprising. We also check that once the deuteron binding energy is included in the fits, the result of $\zeta$ is very close to the empirical value. 

\begin{table*}
\renewcommand{\arraystretch}{1.5}
\centering
\caption{Deuteron properties and $S$-wave parameters calculated with the momentum cutoff $\Lambda =350$ MeV. Notice that the deuteron binding energy is not used as input 
in the fits.} \label{property:deuteron}  
\begin{ruledtabular}
\begin{tabular}{c|ccccc}
                           & $a = 1.97$~fm    & $a = 1.64$~fm    & $a= 1.32$~fm   & $a = 0.99$~fm  & Empirical \\
\hline
$E_d$ (MeV)                &$ 2.26062\pm0.00004 $   &$ 2.21354\pm0.00004 $  &$ 2.21407\pm0.00003 $  &$ 2.21595\pm0.00002$              & 2.224575(9)~\cite{VanDerLeun:1982bhg_D24} \\ 
$A_s(\mathrm{fm}^{-1/2})$  &$ 0.882\pm0.001$      &$ 0.873\pm0.001 $    &$ 0.872\pm0.001 $    &$ 0.872\pm0.001$     & 0.8846(9)~\cite{Ericson:1982ei_D25}       \\
$\zeta$                    &$ 0.02220\pm0.00008$   &$ 0.02189\pm0.00009$  &$ 0.02138\pm0.00009$  &$ 0.02170\pm0.00009$    & 0.0256(4)~\cite{Rodning:1990zz_D26}      \\
$Q_d(\mathrm{fm}^2)$       &$ 0.271\pm0.002 $     &$ 0.275\pm0.002$     &$ 0.276\pm0.002 $    &$ 0.275\pm0.002 $             & 0.2859(3)~\cite{Bishop:1979zz_D27}       \\
$r_d$ (fm)                 &$ 1.9663\pm0.0009 $    &$ 1.9785\pm0.0009 $   &$ 1.9774\pm0.0009$    &$ 1.9733\pm0.0009 $             & 1.97535(85)~\cite{Huber:1998zz_D28}      \\
$a_{^3S_1}$                &$ 5.408\pm0.003$      &$ 5.420\pm0.003 $    &$ 5.421\pm0.003$     &$ 5.421\pm0.003 $    & 5.424(4)~\cite{Dumbrajs:1983jd_D29}      \\
$r_{^3S_1}$                &$ 1.829\pm0.008$       &$ 1.746\pm0.003 $    &$ 1.747\pm0.002 $    &$ 1.746\pm0.002 $    & 1.759(5)~\cite{Dumbrajs:1983jd_D29}       \\
$a_{^1S_0}$                &$-23.69\pm0.06 $     &$-23.71\pm0.06 $    &$ -23.71\pm0.06$    &$ -23.71\pm0.06 $  &-23.748(10)~\cite{Dumbrajs:1983jd_D29}  \\
$r_{^1S_0}$                &$ 2.643\pm0.002 $     &$ 2.646\pm0.002 $    &$ 2.649\pm0.002 $    &$ 2.650\pm0.002$    & 2.75(5)~\cite{Dumbrajs:1983jd_D29}     \\
\end{tabular}
\end{ruledtabular} 
\end{table*}

%%%%%%%%%%%%%%%%%%%%%%%%%%%%%%%%%%%%%%%%%%%%%%%
\subsection{Dependence on the Momentum Cutoff}

\begin{figure*}
\centering
\includegraphics[width=\textwidth]{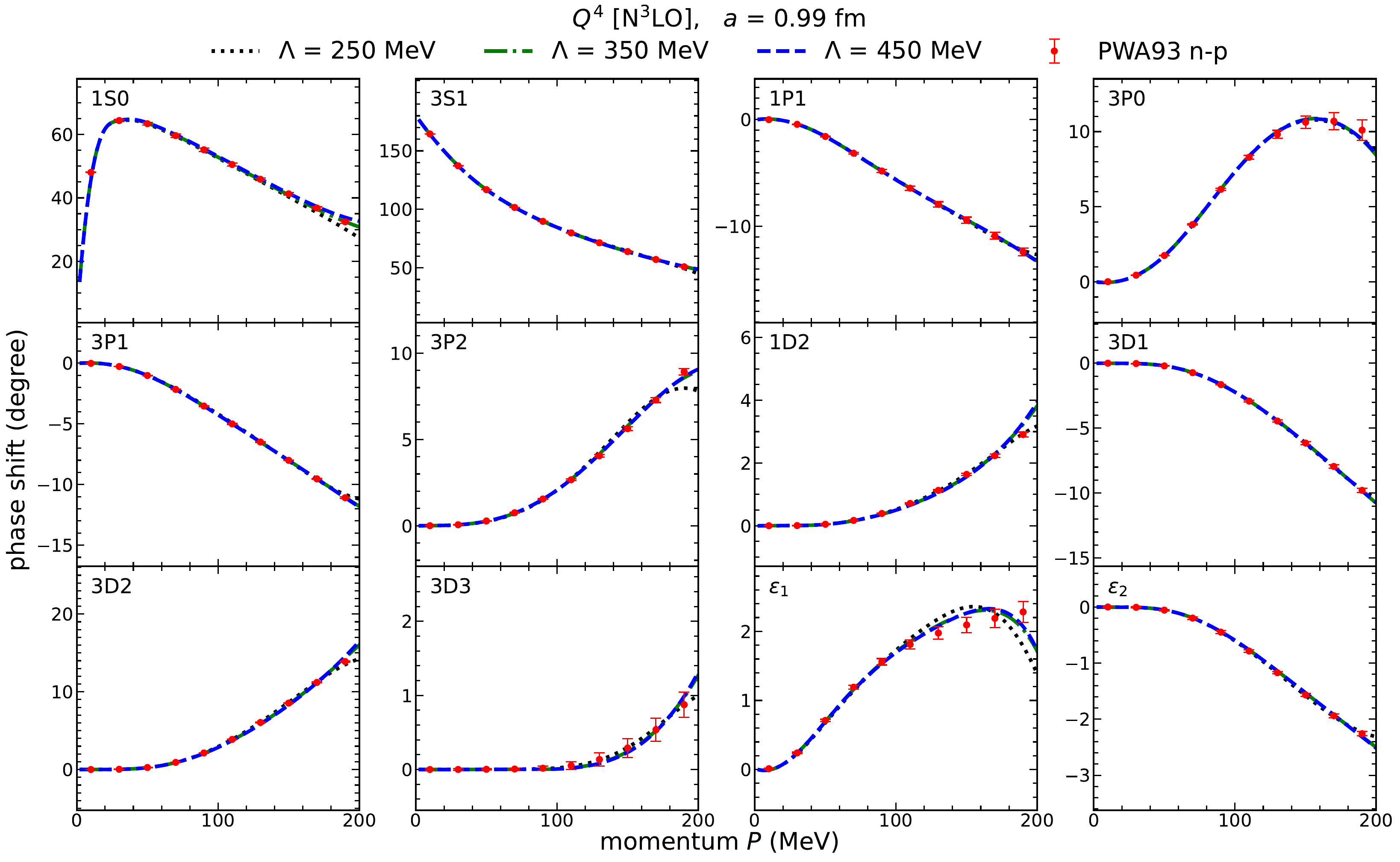} 
\caption{(Color online) Scattering phase shifts and mixing angles \textit{versus} relative momenta $P$ for momentum cutoff
ranging from $250$ to $450$~MeV, i.e., $\Lambda = 250, ~350$ and $450$~MeV. A small lattice spacing $a = 0.99~{\rm fm}$ is used in the 
calculation. } \label{diff_lam}
\end{figure*}

As discussed in section~\ref{NN-interaction}, we introduce a regulator $\exp(-p^6/(2\Lambda^6))$ for each single nucleon, 
with $\Lambda$ the momentum cutoff. To investigate the dependence on $\Lambda$, we perform calculations 
with several cutoffs, i.e., $\Lambda = 250, 350$, and $450$~MeV. In the calculation, the lattice spacing is fixed as $a = 0.99$~fm.
In Fig.~\ref{diff_lam}, we show the plots of the $np$ scattering phase shifts with different cutoffs. One can see that 
except for some channels such as $^1S_0$, $^3P2$, and $\epsilon_1$, 
the results with different cutoffs are consistent with each other, and match the results from Nijmegan partial wave analysis for 
relative momentum $p<200$~MeV. For channels $^1S_0$, $^3P_2$ and $\epsilon_1$, the result of $\Lambda = 250$~MeV has some discrepancies from others for the relative momentum $p$ around $200$~MeV. The difference between $\Lambda=350$ and 450 MeV, however, is quite small, indicting that the dependence on the momentum cutoff is negligible for  $350<\Lambda<450$~MeV.  

%%%%%%%%%%%%%%%%%%%%%%%%%%%%%%%%%%%%%%%%%%%%%%%%%%%%%%%%%%%%%%%%%%%%
\subsection{Contribution of Two-pion-exchange Potential}\label{sec-TPEP}

\begin{figure*}
\centering
\includegraphics[width=\textwidth]{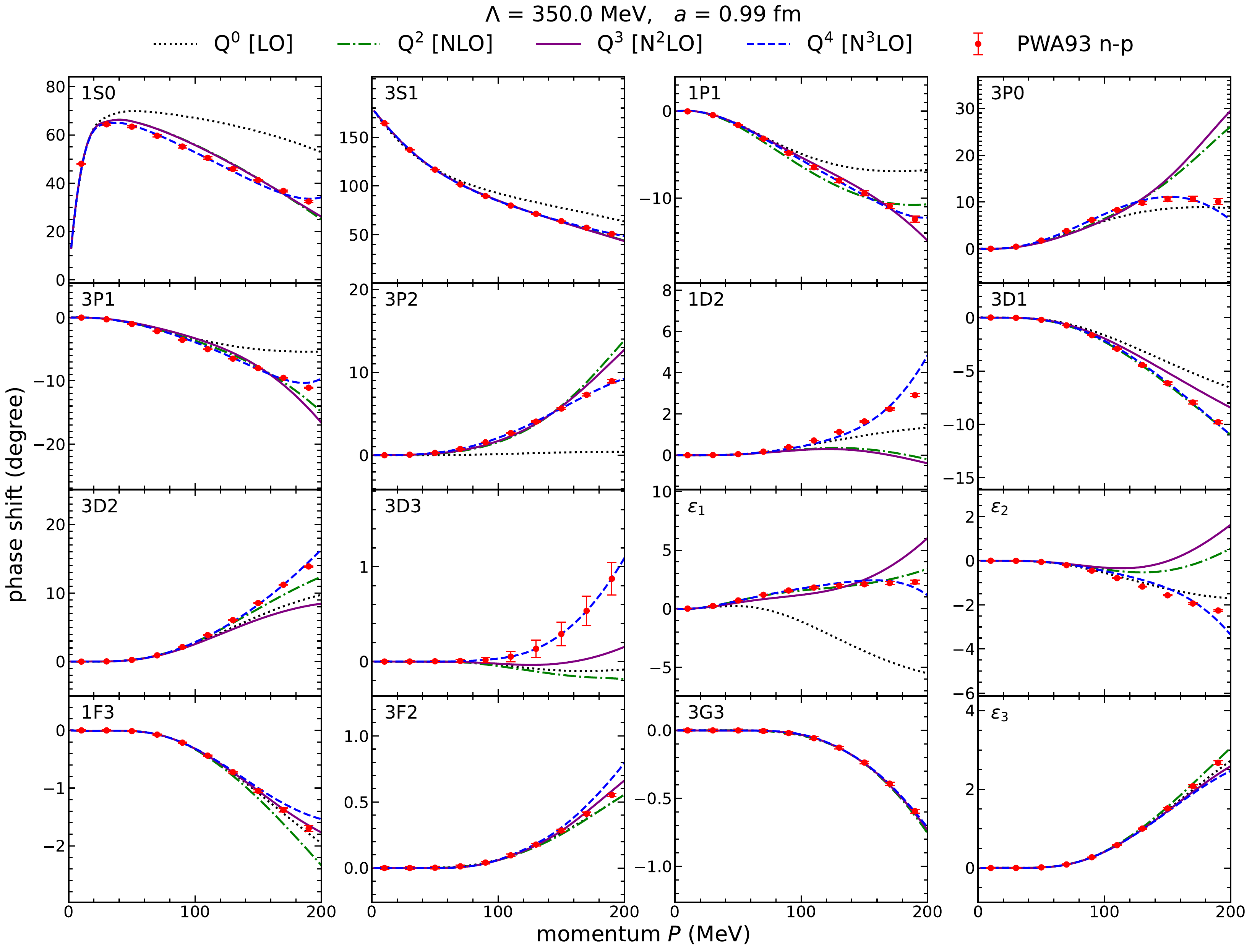} 
\caption{(Color online) The $np$ scattering phase shifts and mixing angles \textit{versus} the relative momenta $P$ between the neutron and proton. The lattice spacing is $a = 0.99~{\rm fm}$ while the momentum cutoff is fixed at  $\Lambda = 350$ MeV. The TPEP is included specifically. } \label{TPEP-0.99} 
\end{figure*}

In NLEFT simulations of the few- and many-body systems, such atomic nuclei, nuclear matter and neutron matter,
the TPEP were not included specifically with the argument that for low momentum, TPEP is analytical and can be absorbed by redefining the LECs associated with the contact terms. In this section, we investigate the contribution of TPEP by comparing the results with and without TPEP. Thus, we repeat the fits with TPEP included specifically. We add TPEP order by order, i.e., the leading TPEP at NLO, the subleading TPEP at N$^2$LO, and next-to-subleading TPEP at 
N$^3$LO. We summarize the LECs in Table~\ref{LECs-TPEP}, and show the plots of the $np$ phase shifts in Fig.~\ref{TPEP-0.99}. 

By comparing the plots in Figs.~\ref{plot-np-0.99} and \ref{TPEP-0.99}, one can see clearly that the results with TPEP are very similar to those without TPEP, except that there are some minor discrepancies for relative momentum $p$ around $200$~MeV for some channels, such as $^1S_0, ^1P_1, ^3P_0, ^3P_1$ and $\epsilon_2$. Therefore, for low-momentum physics around $p= M_\pi$, the major effects of TPEP can be absorbed by redefining the LECs, leaving the contribution of TPEP almost negligible, which is consistent with the statements in Ref.~\cite{Epelbaum:2009rkz, Epelbaum:2009zsa}.  

\begin{table*}
\renewcommand{\arraystretch}{1.5}
\centering
\caption{Low-energy constants determined by N$^3$LO fits using a lattice spacing,  $a = 0.99~{\rm fm}$. The momentum cutoff is set to $\Lambda = 350$ MeV. TPEP has been included.}\label{LECs-TPEP}
\begin{ruledtabular}
\begin{tabular}{c|cccc}
% \hline\hline
 LECs            & LO~($Q^0$)  & NLO~($Q^2$)    & N$^2$LO~($Q^3$) & N$^3$LO~($Q^4$)    \\   
  \hline
 $C_{0,{}^1S_0}$ &$-2.5119\pm0.0004    $&$-2.55\pm0.01      $&$-0.85\pm0.01      $&$ -0.286\pm0.008  $  \\
 $C_{0,{}^3S_1}$ &$-3.2652\pm0.0004    $&$-4.52\pm0.01      $&$-4.23\pm0.01      $&$ -2.79\pm0.06  $  \\
 \hline
 $C_{2,{}^1S_0}$ &$-                   $&$ 1.26\pm0.01      $&$ 1.30\pm0.01      $&$  2.30\pm0.01   $  \\
 $C_{2,{}^3S_1}$ &$-                   $&$ 0.807\pm0.005    $&$ 1.042\pm0.005    $&$  1.04\pm0.04   $  \\
 $C_{2,SD}$      &$-                   $&$ 0.427\pm0.007    $&$ 0.849\pm0.007    $&$  2.341\pm0.03   $  \\
 $C_{2,{}^1P_1}$ &$-                   $&$ 0.008\pm0.017    $&$ 0.82\pm0.02      $&$  1.70\pm0.04   $  \\
 $C_{2,{}^3P_0}$ &$-                   $&$-1.248\pm0.007    $&$-1.569\pm0.007    $&$ -2.60\pm0.02   $  \\ 
 $C_{2,{}^3P_1}$ &$-                   $&$ 1.095\pm0.003    $&$ 1.691\pm0.004    $&$  2.7136\pm0.007  $  \\
 $C_{2,{}^3P_2}$ &$-                   $&$-0.721\pm0.003    $&$-0.621\pm0.003    $&$ -0.56\pm0.01   $  \\
 \hline
 $C_{4,{}^1S_0}$ &$-                   $&$-                 $&$-                 $&$ -0.99\pm0.01   $  \\
 $C_{4,{}^3S_1}$ &$-                   $&$-                 $&$-                 $&$ -0.61\pm0.04   $  \\ 
 $C_{4,SD}$      &$-                   $&$-                 $&$-                 $&$ -1.05\pm0.02   $  \\ 
 $C_{4,{}^1P_1}$ &$-                   $&$-                 $&$-                 $&$ -0.505\pm0.03   $  \\
 $C_{4,{}^3P_0}$ &$-                   $&$-                 $&$-                 $&$  1.01\pm0.04   $  \\ 
 $C_{4,{}^3P_1}$ &$-                   $&$-                 $&$-                 $&$ -0.761\pm0.005  $  \\ 
 $C_{4,{}^3P_2}$ &$-                   $&$-                 $&$-                 $&$  0.082\pm0.008 $  \\ 
 $C_{4,PF}$      &$-                   $&$-                 $&$-                 $&$ -0.243\pm0.002  $  \\ 
 $C_{4,{}^1D_2}$ &$-                   $&$-                 $&$-                 $&$ -0.211\pm0.003 $  \\
 $C_{4,{}^3D_1}$ &$-                   $&$-                 $&$-                 $&$  0.187\pm0.01   $  \\ 
 $C_{4,{}^3D_2}$ &$-                   $&$-                 $&$-                 $&$ -0.262\pm0.001  $  \\ 
 $C_{4,{}^3D_3}$ &$-                   $&$-                 $&$-                 $&$ -0.010\pm0.007  $  \\
 \hline
 $C_{CD}^{pp}$ &$-             $&$-                 $&$-                 $&$ -0.151\pm0.008  $  \\ 
 $C_{CD}^{nn}$ &$-             $&$-                 $&$-                 $&$ -0.224\pm0.008  $  \\
 \end{tabular}  
 \end{ruledtabular}
\end{table*}

%%%%%%%%%%%%%%%%%%%%%%%%%%%%%%%%%%%%%%%%%%%%%%%%%%
\section{Summary and Outlook} \label{conclusion}

In \cite{PhysRevC.98.044002}, we investigated the nucleon-nucleon interaction at N$^3$LO within NLEFT. However, we did not take into account the CIB and CSB effect. To provide high-fidelity nuclear forces which will be applied in NLEFT, we revisit the nucleon-nucleon interaction. We perform calculations on a cubic lattice with spatial lattice spacing $a$, and $N$ lattice points in each of the three dimensions. The periodic boundary condition is adopted. 

Compared to previous calculations in Ref.\cite{PhysRevC.98.044002}, we additionally take into account the CIB and CSB effects. Specifically, we consider the isospin breaking arising from the mass difference between the neutral and charged pions in OPEP, the electromagnetic interaction for $pp$ and two additional contact terms $\mathcal{O}_{CIB}$ and $\mathcal{O}_{CSB}$. We can generate accurately the 
$np$ scattering phase shifts for $S$, $P$, and $D$ waves, and $pp$ phase shifts in the $^1S_0$ partial wave, for relative momentum 
$p <200$~MeV. Additionally, we can reproduce, with high precision, the scattering length $a_{nn}$ and effective range $r_{nn}$ 
for $nn$ scattering in the $^1S_0$ partial wave. The deuteron properties can also be described pretty well. 

For a lattice calculation, the application of nonzero lattice spacing induces lattice artifacts.  
In the present work, we investigate the lattice artifacts by performing calculations using four different lattice spacings, i.e., $a = 1.97, ~1.64,~1.32$ and $0.99$~fm. 
We find that for the coarsest lattice, $a = 1.97$ fm, the $np$ phase shifts of $S$ and $P$ waves can be described pretty well for 
relative momentum $p<200$~MeV. However, for some higher partial waves, such as $^3D_3$, $^3F_2$, and $3G_3$, there are non-negligible 
discrepancy for $p>150$ MeV. Decreasing the lattice spacing to $1.64$ fm, $np$ phase shifts of $S$, $P$, and $D$ waves can be well described in the range $p<200$~MeV. Further decreasing the lattice spacing to $1.32$~fm, and then to $0.99$~fm, 
the results are stable and converge to the empirical values for all $S$, $P$, and $D$ waves for relative momentum $p<200$~MeV. 
This means that lattice spacings of $1.32$ and $0.99$~fm are small enough for the calculation of low-energy nuclear physics.  
 
We regulate the contact terms  nonlocally by introducing a regulator $\exp(-p^6/(2\Lambda^6))$ for each nucleon, with $\Lambda$ a momentum 
cutoff.  We also explore the dependence on $\Lambda$ by performing calculations with different cutoffs ranging from $250$ to $450$ MeV, 
i.e., $\Lambda = 250, 350$, and $450$~MeV. We find that the dependence on the momentum cutoff is negligible for $350 <\Lambda< 450$~MeV, 
whereas the results with $\Lambda = 250$~MeV have some discrepancies from those of the other two when $p$ is round 200~MeV. 
The long-range nuclear force arises from OPEP as well as TPEP up to N$^3$LO. It is argued that for low-energy nuclear physics TPEP is analytical and can be absorbed by redefining the LECs associated with the contact terms. We also investigate the contribution of  TPEP 
by comparing the results with and without specific TPEP. We found that when the lattice spacing is fixed at $0.99$~fm, both results almost match each other,  except for small discrepancies  around $p=200$~MeV. 

To draw a conclusion, we present a high-fidelity two-body nuclear force at N$^3$LO within NLEFT, with which one can precisely generate the $np$, $pp$ and $nn$ scattering data.  In the future, we will apply this interaction to the 
many-body simulations using NLEFT, such as the Hoyle state, as well as structures and properties of $^{16}$O.

\section*{acknowledgments}
The authors acknowledge the members of the international NLEFT collaboration for helpful discussions and suggestions. This work is supported by the Guangdong Basic and Applied Basic Research Foundation (2023A1515011704), and the National Key Research and Development Program of China (No.2023YFA1606000). 
B.L. was supported by 
 NSAF No.U2330401 and National
 Natural Science Foundation of China with Grant No.12275259.
 
 \section*{DATA AVAILABILITY STATEMENT}
 The data supporting the findings of this study is available on Renodo repository~\cite{dataset_citation} or can be obtained directly from the authors upon reasonable request.

\bibliography{Charged_NNforce.bib}% Produces the bibliography via BibTeX.

\providecommand{\noopsort}[1]{}\providecommand{\singleletter}[1]{#1}%
\begin{thebibliography}{111}%
\makeatletter
\providecommand \@ifxundefined [1]{%
 \@ifx{#1\undefined}
}%
\providecommand \@ifnum [1]{%
 \ifnum #1\expandafter \@firstoftwo
 \else \expandafter \@secondoftwo
 \fi
}%
\providecommand \@ifx [1]{%
 \ifx #1\expandafter \@firstoftwo
 \else \expandafter \@secondoftwo
 \fi
}%
\providecommand \natexlab [1]{#1}%
\providecommand \enquote  [1]{``#1''}%
\providecommand \bibnamefont  [1]{#1}%
\providecommand \bibfnamefont [1]{#1}%
\providecommand \citenamefont [1]{#1}%
\providecommand \href@noop [0]{\@secondoftwo}%
\providecommand \href [0]{\begingroup \@sanitize@url \@href}%
\providecommand \@href[1]{\@@startlink{#1}\@@href}%
\providecommand \@@href[1]{\endgroup#1\@@endlink}%
\providecommand \@sanitize@url [0]{\catcode `\\12\catcode `\$12\catcode
  `\&12\catcode `\#12\catcode `\^12\catcode `\_12\catcode `\%12\relax}%
\providecommand \@@startlink[1]{}%
\providecommand \@@endlink[0]{}%
\providecommand \url  [0]{\begingroup\@sanitize@url \@url }%
\providecommand \@url [1]{\endgroup\@href {#1}{\urlprefix }}%
\providecommand \urlprefix  [0]{URL }%
\providecommand \Eprint [0]{\href }%
\providecommand \doibase [0]{https://doi.org/}%
\providecommand \selectlanguage [0]{\@gobble}%
\providecommand \bibinfo  [0]{\@secondoftwo}%
\providecommand \bibfield  [0]{\@secondoftwo}%
\providecommand \translation [1]{[#1]}%
\providecommand \BibitemOpen [0]{}%
\providecommand \bibitemStop [0]{}%
\providecommand \bibitemNoStop [0]{.\EOS\space}%
\providecommand \EOS [0]{\spacefactor3000\relax}%
\providecommand \BibitemShut  [1]{\csname bibitem#1\endcsname}%
\let\auto@bib@innerbib\@empty
%</preamble>
\bibitem [{\citenamefont {Lee}\ \emph {et~al.}(2004)\citenamefont {Lee},
  \citenamefont {Borasoy},\ and\ \citenamefont {Sch\"afer}}]{PRC70-014007}%
  \BibitemOpen
  \bibfield  {author} {\bibinfo {author} {\bibfnamefont {D.}~\bibnamefont
  {Lee}}, \bibinfo {author} {\bibfnamefont {B.}~\bibnamefont {Borasoy}},\ and\
  \bibinfo {author} {\bibfnamefont {T.}~\bibnamefont {Sch\"afer}},\ }\bibfield
  {title} {\bibinfo {title} {{Nuclear lattice simulations with chiral effective
  field theory}},\ }\href {https://doi.org/10.1103/PhysRevC.70.014007}
  {\bibfield  {journal} {\bibinfo  {journal} {Phys. Rev. C}\ }\textbf {\bibinfo
  {volume} {70}},\ \bibinfo {pages} {014007} (\bibinfo {year} {2004})},\
  \Eprint {https://arxiv.org/abs/nucl-th/0402072} {arXiv:nucl-th/0402072}
  \BibitemShut {NoStop}%
\bibitem [{\citenamefont {Lee}\ and\ \citenamefont
  {Sch\"afer}(2005)}]{PRC72-024006}%
  \BibitemOpen
  \bibfield  {author} {\bibinfo {author} {\bibfnamefont {D.}~\bibnamefont
  {Lee}}\ and\ \bibinfo {author} {\bibfnamefont {T.}~\bibnamefont
  {Sch\"afer}},\ }\bibfield  {title} {\bibinfo {title} {{Neutron matter on the
  lattice with pionless effective field theory}},\ }\href
  {https://doi.org/10.1103/PhysRevC.72.024006} {\bibfield  {journal} {\bibinfo
  {journal} {Phys. Rev. C}\ }\textbf {\bibinfo {volume} {72}},\ \bibinfo
  {pages} {024006} (\bibinfo {year} {2005})},\ \Eprint
  {https://arxiv.org/abs/nucl-th/0412002} {arXiv:nucl-th/0412002} \BibitemShut
  {NoStop}%
\bibitem [{\citenamefont {Lee}(2009)}]{PPNP63-117}%
  \BibitemOpen
  \bibfield  {author} {\bibinfo {author} {\bibfnamefont {D.}~\bibnamefont
  {Lee}},\ }\bibfield  {title} {\bibinfo {title} {{Lattice simulations for few-
  and many-body systems}},\ }\href {https://doi.org/10.1016/j.ppnp.2008.12.001}
  {\bibfield  {journal} {\bibinfo  {journal} {Prog. Part. Nucl. Phys.}\
  }\textbf {\bibinfo {volume} {63}},\ \bibinfo {pages} {117} (\bibinfo {year}
  {2009})},\ \Eprint {https://arxiv.org/abs/0804.3501} {arXiv:0804.3501
  [nucl-th]} \BibitemShut {NoStop}%
\bibitem [{\citenamefont {Blankenbecler}\ \emph {et~al.}(1981)\citenamefont
  {Blankenbecler}, \citenamefont {Scalapino},\ and\ \citenamefont
  {Sugar}}]{AFQMC}%
  \BibitemOpen
  \bibfield  {author} {\bibinfo {author} {\bibfnamefont {R.}~\bibnamefont
  {Blankenbecler}}, \bibinfo {author} {\bibfnamefont {D.~J.}\ \bibnamefont
  {Scalapino}},\ and\ \bibinfo {author} {\bibfnamefont {R.~L.}\ \bibnamefont
  {Sugar}},\ }\bibfield  {title} {\bibinfo {title} {Monte carlo calculations of
  coupled boson-fermion systems. i},\ }\href
  {https://doi.org/10.1103/PhysRevD.24.2278} {\bibfield  {journal} {\bibinfo
  {journal} {Phys. Rev. D}\ }\textbf {\bibinfo {volume} {24}},\ \bibinfo
  {pages} {2278} (\bibinfo {year} {1981})}\BibitemShut {NoStop}%
\bibitem [{\citenamefont {Borasoy}\ \emph {et~al.}(2007)\citenamefont
  {Borasoy}, \citenamefont {Epelbaum}, \citenamefont {Krebs}, \citenamefont
  {Lee},\ and\ \citenamefont {Meissner}}]{EPJA31-105}%
  \BibitemOpen
  \bibfield  {author} {\bibinfo {author} {\bibfnamefont {B.}~\bibnamefont
  {Borasoy}}, \bibinfo {author} {\bibfnamefont {E.}~\bibnamefont {Epelbaum}},
  \bibinfo {author} {\bibfnamefont {H.}~\bibnamefont {Krebs}}, \bibinfo
  {author} {\bibfnamefont {D.}~\bibnamefont {Lee}},\ and\ \bibinfo {author}
  {\bibfnamefont {U.-G.}\ \bibnamefont {Meissner}},\ }\bibfield  {title}
  {\bibinfo {title} {{Lattice Simulations for Light Nuclei: Chiral Effective
  Field Theory at Leading Order}},\ }\href
  {https://doi.org/10.1140/epja/i2006-10154-1} {\bibfield  {journal} {\bibinfo
  {journal} {Eur. Phys. J. A}\ }\textbf {\bibinfo {volume} {31}},\ \bibinfo
  {pages} {105} (\bibinfo {year} {2007})},\ \Eprint
  {https://arxiv.org/abs/nucl-th/0611087} {arXiv:nucl-th/0611087} \BibitemShut
  {NoStop}%
\bibitem [{\citenamefont {Epelbaum}\ \emph
  {et~al.}(2010{\natexlab{a}})\citenamefont {Epelbaum}, \citenamefont {Krebs},
  \citenamefont {Lee},\ and\ \citenamefont
  {Meissner}}]{PhysRevLett.104.142501}%
  \BibitemOpen
  \bibfield  {author} {\bibinfo {author} {\bibfnamefont {E.}~\bibnamefont
  {Epelbaum}}, \bibinfo {author} {\bibfnamefont {H.}~\bibnamefont {Krebs}},
  \bibinfo {author} {\bibfnamefont {D.}~\bibnamefont {Lee}},\ and\ \bibinfo
  {author} {\bibfnamefont {U.-G.}\ \bibnamefont {Meissner}},\ }\bibfield
  {title} {\bibinfo {title} {{Lattice effective field theory calculations for A
  = 3,4,6,12 nuclei}},\ }\href {https://doi.org/10.1103/PhysRevLett.104.142501}
  {\bibfield  {journal} {\bibinfo  {journal} {Phys. Rev. Lett.}\ }\textbf
  {\bibinfo {volume} {104}},\ \bibinfo {pages} {142501} (\bibinfo {year}
  {2010}{\natexlab{a}})},\ \Eprint {https://arxiv.org/abs/0912.4195}
  {arXiv:0912.4195 [nucl-th]} \BibitemShut {NoStop}%
\bibitem [{\citenamefont {Epelbaum}\ \emph
  {et~al.}(2010{\natexlab{b}})\citenamefont {Epelbaum}, \citenamefont {Krebs},
  \citenamefont {Lee},\ and\ \citenamefont {Meissner}}]{EPJA45-335}%
  \BibitemOpen
  \bibfield  {author} {\bibinfo {author} {\bibfnamefont {E.}~\bibnamefont
  {Epelbaum}}, \bibinfo {author} {\bibfnamefont {H.}~\bibnamefont {Krebs}},
  \bibinfo {author} {\bibfnamefont {D.}~\bibnamefont {Lee}},\ and\ \bibinfo
  {author} {\bibfnamefont {U.-G.}\ \bibnamefont {Meissner}},\ }\bibfield
  {title} {\bibinfo {title} {{Lattice calculations for A=3,4,6,12 nuclei using
  chiral effective field theory}},\ }\href
  {https://doi.org/10.1140/epja/i2010-11009-x} {\bibfield  {journal} {\bibinfo
  {journal} {Eur. Phys. J. A}\ }\textbf {\bibinfo {volume} {45}},\ \bibinfo
  {pages} {335} (\bibinfo {year} {2010}{\natexlab{b}})},\ \Eprint
  {https://arxiv.org/abs/1003.5697} {arXiv:1003.5697 [nucl-th]} \BibitemShut
  {NoStop}%
\bibitem [{\citenamefont {L\"ahde}\ \emph {et~al.}(2014)\citenamefont
  {L\"ahde}, \citenamefont {Epelbaum}, \citenamefont {Krebs}, \citenamefont
  {Lee}, \citenamefont {Mei\ss{}ner},\ and\ \citenamefont
  {Rupak}}]{PLB732-110}%
  \BibitemOpen
  \bibfield  {author} {\bibinfo {author} {\bibfnamefont {T.~A.}\ \bibnamefont
  {L\"ahde}}, \bibinfo {author} {\bibfnamefont {E.}~\bibnamefont {Epelbaum}},
  \bibinfo {author} {\bibfnamefont {H.}~\bibnamefont {Krebs}}, \bibinfo
  {author} {\bibfnamefont {D.}~\bibnamefont {Lee}}, \bibinfo {author}
  {\bibfnamefont {U.-G.}\ \bibnamefont {Mei\ss{}ner}},\ and\ \bibinfo {author}
  {\bibfnamefont {G.}~\bibnamefont {Rupak}},\ }\bibfield  {title} {\bibinfo
  {title} {{Lattice Effective Field Theory for Medium-Mass Nuclei}},\ }\href
  {https://doi.org/10.1016/j.physletb.2014.03.023} {\bibfield  {journal}
  {\bibinfo  {journal} {Phys. Lett. B}\ }\textbf {\bibinfo {volume} {732}},\
  \bibinfo {pages} {110} (\bibinfo {year} {2014})},\ \Eprint
  {https://arxiv.org/abs/1311.0477} {arXiv:1311.0477 [nucl-th]} \BibitemShut
  {NoStop}%
\bibitem [{\citenamefont {Lu}\ \emph {et~al.}(2019)\citenamefont {Lu},
  \citenamefont {Li}, \citenamefont {Elhatisari}, \citenamefont {Lee},
  \citenamefont {Epelbaum},\ and\ \citenamefont {Mei\ss{}ner}}]{PLB797-134863}%
  \BibitemOpen
  \bibfield  {author} {\bibinfo {author} {\bibfnamefont {B.-N.}\ \bibnamefont
  {Lu}}, \bibinfo {author} {\bibfnamefont {N.}~\bibnamefont {Li}}, \bibinfo
  {author} {\bibfnamefont {S.}~\bibnamefont {Elhatisari}}, \bibinfo {author}
  {\bibfnamefont {D.}~\bibnamefont {Lee}}, \bibinfo {author} {\bibfnamefont
  {E.}~\bibnamefont {Epelbaum}},\ and\ \bibinfo {author} {\bibfnamefont
  {U.-G.}\ \bibnamefont {Mei\ss{}ner}},\ }\bibfield  {title} {\bibinfo {title}
  {{Essential elements for nuclear binding}},\ }\href
  {https://doi.org/10.1016/j.physletb.2019.134863} {\bibfield  {journal}
  {\bibinfo  {journal} {Phys. Lett. B}\ }\textbf {\bibinfo {volume} {797}},\
  \bibinfo {pages} {134863} (\bibinfo {year} {2019})},\ \Eprint
  {https://arxiv.org/abs/1812.10928} {arXiv:1812.10928 [nucl-th]} \BibitemShut
  {NoStop}%
\bibitem [{\citenamefont {Lu}\ \emph {et~al.}(2022{\natexlab{a}})\citenamefont
  {Lu}, \citenamefont {Li}, \citenamefont {Elhatisari}, \citenamefont {Ma},
  \citenamefont {Lee},\ and\ \citenamefont
  {Mei\ss{}ner}}]{PhysRevLett.128.242501}%
  \BibitemOpen
  \bibfield  {author} {\bibinfo {author} {\bibfnamefont {B.-N.}\ \bibnamefont
  {Lu}}, \bibinfo {author} {\bibfnamefont {N.}~\bibnamefont {Li}}, \bibinfo
  {author} {\bibfnamefont {S.}~\bibnamefont {Elhatisari}}, \bibinfo {author}
  {\bibfnamefont {Y.-Z.}\ \bibnamefont {Ma}}, \bibinfo {author} {\bibfnamefont
  {D.}~\bibnamefont {Lee}},\ and\ \bibinfo {author} {\bibfnamefont {U.-G.}\
  \bibnamefont {Mei\ss{}ner}},\ }\bibfield  {title} {\bibinfo {title}
  {{Perturbative Quantum Monte~Carlo Method for Nuclear Physics}},\ }\href
  {https://doi.org/10.1103/PhysRevLett.128.242501} {\bibfield  {journal}
  {\bibinfo  {journal} {Phys. Rev. Lett.}\ }\textbf {\bibinfo {volume} {128}},\
  \bibinfo {pages} {242501} (\bibinfo {year} {2022}{\natexlab{a}})},\ \Eprint
  {https://arxiv.org/abs/2111.14191} {arXiv:2111.14191 [nucl-th]} \BibitemShut
  {NoStop}%
\bibitem [{\citenamefont {Elhatisari}\ \emph {et~al.}(2024)\citenamefont
  {Elhatisari} \emph {et~al.}}]{Nature630-59}%
  \BibitemOpen
  \bibfield  {author} {\bibinfo {author} {\bibfnamefont {S.}~\bibnamefont
  {Elhatisari}} \emph {et~al.},\ }\bibfield  {title} {\bibinfo {title}
  {{Wavefunction matching for solving quantum many-body problems}},\ }\href
  {https://doi.org/10.1038/s41586-024-07422-z} {\bibfield  {journal} {\bibinfo
  {journal} {Nature}\ }\textbf {\bibinfo {volume} {630}},\ \bibinfo {pages}
  {59} (\bibinfo {year} {2024})},\ \Eprint {https://arxiv.org/abs/2210.17488}
  {arXiv:2210.17488 [nucl-th]} \BibitemShut {NoStop}%
\bibitem [{\citenamefont {Epelbaum}\ \emph {et~al.}(2014)\citenamefont
  {Epelbaum}, \citenamefont {Krebs}, \citenamefont {L\"ahde}, \citenamefont
  {Lee}, \citenamefont {Mei\ss{}ner},\ and\ \citenamefont
  {Rupak}}]{PhysRevLett.112.102501}%
  \BibitemOpen
  \bibfield  {author} {\bibinfo {author} {\bibfnamefont {E.}~\bibnamefont
  {Epelbaum}}, \bibinfo {author} {\bibfnamefont {H.}~\bibnamefont {Krebs}},
  \bibinfo {author} {\bibfnamefont {T.~A.}\ \bibnamefont {L\"ahde}}, \bibinfo
  {author} {\bibfnamefont {D.}~\bibnamefont {Lee}}, \bibinfo {author}
  {\bibfnamefont {U.-G.}\ \bibnamefont {Mei\ss{}ner}},\ and\ \bibinfo {author}
  {\bibfnamefont {G.}~\bibnamefont {Rupak}},\ }\bibfield  {title} {\bibinfo
  {title} {{Ab Initio Calculation of the Spectrum and Structure of $^{16}$O}},\
  }\href {https://doi.org/10.1103/PhysRevLett.112.102501} {\bibfield  {journal}
  {\bibinfo  {journal} {Phys. Rev. Lett.}\ }\textbf {\bibinfo {volume} {112}},\
  \bibinfo {pages} {102501} (\bibinfo {year} {2014})},\ \Eprint
  {https://arxiv.org/abs/1312.7703} {arXiv:1312.7703 [nucl-th]} \BibitemShut
  {NoStop}%
\bibitem [{\citenamefont {Shen}\ \emph {et~al.}(2023)\citenamefont {Shen},
  \citenamefont {Elhatisari}, \citenamefont {L\"ahde}, \citenamefont {Lee},
  \citenamefont {Lu},\ and\ \citenamefont {Mei\ss{}ner}}]{Nat.Comm.14-2777}%
  \BibitemOpen
  \bibfield  {author} {\bibinfo {author} {\bibfnamefont {S.}~\bibnamefont
  {Shen}}, \bibinfo {author} {\bibfnamefont {S.}~\bibnamefont {Elhatisari}},
  \bibinfo {author} {\bibfnamefont {T.~A.}\ \bibnamefont {L\"ahde}}, \bibinfo
  {author} {\bibfnamefont {D.}~\bibnamefont {Lee}}, \bibinfo {author}
  {\bibfnamefont {B.-N.}\ \bibnamefont {Lu}},\ and\ \bibinfo {author}
  {\bibfnamefont {U.-G.}\ \bibnamefont {Mei\ss{}ner}},\ }\bibfield  {title}
  {\bibinfo {title} {{Emergent geometry and duality in the carbon nucleus}},\
  }\href {https://doi.org/10.1038/s41467-023-38391-y} {\bibfield  {journal}
  {\bibinfo  {journal} {Nature Commun.}\ }\textbf {\bibinfo {volume} {14}},\
  \bibinfo {pages} {2777} (\bibinfo {year} {2023})},\ \Eprint
  {https://arxiv.org/abs/2202.13596} {arXiv:2202.13596 [nucl-th]} \BibitemShut
  {NoStop}%
\bibitem [{\citenamefont {Mei\ss{}ner}\ \emph {et~al.}(2024)\citenamefont
  {Mei\ss{}ner}, \citenamefont {Shen}, \citenamefont {Elhatisari},\ and\
  \citenamefont {Lee}}]{PhysRevLett.132.062501}%
  \BibitemOpen
  \bibfield  {author} {\bibinfo {author} {\bibfnamefont {U.-G.}\ \bibnamefont
  {Mei\ss{}ner}}, \bibinfo {author} {\bibfnamefont {S.}~\bibnamefont {Shen}},
  \bibinfo {author} {\bibfnamefont {S.}~\bibnamefont {Elhatisari}},\ and\
  \bibinfo {author} {\bibfnamefont {D.}~\bibnamefont {Lee}},\ }\bibfield
  {title} {\bibinfo {title} {{Ab~Initio Calculation of the Alpha-Particle
  Monopole Transition Form Factor}},\ }\href
  {https://doi.org/10.1103/PhysRevLett.132.062501} {\bibfield  {journal}
  {\bibinfo  {journal} {Phys. Rev. Lett.}\ }\textbf {\bibinfo {volume} {132}},\
  \bibinfo {pages} {062501} (\bibinfo {year} {2024})},\ \Eprint
  {https://arxiv.org/abs/2309.01558} {arXiv:2309.01558 [nucl-th]} \BibitemShut
  {NoStop}%
\bibitem [{\citenamefont {Shen}\ \emph {et~al.}(2024)\citenamefont {Shen},
  \citenamefont {Elhatisari}, \citenamefont {Lee}, \citenamefont
  {Mei\ss{}ner},\ and\ \citenamefont {Ren}}]{arxiV2411.14935}%
  \BibitemOpen
  \bibfield  {author} {\bibinfo {author} {\bibfnamefont {S.}~\bibnamefont
  {Shen}}, \bibinfo {author} {\bibfnamefont {S.}~\bibnamefont {Elhatisari}},
  \bibinfo {author} {\bibfnamefont {D.}~\bibnamefont {Lee}}, \bibinfo {author}
  {\bibfnamefont {U.-G.}\ \bibnamefont {Mei\ss{}ner}},\ and\ \bibinfo {author}
  {\bibfnamefont {Z.}~\bibnamefont {Ren}},\ }\bibfield  {title} {\bibinfo
  {title} {{Ab initio study of the beryllium isotopes $^{7}$Be to $^{12}$Be}},\
  }\href@noop {} {\  (\bibinfo {year} {2024})},\ \Eprint
  {https://arxiv.org/abs/2411.14935} {arXiv:2411.14935 [nucl-th]} \BibitemShut
  {NoStop}%
\bibitem [{\citenamefont {Epelbaum}\ \emph {et~al.}(2011)\citenamefont
  {Epelbaum}, \citenamefont {Krebs}, \citenamefont {Lee},\ and\ \citenamefont
  {Meissner}}]{PhysRevLett.106.192501}%
  \BibitemOpen
  \bibfield  {author} {\bibinfo {author} {\bibfnamefont {E.}~\bibnamefont
  {Epelbaum}}, \bibinfo {author} {\bibfnamefont {H.}~\bibnamefont {Krebs}},
  \bibinfo {author} {\bibfnamefont {D.}~\bibnamefont {Lee}},\ and\ \bibinfo
  {author} {\bibfnamefont {U.-G.}\ \bibnamefont {Meissner}},\ }\bibfield
  {title} {\bibinfo {title} {{Ab initio calculation of the Hoyle state}},\
  }\href {https://doi.org/10.1103/PhysRevLett.106.192501} {\bibfield  {journal}
  {\bibinfo  {journal} {Phys. Rev. Lett.}\ }\textbf {\bibinfo {volume} {106}},\
  \bibinfo {pages} {192501} (\bibinfo {year} {2011})},\ \Eprint
  {https://arxiv.org/abs/1101.2547} {arXiv:1101.2547 [nucl-th]} \BibitemShut
  {NoStop}%
\bibitem [{\citenamefont {Epelbaum}\ \emph {et~al.}(2012)\citenamefont
  {Epelbaum}, \citenamefont {Krebs}, \citenamefont {Lahde}, \citenamefont
  {Lee},\ and\ \citenamefont {Meissner}}]{PhysRevLett.109.252501}%
  \BibitemOpen
  \bibfield  {author} {\bibinfo {author} {\bibfnamefont {E.}~\bibnamefont
  {Epelbaum}}, \bibinfo {author} {\bibfnamefont {H.}~\bibnamefont {Krebs}},
  \bibinfo {author} {\bibfnamefont {T.~A.}\ \bibnamefont {Lahde}}, \bibinfo
  {author} {\bibfnamefont {D.}~\bibnamefont {Lee}},\ and\ \bibinfo {author}
  {\bibfnamefont {U.-G.}\ \bibnamefont {Meissner}},\ }\bibfield  {title}
  {\bibinfo {title} {{Structure and rotations of the Hoyle state}},\ }\href
  {https://doi.org/10.1103/PhysRevLett.109.252501} {\bibfield  {journal}
  {\bibinfo  {journal} {Phys. Rev. Lett.}\ }\textbf {\bibinfo {volume} {109}},\
  \bibinfo {pages} {252501} (\bibinfo {year} {2012})},\ \Eprint
  {https://arxiv.org/abs/1208.1328} {arXiv:1208.1328 [nucl-th]} \BibitemShut
  {NoStop}%
\bibitem [{\citenamefont {Epelbaum}\ \emph {et~al.}(2013)\citenamefont
  {Epelbaum}, \citenamefont {Krebs}, \citenamefont {L\"ahde}, \citenamefont
  {Lee},\ and\ \citenamefont {Mei\ss{}ner}}]{PhysRevLett.110.112502}%
  \BibitemOpen
  \bibfield  {author} {\bibinfo {author} {\bibfnamefont {E.}~\bibnamefont
  {Epelbaum}}, \bibinfo {author} {\bibfnamefont {H.}~\bibnamefont {Krebs}},
  \bibinfo {author} {\bibfnamefont {T.~A.}\ \bibnamefont {L\"ahde}}, \bibinfo
  {author} {\bibfnamefont {D.}~\bibnamefont {Lee}},\ and\ \bibinfo {author}
  {\bibfnamefont {U.-G.}\ \bibnamefont {Mei\ss{}ner}},\ }\bibfield  {title}
  {\bibinfo {title} {{Viability of Carbon-Based Life as a Function of the Light
  Quark Mass}},\ }\href {https://doi.org/10.1103/PhysRevLett.110.112502}
  {\bibfield  {journal} {\bibinfo  {journal} {Phys. Rev. Lett.}\ }\textbf
  {\bibinfo {volume} {110}},\ \bibinfo {pages} {112502} (\bibinfo {year}
  {2013})},\ \Eprint {https://arxiv.org/abs/1212.4181} {arXiv:1212.4181
  [nucl-th]} \BibitemShut {NoStop}%
\bibitem [{\citenamefont {Elhatisari}\ \emph {et~al.}(2017)\citenamefont
  {Elhatisari}, \citenamefont {Epelbaum}, \citenamefont {Krebs}, \citenamefont
  {L\"ahde}, \citenamefont {Lee}, \citenamefont {Li}, \citenamefont {Lu},
  \citenamefont {Mei\ss{}ner},\ and\ \citenamefont
  {Rupak}}]{PhysRevLett.119.222505}%
  \BibitemOpen
  \bibfield  {author} {\bibinfo {author} {\bibfnamefont {S.}~\bibnamefont
  {Elhatisari}}, \bibinfo {author} {\bibfnamefont {E.}~\bibnamefont
  {Epelbaum}}, \bibinfo {author} {\bibfnamefont {H.}~\bibnamefont {Krebs}},
  \bibinfo {author} {\bibfnamefont {T.~A.}\ \bibnamefont {L\"ahde}}, \bibinfo
  {author} {\bibfnamefont {D.}~\bibnamefont {Lee}}, \bibinfo {author}
  {\bibfnamefont {N.}~\bibnamefont {Li}}, \bibinfo {author} {\bibfnamefont
  {B.-n.}\ \bibnamefont {Lu}}, \bibinfo {author} {\bibfnamefont {U.-G.}\
  \bibnamefont {Mei\ss{}ner}},\ and\ \bibinfo {author} {\bibfnamefont
  {G.}~\bibnamefont {Rupak}},\ }\bibfield  {title} {\bibinfo {title} {{Ab
  initio Calculations of the Isotopic Dependence of Nuclear Clustering}},\
  }\href {https://doi.org/10.1103/PhysRevLett.119.222505} {\bibfield  {journal}
  {\bibinfo  {journal} {Phys. Rev. Lett.}\ }\textbf {\bibinfo {volume} {119}},\
  \bibinfo {pages} {222505} (\bibinfo {year} {2017})},\ \Eprint
  {https://arxiv.org/abs/1702.05177} {arXiv:1702.05177 [nucl-th]} \BibitemShut
  {NoStop}%
\bibitem [{\citenamefont {Zhang}\ \emph {et~al.}(2024)\citenamefont {Zhang},
  \citenamefont {Elhatisari}, \citenamefont {Mei\ss{}ner},\ and\ \citenamefont
  {Shen}}]{arxiV2411.17462}%
  \BibitemOpen
  \bibfield  {author} {\bibinfo {author} {\bibfnamefont {S.}~\bibnamefont
  {Zhang}}, \bibinfo {author} {\bibfnamefont {S.}~\bibnamefont {Elhatisari}},
  \bibinfo {author} {\bibfnamefont {U.-G.}\ \bibnamefont {Mei\ss{}ner}},\ and\
  \bibinfo {author} {\bibfnamefont {S.}~\bibnamefont {Shen}},\ }\bibfield
  {title} {\bibinfo {title} {{Lattice simulation of nucleon distribution and
  shell closure in the proton-rich nucleus $^{22}$Si}},\ }\href@noop {} {\
  (\bibinfo {year} {2024})},\ \Eprint {https://arxiv.org/abs/2411.17462}
  {arXiv:2411.17462 [nucl-th]} \BibitemShut {NoStop}%
\bibitem [{\citenamefont {Bour}\ \emph {et~al.}(2012)\citenamefont {Bour},
  \citenamefont {Hammer}, \citenamefont {Lee},\ and\ \citenamefont
  {Meissner}}]{PhysRevC.86.034003}%
  \BibitemOpen
  \bibfield  {author} {\bibinfo {author} {\bibfnamefont {S.}~\bibnamefont
  {Bour}}, \bibinfo {author} {\bibfnamefont {H.~W.}\ \bibnamefont {Hammer}},
  \bibinfo {author} {\bibfnamefont {D.}~\bibnamefont {Lee}},\ and\ \bibinfo
  {author} {\bibfnamefont {U.-G.}\ \bibnamefont {Meissner}},\ }\bibfield
  {title} {\bibinfo {title} {{Benchmark calculations for elastic fermion-dimer
  scattering}},\ }\href {https://doi.org/10.1103/PhysRevC.86.034003} {\bibfield
   {journal} {\bibinfo  {journal} {Phys. Rev. C}\ }\textbf {\bibinfo {volume}
  {86}},\ \bibinfo {pages} {034003} (\bibinfo {year} {2012})},\ \Eprint
  {https://arxiv.org/abs/1206.1765} {arXiv:1206.1765 [nucl-th]} \BibitemShut
  {NoStop}%
\bibitem [{\citenamefont {Elhatisari}\ \emph {et~al.}(2015)\citenamefont
  {Elhatisari}, \citenamefont {Lee}, \citenamefont {Rupak}, \citenamefont
  {Epelbaum}, \citenamefont {Krebs}, \citenamefont {L\"ahde}, \citenamefont
  {Luu},\ and\ \citenamefont {Mei\ss{}ner}}]{Nature528-111}%
  \BibitemOpen
  \bibfield  {author} {\bibinfo {author} {\bibfnamefont {S.}~\bibnamefont
  {Elhatisari}}, \bibinfo {author} {\bibfnamefont {D.}~\bibnamefont {Lee}},
  \bibinfo {author} {\bibfnamefont {G.}~\bibnamefont {Rupak}}, \bibinfo
  {author} {\bibfnamefont {E.}~\bibnamefont {Epelbaum}}, \bibinfo {author}
  {\bibfnamefont {H.}~\bibnamefont {Krebs}}, \bibinfo {author} {\bibfnamefont
  {T.~A.}\ \bibnamefont {L\"ahde}}, \bibinfo {author} {\bibfnamefont
  {T.}~\bibnamefont {Luu}},\ and\ \bibinfo {author} {\bibfnamefont {U.-G.}\
  \bibnamefont {Mei\ss{}ner}},\ }\bibfield  {title} {\bibinfo {title} {{Ab
  initio alpha-alpha scattering}},\ }\href
  {https://doi.org/10.1038/nature16067} {\bibfield  {journal} {\bibinfo
  {journal} {Nature}\ }\textbf {\bibinfo {volume} {528}},\ \bibinfo {pages}
  {111} (\bibinfo {year} {2015})},\ \Eprint {https://arxiv.org/abs/1506.03513}
  {arXiv:1506.03513 [nucl-th]} \BibitemShut {NoStop}%
\bibitem [{\citenamefont {Elhatisari}\ \emph
  {et~al.}(2016{\natexlab{a}})\citenamefont {Elhatisari} \emph
  {et~al.}}]{PhysRevLett.117.132501}%
  \BibitemOpen
  \bibfield  {author} {\bibinfo {author} {\bibfnamefont {S.}~\bibnamefont
  {Elhatisari}} \emph {et~al.},\ }\bibfield  {title} {\bibinfo {title}
  {{Nuclear binding near a quantum phase transition}},\ }\href
  {https://doi.org/10.1103/PhysRevLett.117.132501} {\bibfield  {journal}
  {\bibinfo  {journal} {Phys. Rev. Lett.}\ }\textbf {\bibinfo {volume} {117}},\
  \bibinfo {pages} {132501} (\bibinfo {year} {2016}{\natexlab{a}})},\ \Eprint
  {https://arxiv.org/abs/1602.04539} {arXiv:1602.04539 [nucl-th]} \BibitemShut
  {NoStop}%
\bibitem [{\citenamefont {Lu}\ \emph {et~al.}(2020)\citenamefont {Lu},
  \citenamefont {Li}, \citenamefont {Elhatisari}, \citenamefont {Lee},
  \citenamefont {Drut}, \citenamefont {L\"ahde}, \citenamefont {Epelbaum},\
  and\ \citenamefont {Mei\ss{}ner}}]{PhysRevLett.125.192502}%
  \BibitemOpen
  \bibfield  {author} {\bibinfo {author} {\bibfnamefont {B.-N.}\ \bibnamefont
  {Lu}}, \bibinfo {author} {\bibfnamefont {N.}~\bibnamefont {Li}}, \bibinfo
  {author} {\bibfnamefont {S.}~\bibnamefont {Elhatisari}}, \bibinfo {author}
  {\bibfnamefont {D.}~\bibnamefont {Lee}}, \bibinfo {author} {\bibfnamefont
  {J.~E.}\ \bibnamefont {Drut}}, \bibinfo {author} {\bibfnamefont {T.~A.}\
  \bibnamefont {L\"ahde}}, \bibinfo {author} {\bibfnamefont {E.}~\bibnamefont
  {Epelbaum}},\ and\ \bibinfo {author} {\bibfnamefont {U.-G.}\ \bibnamefont
  {Mei\ss{}ner}},\ }\bibfield  {title} {\bibinfo {title} {{$Ab Initio$ Nuclear
  Thermodynamics}},\ }\href {https://doi.org/10.1103/PhysRevLett.125.192502}
  {\bibfield  {journal} {\bibinfo  {journal} {Phys. Rev. Lett.}\ }\textbf
  {\bibinfo {volume} {125}},\ \bibinfo {pages} {192502} (\bibinfo {year}
  {2020})},\ \Eprint {https://arxiv.org/abs/1912.05105} {arXiv:1912.05105
  [nucl-th]} \BibitemShut {NoStop}%
\bibitem [{\citenamefont {Ren}\ \emph {et~al.}(2024)\citenamefont {Ren},
  \citenamefont {Elhatisari}, \citenamefont {L\"ahde}, \citenamefont {Lee},\
  and\ \citenamefont {Mei\ss{}ner}}]{PLB850-138463}%
  \BibitemOpen
  \bibfield  {author} {\bibinfo {author} {\bibfnamefont {Z.}~\bibnamefont
  {Ren}}, \bibinfo {author} {\bibfnamefont {S.}~\bibnamefont {Elhatisari}},
  \bibinfo {author} {\bibfnamefont {T.~A.}\ \bibnamefont {L\"ahde}}, \bibinfo
  {author} {\bibfnamefont {D.}~\bibnamefont {Lee}},\ and\ \bibinfo {author}
  {\bibfnamefont {U.-G.}\ \bibnamefont {Mei\ss{}ner}},\ }\bibfield  {title}
  {\bibinfo {title} {{Ab initio study of nuclear clustering in hot dilute
  nuclear matter}},\ }\href {https://doi.org/10.1016/j.physletb.2024.138463}
  {\bibfield  {journal} {\bibinfo  {journal} {Phys. Lett. B}\ }\textbf
  {\bibinfo {volume} {850}},\ \bibinfo {pages} {138463} (\bibinfo {year}
  {2024})},\ \Eprint {https://arxiv.org/abs/2305.15037} {arXiv:2305.15037
  [nucl-th]} \BibitemShut {NoStop}%
\bibitem [{\citenamefont {Ma}\ \emph {et~al.}(2024)\citenamefont {Ma},
  \citenamefont {Lin}, \citenamefont {Lu}, \citenamefont {Elhatisari},
  \citenamefont {Lee}, \citenamefont {Li}, \citenamefont {Mei\ss{}ner},
  \citenamefont {Steiner},\ and\ \citenamefont
  {Wang}}]{PhysRevLett.132.232502}%
  \BibitemOpen
  \bibfield  {author} {\bibinfo {author} {\bibfnamefont {Y.-Z.}\ \bibnamefont
  {Ma}}, \bibinfo {author} {\bibfnamefont {Z.}~\bibnamefont {Lin}}, \bibinfo
  {author} {\bibfnamefont {B.-N.}\ \bibnamefont {Lu}}, \bibinfo {author}
  {\bibfnamefont {S.}~\bibnamefont {Elhatisari}}, \bibinfo {author}
  {\bibfnamefont {D.}~\bibnamefont {Lee}}, \bibinfo {author} {\bibfnamefont
  {N.}~\bibnamefont {Li}}, \bibinfo {author} {\bibfnamefont {U.-G.}\
  \bibnamefont {Mei\ss{}ner}}, \bibinfo {author} {\bibfnamefont {A.~W.}\
  \bibnamefont {Steiner}},\ and\ \bibinfo {author} {\bibfnamefont
  {Q.}~\bibnamefont {Wang}},\ }\bibfield  {title} {\bibinfo {title} {{Structure
  Factors for Hot Neutron Matter from Ab~Initio Lattice Simulations with
  High-Fidelity Chiral Interactions}},\ }\href
  {https://doi.org/10.1103/PhysRevLett.132.232502} {\bibfield  {journal}
  {\bibinfo  {journal} {Phys. Rev. Lett.}\ }\textbf {\bibinfo {volume} {132}},\
  \bibinfo {pages} {232502} (\bibinfo {year} {2024})},\ \Eprint
  {https://arxiv.org/abs/2306.04500} {arXiv:2306.04500 [nucl-th]} \BibitemShut
  {NoStop}%
\bibitem [{\citenamefont {Bour}\ \emph {et~al.}(2015)\citenamefont {Bour},
  \citenamefont {Lee}, \citenamefont {Hammer},\ and\ \citenamefont
  {Mei\ss{}ner}}]{PhysRevLett.115.185301}%
  \BibitemOpen
  \bibfield  {author} {\bibinfo {author} {\bibfnamefont {S.}~\bibnamefont
  {Bour}}, \bibinfo {author} {\bibfnamefont {D.}~\bibnamefont {Lee}}, \bibinfo
  {author} {\bibfnamefont {H.~W.}\ \bibnamefont {Hammer}},\ and\ \bibinfo
  {author} {\bibfnamefont {U.-G.}\ \bibnamefont {Mei\ss{}ner}},\ }\bibfield
  {title} {\bibinfo {title} {{Ab initio Lattice Results for Fermi Polarons in
  Two Dimensions}},\ }\href {https://doi.org/10.1103/PhysRevLett.115.185301}
  {\bibfield  {journal} {\bibinfo  {journal} {Phys. Rev. Lett.}\ }\textbf
  {\bibinfo {volume} {115}},\ \bibinfo {pages} {185301} (\bibinfo {year}
  {2015})},\ \Eprint {https://arxiv.org/abs/1412.8175} {arXiv:1412.8175
  [cond-mat.quant-gas]} \BibitemShut {NoStop}%
\bibitem [{\citenamefont {Scarduelli}\ \emph {et~al.}(2020)\citenamefont
  {Scarduelli}, \citenamefont {Gasques}, \citenamefont {Chamon},\ and\
  \citenamefont {L\'epine-Szily}}]{EPJA56-24}%
  \BibitemOpen
  \bibfield  {author} {\bibinfo {author} {\bibfnamefont {V.}~\bibnamefont
  {Scarduelli}}, \bibinfo {author} {\bibfnamefont {L.~R.}\ \bibnamefont
  {Gasques}}, \bibinfo {author} {\bibfnamefont {L.~C.}\ \bibnamefont
  {Chamon}},\ and\ \bibinfo {author} {\bibfnamefont {A.}~\bibnamefont
  {L\'epine-Szily}},\ }\bibfield  {title} {\bibinfo {title} {{A method to
  optimize mass discrimination of particles identified in $\Delta
  E$\textendash{}E silicon surface barrier detector systems}},\ }\href
  {https://doi.org/10.1140/epja/s10050-020-00021-2} {\bibfield  {journal}
  {\bibinfo  {journal} {Eur. Phys. J. A}\ }\textbf {\bibinfo {volume} {56}},\
  \bibinfo {pages} {24} (\bibinfo {year} {2020})}\BibitemShut {NoStop}%
\bibitem [{\citenamefont {Hildenbrand}\ \emph {et~al.}(2024)\citenamefont
  {Hildenbrand}, \citenamefont {Elhatisari}, \citenamefont {Ren},\ and\
  \citenamefont {Mei\ss{}ner}}]{EPJA60-215}%
  \BibitemOpen
  \bibfield  {author} {\bibinfo {author} {\bibfnamefont {F.}~\bibnamefont
  {Hildenbrand}}, \bibinfo {author} {\bibfnamefont {S.}~\bibnamefont
  {Elhatisari}}, \bibinfo {author} {\bibfnamefont {Z.}~\bibnamefont {Ren}},\
  and\ \bibinfo {author} {\bibfnamefont {U.-G.}\ \bibnamefont {Mei\ss{}ner}},\
  }\bibfield  {title} {\bibinfo {title} {{Towards hypernuclei from nuclear
  lattice effective field theory}},\ }\href
  {https://doi.org/10.1140/epja/s10050-024-01427-y} {\bibfield  {journal}
  {\bibinfo  {journal} {Eur. Phys. J. A}\ }\textbf {\bibinfo {volume} {60}},\
  \bibinfo {pages} {215} (\bibinfo {year} {2024})},\ \Eprint
  {https://arxiv.org/abs/2406.17638} {arXiv:2406.17638 [nucl-th]} \BibitemShut
  {NoStop}%
\bibitem [{\citenamefont {Wiringa}\ \emph {et~al.}(1995)\citenamefont
  {Wiringa}, \citenamefont {Stoks},\ and\ \citenamefont
  {Schiavilla}}]{PRC51-38}%
  \BibitemOpen
  \bibfield  {author} {\bibinfo {author} {\bibfnamefont {R.~B.}\ \bibnamefont
  {Wiringa}}, \bibinfo {author} {\bibfnamefont {V.~G.~J.}\ \bibnamefont
  {Stoks}},\ and\ \bibinfo {author} {\bibfnamefont {R.}~\bibnamefont
  {Schiavilla}},\ }\bibfield  {title} {\bibinfo {title} {{An Accurate
  nucleon-nucleon potential with charge independence breaking}},\ }\href
  {https://doi.org/10.1103/PhysRevC.51.38} {\bibfield  {journal} {\bibinfo
  {journal} {Phys. Rev. C}\ }\textbf {\bibinfo {volume} {51}},\ \bibinfo
  {pages} {38} (\bibinfo {year} {1995})},\ \Eprint
  {https://arxiv.org/abs/nucl-th/9408016} {arXiv:nucl-th/9408016} \BibitemShut
  {NoStop}%
\bibitem [{\citenamefont {Machleidt}(2001)}]{PRC63-024001}%
  \BibitemOpen
  \bibfield  {author} {\bibinfo {author} {\bibfnamefont {R.}~\bibnamefont
  {Machleidt}},\ }\bibfield  {title} {\bibinfo {title} {{The High precision,
  charge dependent Bonn nucleon-nucleon potential (CD-Bonn)}},\ }\href
  {https://doi.org/10.1103/PhysRevC.63.024001} {\bibfield  {journal} {\bibinfo
  {journal} {Phys. Rev. C}\ }\textbf {\bibinfo {volume} {63}},\ \bibinfo
  {pages} {024001} (\bibinfo {year} {2001})},\ \Eprint
  {https://arxiv.org/abs/nucl-th/0006014} {arXiv:nucl-th/0006014} \BibitemShut
  {NoStop}%
\bibitem [{\citenamefont {Entem}\ and\ \citenamefont
  {Machleidt}(2003)}]{PRC68-041001}%
  \BibitemOpen
  \bibfield  {author} {\bibinfo {author} {\bibfnamefont {D.~R.}\ \bibnamefont
  {Entem}}\ and\ \bibinfo {author} {\bibfnamefont {R.}~\bibnamefont
  {Machleidt}},\ }\bibfield  {title} {\bibinfo {title} {{Accurate charge
  dependent nucleon nucleon potential at fourth order of chiral perturbation
  theory}},\ }\href {https://doi.org/10.1103/PhysRevC.68.041001} {\bibfield
  {journal} {\bibinfo  {journal} {Phys. Rev. C}\ }\textbf {\bibinfo {volume}
  {68}},\ \bibinfo {pages} {041001} (\bibinfo {year} {2003})},\ \Eprint
  {https://arxiv.org/abs/nucl-th/0304018} {arXiv:nucl-th/0304018} \BibitemShut
  {NoStop}%
\bibitem [{\citenamefont {Epelbaum}\ \emph {et~al.}(2005)\citenamefont
  {Epelbaum}, \citenamefont {Glockle},\ and\ \citenamefont
  {Meissner}}]{NPA747-362}%
  \BibitemOpen
  \bibfield  {author} {\bibinfo {author} {\bibfnamefont {E.}~\bibnamefont
  {Epelbaum}}, \bibinfo {author} {\bibfnamefont {W.}~\bibnamefont {Glockle}},\
  and\ \bibinfo {author} {\bibfnamefont {U.-G.}\ \bibnamefont {Meissner}},\
  }\bibfield  {title} {\bibinfo {title} {{The Two-nucleon system at
  next-to-next-to-next-to-leading order}},\ }\href
  {https://doi.org/10.1016/j.nuclphysa.2004.09.107} {\bibfield  {journal}
  {\bibinfo  {journal} {Nucl. Phys. A}\ }\textbf {\bibinfo {volume} {747}},\
  \bibinfo {pages} {362} (\bibinfo {year} {2005})},\ \Eprint
  {https://arxiv.org/abs/nucl-th/0405048} {arXiv:nucl-th/0405048} \BibitemShut
  {NoStop}%
\bibitem [{\citenamefont {Lu}\ \emph {et~al.}(2022{\natexlab{b}})\citenamefont
  {Lu}, \citenamefont {Wang}, \citenamefont {Xiao}, \citenamefont {Geng},
  \citenamefont {Meng},\ and\ \citenamefont {Ring}}]{PRL128-142002}%
  \BibitemOpen
  \bibfield  {author} {\bibinfo {author} {\bibfnamefont {J.-X.}\ \bibnamefont
  {Lu}}, \bibinfo {author} {\bibfnamefont {C.-X.}\ \bibnamefont {Wang}},
  \bibinfo {author} {\bibfnamefont {Y.}~\bibnamefont {Xiao}}, \bibinfo {author}
  {\bibfnamefont {L.-S.}\ \bibnamefont {Geng}}, \bibinfo {author}
  {\bibfnamefont {J.}~\bibnamefont {Meng}},\ and\ \bibinfo {author}
  {\bibfnamefont {P.}~\bibnamefont {Ring}},\ }\bibfield  {title} {\bibinfo
  {title} {{Accurate Relativistic Chiral Nucleon-Nucleon Interaction up to
  Next-to-Next-to-Leading Order}},\ }\href
  {https://doi.org/10.1103/PhysRevLett.128.142002} {\bibfield  {journal}
  {\bibinfo  {journal} {Phys. Rev. Lett.}\ }\textbf {\bibinfo {volume} {128}},\
  \bibinfo {pages} {142002} (\bibinfo {year} {2022}{\natexlab{b}})},\ \Eprint
  {https://arxiv.org/abs/2111.07766} {arXiv:2111.07766 [nucl-th]} \BibitemShut
  {NoStop}%
\bibitem [{\citenamefont {Lu}\ \emph {et~al.}(2014)\citenamefont {Lu},
  \citenamefont {L\"ahde}, \citenamefont {Lee},\ and\ \citenamefont
  {Mei\ss{}ner}}]{PRD90-034507}%
  \BibitemOpen
  \bibfield  {author} {\bibinfo {author} {\bibfnamefont {B.-N.}\ \bibnamefont
  {Lu}}, \bibinfo {author} {\bibfnamefont {T.~A.}\ \bibnamefont {L\"ahde}},
  \bibinfo {author} {\bibfnamefont {D.}~\bibnamefont {Lee}},\ and\ \bibinfo
  {author} {\bibfnamefont {U.-G.}\ \bibnamefont {Mei\ss{}ner}},\ }\bibfield
  {title} {\bibinfo {title} {{Breaking and restoration of rotational symmetry
  on the lattice for bound state multiplets}},\ }\href
  {https://doi.org/10.1103/PhysRevD.90.034507} {\bibfield  {journal} {\bibinfo
  {journal} {Phys. Rev. D}\ }\textbf {\bibinfo {volume} {90}},\ \bibinfo
  {pages} {034507} (\bibinfo {year} {2014})},\ \Eprint
  {https://arxiv.org/abs/1403.8056} {arXiv:1403.8056 [nucl-th]} \BibitemShut
  {NoStop}%
\bibitem [{\citenamefont {Lu}\ \emph {et~al.}(2015)\citenamefont {Lu},
  \citenamefont {L\"ahde}, \citenamefont {Lee},\ and\ \citenamefont
  {Mei\ss{}ner}}]{PRD92-014506}%
  \BibitemOpen
  \bibfield  {author} {\bibinfo {author} {\bibfnamefont {B.-N.}\ \bibnamefont
  {Lu}}, \bibinfo {author} {\bibfnamefont {T.~A.}\ \bibnamefont {L\"ahde}},
  \bibinfo {author} {\bibfnamefont {D.}~\bibnamefont {Lee}},\ and\ \bibinfo
  {author} {\bibfnamefont {U.-G.}\ \bibnamefont {Mei\ss{}ner}},\ }\bibfield
  {title} {\bibinfo {title} {{Breaking and restoration of rotational symmetry
  for irreducible tensor operators on the lattice}},\ }\href
  {https://doi.org/10.1103/PhysRevD.92.014506} {\bibfield  {journal} {\bibinfo
  {journal} {Phys. Rev. D}\ }\textbf {\bibinfo {volume} {92}},\ \bibinfo
  {pages} {014506} (\bibinfo {year} {2015})},\ \Eprint
  {https://arxiv.org/abs/1504.01685} {arXiv:1504.01685 [nucl-th]} \BibitemShut
  {NoStop}%
\bibitem [{\citenamefont {Li}\ \emph {et~al.}(2019)\citenamefont {Li},
  \citenamefont {Elhatisari}, \citenamefont {Epelbaum}, \citenamefont {Lee},
  \citenamefont {Lu},\ and\ \citenamefont {Mei\ss{}ner}}]{PRC99-064001}%
  \BibitemOpen
  \bibfield  {author} {\bibinfo {author} {\bibfnamefont {N.}~\bibnamefont
  {Li}}, \bibinfo {author} {\bibfnamefont {S.}~\bibnamefont {Elhatisari}},
  \bibinfo {author} {\bibfnamefont {E.}~\bibnamefont {Epelbaum}}, \bibinfo
  {author} {\bibfnamefont {D.}~\bibnamefont {Lee}}, \bibinfo {author}
  {\bibfnamefont {B.}~\bibnamefont {Lu}},\ and\ \bibinfo {author}
  {\bibfnamefont {U.-G.}\ \bibnamefont {Mei\ss{}ner}},\ }\bibfield  {title}
  {\bibinfo {title} {{Galilean invariance restoration on the lattice}},\ }\href
  {https://doi.org/10.1103/PhysRevC.99.064001} {\bibfield  {journal} {\bibinfo
  {journal} {Phys. Rev. C}\ }\textbf {\bibinfo {volume} {99}},\ \bibinfo
  {pages} {064001} (\bibinfo {year} {2019})},\ \Eprint
  {https://arxiv.org/abs/1902.01295} {arXiv:1902.01295 [nucl-th]} \BibitemShut
  {NoStop}%
\bibitem [{\citenamefont {Symanzik}(1983{\natexlab{a}})}]{NPB226-187}%
  \BibitemOpen
  \bibfield  {author} {\bibinfo {author} {\bibfnamefont {K.}~\bibnamefont
  {Symanzik}},\ }\bibfield  {title} {\bibinfo {title} {{Continuum Limit and
  Improved Action in Lattice Theories. 1. Principles and $\varphi^4$ Theory}},\
  }\href {https://doi.org/10.1016/0550-3213(83)90468-6} {\bibfield  {journal}
  {\bibinfo  {journal} {Nucl. Phys. B}\ }\textbf {\bibinfo {volume} {226}},\
  \bibinfo {pages} {187} (\bibinfo {year} {1983}{\natexlab{a}})}\BibitemShut
  {NoStop}%
\bibitem [{\citenamefont {Symanzik}(1983{\natexlab{b}})}]{NPB226-205}%
  \BibitemOpen
  \bibfield  {author} {\bibinfo {author} {\bibfnamefont {K.}~\bibnamefont
  {Symanzik}},\ }\bibfield  {title} {\bibinfo {title} {{Continuum Limit and
  Improved Action in Lattice Theories. 2. O(N) Nonlinear Sigma Model in
  Perturbation Theory}},\ }\href {https://doi.org/10.1016/0550-3213(83)90469-8}
  {\bibfield  {journal} {\bibinfo  {journal} {Nucl. Phys. B}\ }\textbf
  {\bibinfo {volume} {226}},\ \bibinfo {pages} {205} (\bibinfo {year}
  {1983}{\natexlab{b}})}\BibitemShut {NoStop}%
\bibitem [{\citenamefont {Luscher}(1986)}]{Comm.Math.Phys.105-153}%
  \BibitemOpen
  \bibfield  {author} {\bibinfo {author} {\bibfnamefont {M.}~\bibnamefont
  {Luscher}},\ }\bibfield  {title} {\bibinfo {title} {{Volume Dependence of the
  Energy Spectrum in Massive Quantum Field Theories. 2. Scattering States}},\
  }\href {https://doi.org/10.1007/BF01211097} {\bibfield  {journal} {\bibinfo
  {journal} {Commun. Math. Phys.}\ }\textbf {\bibinfo {volume} {105}},\
  \bibinfo {pages} {153} (\bibinfo {year} {1986})}\BibitemShut {NoStop}%
\bibitem [{\citenamefont {Luscher}\ and\ \citenamefont
  {Wolff}(1990)}]{NPB339-222}%
  \BibitemOpen
  \bibfield  {author} {\bibinfo {author} {\bibfnamefont {M.}~\bibnamefont
  {Luscher}}\ and\ \bibinfo {author} {\bibfnamefont {U.}~\bibnamefont
  {Wolff}},\ }\bibfield  {title} {\bibinfo {title} {{How to Calculate the
  Elastic Scattering Matrix in Two-dimensional Quantum Field Theories by
  Numerical Simulation}},\ }\href
  {https://doi.org/10.1016/0550-3213(90)90540-T} {\bibfield  {journal}
  {\bibinfo  {journal} {Nucl. Phys. B}\ }\textbf {\bibinfo {volume} {339}},\
  \bibinfo {pages} {222} (\bibinfo {year} {1990})}\BibitemShut {NoStop}%
\bibitem [{\citenamefont {Luscher}(1991{\natexlab{a}})}]{NPB354-531}%
  \BibitemOpen
  \bibfield  {author} {\bibinfo {author} {\bibfnamefont {M.}~\bibnamefont
  {Luscher}},\ }\bibfield  {title} {\bibinfo {title} {{Two particle states on a
  torus and their relation to the scattering matrix}},\ }\href
  {https://doi.org/10.1016/0550-3213(91)90366-6} {\bibfield  {journal}
  {\bibinfo  {journal} {Nucl. Phys. B}\ }\textbf {\bibinfo {volume} {354}},\
  \bibinfo {pages} {531} (\bibinfo {year} {1991}{\natexlab{a}})}\BibitemShut
  {NoStop}%
\bibitem [{\citenamefont {Luscher}(1991{\natexlab{b}})}]{NPB364-237}%
  \BibitemOpen
  \bibfield  {author} {\bibinfo {author} {\bibfnamefont {M.}~\bibnamefont
  {Luscher}},\ }\bibfield  {title} {\bibinfo {title} {{Signatures of unstable
  particles in finite volume}},\ }\href
  {https://doi.org/10.1016/0550-3213(91)90584-K} {\bibfield  {journal}
  {\bibinfo  {journal} {Nucl. Phys. B}\ }\textbf {\bibinfo {volume} {364}},\
  \bibinfo {pages} {237} (\bibinfo {year} {1991}{\natexlab{b}})}\BibitemShut
  {NoStop}%
\bibitem [{\citenamefont {Lu}\ \emph {et~al.}(2016)\citenamefont {Lu},
  \citenamefont {L\"ahde}, \citenamefont {Lee},\ and\ \citenamefont
  {Mei\ss{}ner}}]{PLB760-309}%
  \BibitemOpen
  \bibfield  {author} {\bibinfo {author} {\bibfnamefont {B.-N.}\ \bibnamefont
  {Lu}}, \bibinfo {author} {\bibfnamefont {T.~A.}\ \bibnamefont {L\"ahde}},
  \bibinfo {author} {\bibfnamefont {D.}~\bibnamefont {Lee}},\ and\ \bibinfo
  {author} {\bibfnamefont {U.-G.}\ \bibnamefont {Mei\ss{}ner}},\ }\bibfield
  {title} {\bibinfo {title} {{Precise determination of lattice phase shifts and
  mixing angles}},\ }\href {https://doi.org/10.1016/j.physletb.2016.06.081}
  {\bibfield  {journal} {\bibinfo  {journal} {Phys. Lett. B}\ }\textbf
  {\bibinfo {volume} {760}},\ \bibinfo {pages} {309} (\bibinfo {year}
  {2016})},\ \Eprint {https://arxiv.org/abs/1506.05652} {arXiv:1506.05652
  [nucl-th]} \BibitemShut {NoStop}%
\bibitem [{\citenamefont {K\"onig}\ \emph {et~al.}(2017)\citenamefont
  {K\"onig}, \citenamefont {Grie\ss{}hammer}, \citenamefont {Hammer},\ and\
  \citenamefont {van Kolck}}]{PRL118-202501}%
  \BibitemOpen
  \bibfield  {author} {\bibinfo {author} {\bibfnamefont {S.}~\bibnamefont
  {K\"onig}}, \bibinfo {author} {\bibfnamefont {H.~W.}\ \bibnamefont
  {Grie\ss{}hammer}}, \bibinfo {author} {\bibfnamefont {H.~W.}\ \bibnamefont
  {Hammer}},\ and\ \bibinfo {author} {\bibfnamefont {U.}~\bibnamefont {van
  Kolck}},\ }\bibfield  {title} {\bibinfo {title} {{Nuclear Physics Around the
  Unitarity Limit}},\ }\href {https://doi.org/10.1103/PhysRevLett.118.202501}
  {\bibfield  {journal} {\bibinfo  {journal} {Phys. Rev. Lett.}\ }\textbf
  {\bibinfo {volume} {118}},\ \bibinfo {pages} {202501} (\bibinfo {year}
  {2017})},\ \Eprint {https://arxiv.org/abs/1607.04623} {arXiv:1607.04623
  [nucl-th]} \BibitemShut {NoStop}%
\bibitem [{\citenamefont {K\"onig}(2020)}]{EPJA56-113}%
  \BibitemOpen
  \bibfield  {author} {\bibinfo {author} {\bibfnamefont {S.}~\bibnamefont
  {K\"onig}},\ }\bibfield  {title} {\bibinfo {title} {{Energies and radii of
  light nuclei around unitarity}},\ }\href
  {https://doi.org/10.1140/epja/s10050-020-00098-9} {\bibfield  {journal}
  {\bibinfo  {journal} {Eur. Phys. J. A}\ }\textbf {\bibinfo {volume} {56}},\
  \bibinfo {pages} {113} (\bibinfo {year} {2020})},\ \Eprint
  {https://arxiv.org/abs/1910.12627} {arXiv:1910.12627 [nucl-th]} \BibitemShut
  {NoStop}%
\bibitem [{\citenamefont {Vanasse}\ and\ \citenamefont
  {Phillips}(2017)}]{Few-Body.Syst.58-26}%
  \BibitemOpen
  \bibfield  {author} {\bibinfo {author} {\bibfnamefont {J.}~\bibnamefont
  {Vanasse}}\ and\ \bibinfo {author} {\bibfnamefont {D.~R.}\ \bibnamefont
  {Phillips}},\ }\bibfield  {title} {\bibinfo {title} {{Three-nucleon bound
  states and the Wigner-SU(4) limit}},\ }\href
  {https://doi.org/10.1007/s00601-016-1173-2} {\bibfield  {journal} {\bibinfo
  {journal} {Few Body Syst.}\ }\textbf {\bibinfo {volume} {58}},\ \bibinfo
  {pages} {26} (\bibinfo {year} {2017})},\ \Eprint
  {https://arxiv.org/abs/1607.08585} {arXiv:1607.08585 [nucl-th]} \BibitemShut
  {NoStop}%
\bibitem [{\citenamefont {L\"ahde}\ \emph {et~al.}(2015)\citenamefont
  {L\"ahde}, \citenamefont {Luu}, \citenamefont {Lee}, \citenamefont
  {Mei\ss{}ner}, \citenamefont {Epelbaum}, \citenamefont {Krebs},\ and\
  \citenamefont {Rupak}}]{EPJA51-92}%
  \BibitemOpen
  \bibfield  {author} {\bibinfo {author} {\bibfnamefont {T.~A.}\ \bibnamefont
  {L\"ahde}}, \bibinfo {author} {\bibfnamefont {T.}~\bibnamefont {Luu}},
  \bibinfo {author} {\bibfnamefont {D.}~\bibnamefont {Lee}}, \bibinfo {author}
  {\bibfnamefont {U.-G.}\ \bibnamefont {Mei\ss{}ner}}, \bibinfo {author}
  {\bibfnamefont {E.}~\bibnamefont {Epelbaum}}, \bibinfo {author}
  {\bibfnamefont {H.}~\bibnamefont {Krebs}},\ and\ \bibinfo {author}
  {\bibfnamefont {G.}~\bibnamefont {Rupak}},\ }\bibfield  {title} {\bibinfo
  {title} {{Nuclear Lattice Simulations using Symmetry-Sign Extrapolation}},\
  }\href {https://doi.org/10.1140/epja/i2015-15092-1} {\bibfield  {journal}
  {\bibinfo  {journal} {Eur. Phys. J. A}\ }\textbf {\bibinfo {volume} {51}},\
  \bibinfo {pages} {92} (\bibinfo {year} {2015})},\ \Eprint
  {https://arxiv.org/abs/1502.06787} {arXiv:1502.06787 [nucl-th]} \BibitemShut
  {NoStop}%
\bibitem [{\citenamefont {Barrett}\ \emph {et~al.}(2013)\citenamefont
  {Barrett}, \citenamefont {Navratil},\ and\ \citenamefont
  {Vary}}]{PPNP69-131}%
  \BibitemOpen
  \bibfield  {author} {\bibinfo {author} {\bibfnamefont {B.~R.}\ \bibnamefont
  {Barrett}}, \bibinfo {author} {\bibfnamefont {P.}~\bibnamefont {Navratil}},\
  and\ \bibinfo {author} {\bibfnamefont {J.~P.}\ \bibnamefont {Vary}},\
  }\bibfield  {title} {\bibinfo {title} {{Ab initio no core shell model}},\
  }\href {https://doi.org/10.1016/j.ppnp.2012.10.003} {\bibfield  {journal}
  {\bibinfo  {journal} {Prog. Part. Nucl. Phys.}\ }\textbf {\bibinfo {volume}
  {69}},\ \bibinfo {pages} {131} (\bibinfo {year} {2013})}\BibitemShut
  {NoStop}%
\bibitem [{\citenamefont {Hagen}\ \emph {et~al.}(2014)\citenamefont {Hagen},
  \citenamefont {Papenbrock}, \citenamefont {Hjorth-Jensen},\ and\
  \citenamefont {Dean}}]{Rept.Prog.Phys.77-096302}%
  \BibitemOpen
  \bibfield  {author} {\bibinfo {author} {\bibfnamefont {G.}~\bibnamefont
  {Hagen}}, \bibinfo {author} {\bibfnamefont {T.}~\bibnamefont {Papenbrock}},
  \bibinfo {author} {\bibfnamefont {M.}~\bibnamefont {Hjorth-Jensen}},\ and\
  \bibinfo {author} {\bibfnamefont {D.~J.}\ \bibnamefont {Dean}},\ }\bibfield
  {title} {\bibinfo {title} {{Coupled-cluster computations of atomic nuclei}},\
  }\href {https://doi.org/10.1088/0034-4885/77/9/096302} {\bibfield  {journal}
  {\bibinfo  {journal} {Rept. Prog. Phys.}\ }\textbf {\bibinfo {volume} {77}},\
  \bibinfo {pages} {096302} (\bibinfo {year} {2014})},\ \Eprint
  {https://arxiv.org/abs/1312.7872} {arXiv:1312.7872 [nucl-th]} \BibitemShut
  {NoStop}%
\bibitem [{\citenamefont {Liu}\ \emph {et~al.}(2025)\citenamefont {Liu},
  \citenamefont {Wang},\ and\ \citenamefont {Lu}}]{arXiv2502.13565}%
  \BibitemOpen
  \bibfield  {author} {\bibinfo {author} {\bibfnamefont {J.}~\bibnamefont
  {Liu}}, \bibinfo {author} {\bibfnamefont {T.}~\bibnamefont {Wang}},\ and\
  \bibinfo {author} {\bibfnamefont {B.-N.}\ \bibnamefont {Lu}},\ }\bibfield
  {title} {\bibinfo {title} {{Perturbative quantum Monte Carlo calculation with
  high-fidelity nuclear forces}},\ }\href@noop {} {\  (\bibinfo {year}
  {2025})},\ \Eprint {https://arxiv.org/abs/2502.13565} {arXiv:2502.13565
  [nucl-th]} \BibitemShut {NoStop}%
\bibitem [{\citenamefont {Alarc\'on}\ \emph {et~al.}(2017)\citenamefont
  {Alarc\'on}, \citenamefont {Du}, \citenamefont {Klein}, \citenamefont
  {L\"ahde}, \citenamefont {Lee}, \citenamefont {Li}, \citenamefont {Lu},
  \citenamefont {Luu},\ and\ \citenamefont {Mei\ss{}ner}}]{EPJA53-83}%
  \BibitemOpen
  \bibfield  {author} {\bibinfo {author} {\bibfnamefont {J.~M.}\ \bibnamefont
  {Alarc\'on}}, \bibinfo {author} {\bibfnamefont {D.}~\bibnamefont {Du}},
  \bibinfo {author} {\bibfnamefont {N.}~\bibnamefont {Klein}}, \bibinfo
  {author} {\bibfnamefont {T.~A.}\ \bibnamefont {L\"ahde}}, \bibinfo {author}
  {\bibfnamefont {D.}~\bibnamefont {Lee}}, \bibinfo {author} {\bibfnamefont
  {N.}~\bibnamefont {Li}}, \bibinfo {author} {\bibfnamefont {B.-N.}\
  \bibnamefont {Lu}}, \bibinfo {author} {\bibfnamefont {T.}~\bibnamefont
  {Luu}},\ and\ \bibinfo {author} {\bibfnamefont {U.-G.}\ \bibnamefont
  {Mei\ss{}ner}},\ }\bibfield  {title} {\bibinfo {title} {{Neutron-proton
  scattering at next-to-next-to-leading order in Nuclear Lattice Effective
  Field Theory}},\ }\href {https://doi.org/10.1140/epja/i2017-12273-x}
  {\bibfield  {journal} {\bibinfo  {journal} {Eur. Phys. J. A}\ }\textbf
  {\bibinfo {volume} {53}},\ \bibinfo {pages} {83} (\bibinfo {year} {2017})},\
  \Eprint {https://arxiv.org/abs/1702.05319} {arXiv:1702.05319 [nucl-th]}
  \BibitemShut {NoStop}%
\bibitem [{\citenamefont {Li}\ \emph {et~al.}(2018)\citenamefont {Li},
  \citenamefont {Elhatisari}, \citenamefont {Epelbaum}, \citenamefont {Lee},
  \citenamefont {Lu},\ and\ \citenamefont {Mei\ss{}ner}}]{PhysRevC.98.044002}%
  \BibitemOpen
  \bibfield  {author} {\bibinfo {author} {\bibfnamefont {N.}~\bibnamefont
  {Li}}, \bibinfo {author} {\bibfnamefont {S.}~\bibnamefont {Elhatisari}},
  \bibinfo {author} {\bibfnamefont {E.}~\bibnamefont {Epelbaum}}, \bibinfo
  {author} {\bibfnamefont {D.}~\bibnamefont {Lee}}, \bibinfo {author}
  {\bibfnamefont {B.-N.}\ \bibnamefont {Lu}},\ and\ \bibinfo {author}
  {\bibfnamefont {U.-G.}\ \bibnamefont {Mei\ss{}ner}},\ }\bibfield  {title}
  {\bibinfo {title} {{Neutron-proton scattering with lattice chiral effective
  field theory at next-to-next-to-next-to-leading order}},\ }\href
  {https://doi.org/10.1103/PhysRevC.98.044002} {\bibfield  {journal} {\bibinfo
  {journal} {Phys. Rev. C}\ }\textbf {\bibinfo {volume} {98}},\ \bibinfo
  {pages} {044002} (\bibinfo {year} {2018})},\ \Eprint
  {https://arxiv.org/abs/1806.07994} {arXiv:1806.07994 [nucl-th]} \BibitemShut
  {NoStop}%
\bibitem [{\citenamefont {Zuker}\ \emph {et~al.}(2002)\citenamefont {Zuker},
  \citenamefont {Lenzi}, \citenamefont {Martinez-Pinedo},\ and\ \citenamefont
  {Poves}}]{PRL89-142502}%
  \BibitemOpen
  \bibfield  {author} {\bibinfo {author} {\bibfnamefont {A.~P.}\ \bibnamefont
  {Zuker}}, \bibinfo {author} {\bibfnamefont {S.~M.}\ \bibnamefont {Lenzi}},
  \bibinfo {author} {\bibfnamefont {G.}~\bibnamefont {Martinez-Pinedo}},\ and\
  \bibinfo {author} {\bibfnamefont {A.}~\bibnamefont {Poves}},\ }\bibfield
  {title} {\bibinfo {title} {{Isobaric multiplet yrast energies and isospin
  nonconserving forces}},\ }\href
  {https://doi.org/10.1103/PhysRevLett.89.142502} {\bibfield  {journal}
  {\bibinfo  {journal} {Phys. Rev. Lett.}\ }\textbf {\bibinfo {volume} {89}},\
  \bibinfo {pages} {142502} (\bibinfo {year} {2002})},\ \Eprint
  {https://arxiv.org/abs/nucl-th/0204053} {arXiv:nucl-th/0204053} \BibitemShut
  {NoStop}%
\bibitem [{\citenamefont {Suzuki}\ and\ \citenamefont
  {Otsuka}(1998)}]{Suzuki:1998jgw}%
  \BibitemOpen
  \bibfield  {author} {\bibinfo {author} {\bibfnamefont {T.}~\bibnamefont
  {Suzuki}}\ and\ \bibinfo {author} {\bibfnamefont {T.}~\bibnamefont
  {Otsuka}},\ }\bibfield  {title} {\bibinfo {title} {{Effects of halo on the
  Coulomb displacement energy of the isobaric analog state of $^{11}$Li}},\
  }\href {https://doi.org/10.1016/S0375-9474(98)00177-8} {\bibfield  {journal}
  {\bibinfo  {journal} {Nucl. Phys. A}\ }\textbf {\bibinfo {volume} {635}},\
  \bibinfo {pages} {86} (\bibinfo {year} {1998})},\ \bibinfo {note} {[Erratum:
  Nucl.Phys.A 640, 507--507 (1998)]}\BibitemShut {NoStop}%
\bibitem [{\citenamefont {Suzuki}\ \emph {et~al.}(1993)\citenamefont {Suzuki},
  \citenamefont {Sagawa},\ and\ \citenamefont {Van~Giai}}]{Suzuki:1993zz}%
  \BibitemOpen
  \bibfield  {author} {\bibinfo {author} {\bibfnamefont {T.}~\bibnamefont
  {Suzuki}}, \bibinfo {author} {\bibfnamefont {H.}~\bibnamefont {Sagawa}},\
  and\ \bibinfo {author} {\bibfnamefont {N.}~\bibnamefont {Van~Giai}},\
  }\bibfield  {title} {\bibinfo {title} {{Charge independence and charge
  symmetry breaking interactions and the Coulomb energy anomaly in isobaric
  analog states}},\ }\href {https://doi.org/10.1103/PhysRevC.47.R1360}
  {\bibfield  {journal} {\bibinfo  {journal} {Phys. Rev. C}\ }\textbf {\bibinfo
  {volume} {47}},\ \bibinfo {pages} {R1360} (\bibinfo {year}
  {1993})}\BibitemShut {NoStop}%
\bibitem [{\citenamefont {Kaneko}\ \emph
  {et~al.}(2017{\natexlab{a}})\citenamefont {Kaneko}, \citenamefont {Sun},
  \citenamefont {Mizusaki}, \citenamefont {Tazaki},\ and\ \citenamefont
  {Ghorui}}]{PLB773-521}%
  \BibitemOpen
  \bibfield  {author} {\bibinfo {author} {\bibfnamefont {K.}~\bibnamefont
  {Kaneko}}, \bibinfo {author} {\bibfnamefont {Y.}~\bibnamefont {Sun}},
  \bibinfo {author} {\bibfnamefont {T.}~\bibnamefont {Mizusaki}}, \bibinfo
  {author} {\bibfnamefont {S.}~\bibnamefont {Tazaki}},\ and\ \bibinfo {author}
  {\bibfnamefont {S.~K.}\ \bibnamefont {Ghorui}},\ }\bibfield  {title}
  {\bibinfo {title} {{Isospin-symmetry breaking in superallowed Fermi
  $\beta$-decay due to isospin-nonconserving forces}},\ }\href
  {https://doi.org/10.1016/j.physletb.2017.08.056} {\bibfield  {journal}
  {\bibinfo  {journal} {Phys. Lett. B}\ }\textbf {\bibinfo {volume} {773}},\
  \bibinfo {pages} {521} (\bibinfo {year} {2017}{\natexlab{a}})},\ \Eprint
  {https://arxiv.org/abs/1708.08627} {arXiv:1708.08627 [nucl-th]} \BibitemShut
  {NoStop}%
\bibitem [{\citenamefont {Sagawa}\ \emph {et~al.}(1996)\citenamefont {Sagawa},
  \citenamefont {Van~Giai},\ and\ \citenamefont {Suzuki}}]{Sagawa:1996zz}%
  \BibitemOpen
  \bibfield  {author} {\bibinfo {author} {\bibfnamefont {H.}~\bibnamefont
  {Sagawa}}, \bibinfo {author} {\bibfnamefont {N.}~\bibnamefont {Van~Giai}},\
  and\ \bibinfo {author} {\bibfnamefont {T.}~\bibnamefont {Suzuki}},\
  }\bibfield  {title} {\bibinfo {title} {{Effect of isospin mixing on
  superallowed Fermi beta decay}},\ }\href
  {https://doi.org/10.1103/PhysRevC.53.2163} {\bibfield  {journal} {\bibinfo
  {journal} {Phys. Rev. C}\ }\textbf {\bibinfo {volume} {53}},\ \bibinfo
  {pages} {2163} (\bibinfo {year} {1996})}\BibitemShut {NoStop}%
\bibitem [{\citenamefont {Sagawa}\ \emph {et~al.}(1995)\citenamefont {Sagawa},
  \citenamefont {Nguyen Van~Giai},\ and\ \citenamefont
  {Suzuki}}]{Sagawa:1995qpt}%
  \BibitemOpen
  \bibfield  {author} {\bibinfo {author} {\bibfnamefont {H.}~\bibnamefont
  {Sagawa}}, \bibinfo {author} {\bibfnamefont {N.}~\bibnamefont {Nguyen
  Van~Giai}},\ and\ \bibinfo {author} {\bibfnamefont {T.}~\bibnamefont
  {Suzuki}},\ }\bibfield  {title} {\bibinfo {title} {{Isospin mixing and the
  sum rule of super-allowed fermi \ensuremath{\beta} decay}},\ }\href
  {https://doi.org/10.1016/0370-2693(95)00498-A} {\bibfield  {journal}
  {\bibinfo  {journal} {Phys. Lett. B}\ }\textbf {\bibinfo {volume} {353}},\
  \bibinfo {pages} {7} (\bibinfo {year} {1995})}\BibitemShut {NoStop}%
\bibitem [{\citenamefont {Kaneko}\ \emph
  {et~al.}(2017{\natexlab{b}})\citenamefont {Kaneko}, \citenamefont {Sun},
  \citenamefont {Mizusaki}, \citenamefont {Tazaki},\ and\ \citenamefont
  {Ghorui}}]{Kaneko:2017lvv}%
  \BibitemOpen
  \bibfield  {author} {\bibinfo {author} {\bibfnamefont {K.}~\bibnamefont
  {Kaneko}}, \bibinfo {author} {\bibfnamefont {Y.}~\bibnamefont {Sun}},
  \bibinfo {author} {\bibfnamefont {T.}~\bibnamefont {Mizusaki}}, \bibinfo
  {author} {\bibfnamefont {S.}~\bibnamefont {Tazaki}},\ and\ \bibinfo {author}
  {\bibfnamefont {S.~K.}\ \bibnamefont {Ghorui}},\ }\bibfield  {title}
  {\bibinfo {title} {{Isospin-symmetry breaking in superallowed Fermi
  $\beta$-decay due to isospin-nonconserving forces}},\ }\href
  {https://doi.org/10.1016/j.physletb.2017.08.056} {\bibfield  {journal}
  {\bibinfo  {journal} {Phys. Lett. B}\ }\textbf {\bibinfo {volume} {773}},\
  \bibinfo {pages} {521} (\bibinfo {year} {2017}{\natexlab{b}})},\ \Eprint
  {https://arxiv.org/abs/1708.08627} {arXiv:1708.08627 [nucl-th]} \BibitemShut
  {NoStop}%
\bibitem [{\citenamefont {Ekman}\ \emph {et~al.}(2004)\citenamefont {Ekman}
  \emph {et~al.}}]{PRL92-132502}%
  \BibitemOpen
  \bibfield  {author} {\bibinfo {author} {\bibfnamefont {J.}~\bibnamefont
  {Ekman}} \emph {et~al.},\ }\bibfield  {title} {\bibinfo {title} {{Unusual
  Isospin-Breaking and Isospin-Mixing Effects in the A=35 Mirror Nuclei}},\
  }\href {https://doi.org/10.1103/PhysRevLett.92.132502} {\bibfield  {journal}
  {\bibinfo  {journal} {Phys. Rev. Lett.}\ }\textbf {\bibinfo {volume} {92}},\
  \bibinfo {pages} {132502} (\bibinfo {year} {2004})}\BibitemShut {NoStop}%
\bibitem [{\citenamefont {Bentley}\ \emph {et~al.}(2006)\citenamefont
  {Bentley}, \citenamefont {Chandler}, \citenamefont {Taylor}, \citenamefont
  {Brown}, \citenamefont {Carpenter}, \citenamefont {Davids}, \citenamefont
  {Ekman}, \citenamefont {Freeman}, \citenamefont {Garrett}, \citenamefont
  {Hammond}, \citenamefont {Janssens}, \citenamefont {Lenzi}, \citenamefont
  {Lister}, \citenamefont {du~Rietz},\ and\ \citenamefont
  {Seweryniak}}]{PRL97-132501}%
  \BibitemOpen
  \bibfield  {author} {\bibinfo {author} {\bibfnamefont {M.~A.}\ \bibnamefont
  {Bentley}}, \bibinfo {author} {\bibfnamefont {C.}~\bibnamefont {Chandler}},
  \bibinfo {author} {\bibfnamefont {M.~J.}\ \bibnamefont {Taylor}}, \bibinfo
  {author} {\bibfnamefont {J.~R.}\ \bibnamefont {Brown}}, \bibinfo {author}
  {\bibfnamefont {M.~P.}\ \bibnamefont {Carpenter}}, \bibinfo {author}
  {\bibfnamefont {C.}~\bibnamefont {Davids}}, \bibinfo {author} {\bibfnamefont
  {J.}~\bibnamefont {Ekman}}, \bibinfo {author} {\bibfnamefont {S.~J.}\
  \bibnamefont {Freeman}}, \bibinfo {author} {\bibfnamefont {P.~E.}\
  \bibnamefont {Garrett}}, \bibinfo {author} {\bibfnamefont {G.}~\bibnamefont
  {Hammond}}, \bibinfo {author} {\bibfnamefont {R.~V.~F.}\ \bibnamefont
  {Janssens}}, \bibinfo {author} {\bibfnamefont {S.~M.}\ \bibnamefont {Lenzi}},
  \bibinfo {author} {\bibfnamefont {C.~J.}\ \bibnamefont {Lister}}, \bibinfo
  {author} {\bibfnamefont {R.}~\bibnamefont {du~Rietz}},\ and\ \bibinfo
  {author} {\bibfnamefont {D.}~\bibnamefont {Seweryniak}},\ }\bibfield  {title}
  {\bibinfo {title} {Isospin symmetry of odd-odd mirror nuclei: Identification
  of excited states in $n=z\ensuremath{-}2$ $^{48}\mathrm{Mn}$},\ }\href
  {https://doi.org/10.1103/PhysRevLett.97.132501} {\bibfield  {journal}
  {\bibinfo  {journal} {Phys. Rev. Lett.}\ }\textbf {\bibinfo {volume} {97}},\
  \bibinfo {pages} {132501} (\bibinfo {year} {2006})}\BibitemShut {NoStop}%
\bibitem [{\citenamefont {Gadea}\ \emph {et~al.}(2006)\citenamefont {Gadea}
  \emph {et~al.}}]{PRL97-152501}%
  \BibitemOpen
  \bibfield  {author} {\bibinfo {author} {\bibfnamefont {A.}~\bibnamefont
  {Gadea}} \emph {et~al.},\ }\bibfield  {title} {\bibinfo {title} {{Observation
  of Ni-54: Cross-Conjugate Symmetry in f7/2 Mirror Energy Differences}},\
  }\href {https://doi.org/10.1103/PhysRevLett.97.152501} {\bibfield  {journal}
  {\bibinfo  {journal} {Phys. Rev. Lett.}\ }\textbf {\bibinfo {volume} {97}},\
  \bibinfo {pages} {152501} (\bibinfo {year} {2006})}\BibitemShut {NoStop}%
\bibitem [{\citenamefont {WANG}\ \emph {et~al.}(2024)\citenamefont {WANG},
  \citenamefont {LI},\ and\ \citenamefont {LI}}]{NPR41-233}%
  \BibitemOpen
  \bibfield  {author} {\bibinfo {author} {\bibfnamefont {X.}~\bibnamefont
  {WANG}}, \bibinfo {author} {\bibfnamefont {H.}~\bibnamefont {LI}},\ and\
  \bibinfo {author} {\bibfnamefont {J.}~\bibnamefont {LI}},\ }\bibfield
  {title} {\bibinfo {title} {Ab initio calculations for isospin symmetry
  breaking in mirror nuclei},\ }\href
  {https://doi.org/10.11804/NuclPhysRev.41.2023CNPC34} {\bibfield  {journal}
  {\bibinfo  {journal} {Nuclear Physics Review}\ }\textbf {\bibinfo {volume}
  {41}},\ \bibinfo {pages} {233} (\bibinfo {year} {2024})}\BibitemShut
  {NoStop}%
\bibitem [{\citenamefont {Naito}\ \emph {et~al.}(2025)\citenamefont {Naito},
  \citenamefont {Hijikata}, \citenamefont {Zenihiro}, \citenamefont {Col\`o},\
  and\ \citenamefont {Sagawa}}]{Naito:2025qub}%
  \BibitemOpen
  \bibfield  {author} {\bibinfo {author} {\bibfnamefont {T.}~\bibnamefont
  {Naito}}, \bibinfo {author} {\bibfnamefont {Y.}~\bibnamefont {Hijikata}},
  \bibinfo {author} {\bibfnamefont {J.}~\bibnamefont {Zenihiro}}, \bibinfo
  {author} {\bibfnamefont {G.}~\bibnamefont {Col\`o}},\ and\ \bibinfo {author}
  {\bibfnamefont {H.}~\bibnamefont {Sagawa}},\ }\bibfield  {title} {\bibinfo
  {title} {{Mirror-skin thickness: A possible observable sensitive to the
  charge symmetry breaking energy density functional}},\ }\href@noop {} {\
  (\bibinfo {year} {2025})},\ \Eprint {https://arxiv.org/abs/2503.05147}
  {arXiv:2503.05147 [nucl-th]} \BibitemShut {NoStop}%
\bibitem [{\citenamefont {Sarma}\ \emph {et~al.}(2024)\citenamefont {Sarma},
  \citenamefont {Srivastava}, \citenamefont {Suzuki},\ and\ \citenamefont
  {Shimizu}}]{Sarma:2024bow}%
  \BibitemOpen
  \bibfield  {author} {\bibinfo {author} {\bibfnamefont {C.}~\bibnamefont
  {Sarma}}, \bibinfo {author} {\bibfnamefont {P.~C.}\ \bibnamefont
  {Srivastava}}, \bibinfo {author} {\bibfnamefont {T.}~\bibnamefont {Suzuki}},\
  and\ \bibinfo {author} {\bibfnamefont {N.}~\bibnamefont {Shimizu}},\
  }\bibfield  {title} {\bibinfo {title} {{Mirror and triplet energy differences
  in $sd$-shell nuclei using microscopic interactions with isospin-symmetry
  breaking effects}},\ }\href@noop {} {\  (\bibinfo {year} {2024})},\ \Eprint
  {https://arxiv.org/abs/2410.00765} {arXiv:2410.00765 [nucl-th]} \BibitemShut
  {NoStop}%
\bibitem [{\citenamefont {Li}\ \emph {et~al.}(2023)\citenamefont {Li},
  \citenamefont {Yuan}, \citenamefont {Li}, \citenamefont {Xie}, \citenamefont
  {Zhang}, \citenamefont {Zhang}, \citenamefont {Xu}, \citenamefont {Michel},
  \citenamefont {Xu},\ and\ \citenamefont {Zuo}}]{Li:2023hky}%
  \BibitemOpen
  \bibfield  {author} {\bibinfo {author} {\bibfnamefont {H.~H.}\ \bibnamefont
  {Li}}, \bibinfo {author} {\bibfnamefont {Q.}~\bibnamefont {Yuan}}, \bibinfo
  {author} {\bibfnamefont {J.~G.}\ \bibnamefont {Li}}, \bibinfo {author}
  {\bibfnamefont {M.~R.}\ \bibnamefont {Xie}}, \bibinfo {author} {\bibfnamefont
  {S.}~\bibnamefont {Zhang}}, \bibinfo {author} {\bibfnamefont {Y.~H.}\
  \bibnamefont {Zhang}}, \bibinfo {author} {\bibfnamefont {X.~X.}\ \bibnamefont
  {Xu}}, \bibinfo {author} {\bibfnamefont {N.}~\bibnamefont {Michel}}, \bibinfo
  {author} {\bibfnamefont {F.~R.}\ \bibnamefont {Xu}},\ and\ \bibinfo {author}
  {\bibfnamefont {W.}~\bibnamefont {Zuo}},\ }\bibfield  {title} {\bibinfo
  {title} {{Investigation of isospin-symmetry breaking in mirror energy
  difference and nuclear mass with ab initio calculations}},\ }\href
  {https://doi.org/10.1103/PhysRevC.107.014302} {\bibfield  {journal} {\bibinfo
   {journal} {Phys. Rev. C}\ }\textbf {\bibinfo {volume} {107}},\ \bibinfo
  {pages} {014302} (\bibinfo {year} {2023})},\ \Eprint
  {https://arxiv.org/abs/2303.11022} {arXiv:2303.11022 [nucl-th]} \BibitemShut
  {NoStop}%
\bibitem [{\citenamefont {Stoks}\ \emph {et~al.}(1988)\citenamefont {Stoks},
  \citenamefont {Van~Campen}, \citenamefont {Rijken},\ and\ \citenamefont
  {De~Swart}}]{Stoks:1988vn}%
  \BibitemOpen
  \bibfield  {author} {\bibinfo {author} {\bibfnamefont {V.~G.~J.}\
  \bibnamefont {Stoks}}, \bibinfo {author} {\bibfnamefont {P.~C.}\ \bibnamefont
  {Van~Campen}}, \bibinfo {author} {\bibfnamefont {T.~A.}\ \bibnamefont
  {Rijken}},\ and\ \bibinfo {author} {\bibfnamefont {J.~J.}\ \bibnamefont
  {De~Swart}},\ }\bibfield  {title} {\bibinfo {title} {{Evidence for a Large
  Breaking of Charge Independence in the $N N$ Interaction}},\ }\href
  {https://doi.org/10.1103/PhysRevLett.61.1702} {\bibfield  {journal} {\bibinfo
   {journal} {Phys. Rev. Lett.}\ }\textbf {\bibinfo {volume} {61}},\ \bibinfo
  {pages} {1702} (\bibinfo {year} {1988})}\BibitemShut {NoStop}%
\bibitem [{\citenamefont {Stoks}\ \emph {et~al.}(1993)\citenamefont {Stoks},
  \citenamefont {Klomp}, \citenamefont {Rentmeester},\ and\ \citenamefont
  {de~Swart}}]{Stoks:1993tb_Nij}%
  \BibitemOpen
  \bibfield  {author} {\bibinfo {author} {\bibfnamefont {V.~G.~J.}\
  \bibnamefont {Stoks}}, \bibinfo {author} {\bibfnamefont {R.~A.~M.}\
  \bibnamefont {Klomp}}, \bibinfo {author} {\bibfnamefont {M.~C.~M.}\
  \bibnamefont {Rentmeester}},\ and\ \bibinfo {author} {\bibfnamefont {J.~J.}\
  \bibnamefont {de~Swart}},\ }\bibfield  {title} {\bibinfo {title} {{Partial
  wave analaysis of all nucleon-nucleon scattering data below 350-MeV}},\
  }\href {https://doi.org/10.1103/PhysRevC.48.792} {\bibfield  {journal}
  {\bibinfo  {journal} {Phys. Rev. C}\ }\textbf {\bibinfo {volume} {48}},\
  \bibinfo {pages} {792} (\bibinfo {year} {1993})}\BibitemShut {NoStop}%
\bibitem [{\citenamefont {Bergervoet}\ \emph {et~al.}(1988)\citenamefont
  {Bergervoet}, \citenamefont {van Campen}, \citenamefont {van~der Sanden},\
  and\ \citenamefont {de~Swart}}]{Bergervoet:1988zz_Nij}%
  \BibitemOpen
  \bibfield  {author} {\bibinfo {author} {\bibfnamefont {J.~R.}\ \bibnamefont
  {Bergervoet}}, \bibinfo {author} {\bibfnamefont {P.~C.}\ \bibnamefont {van
  Campen}}, \bibinfo {author} {\bibfnamefont {W.~A.}\ \bibnamefont {van~der
  Sanden}},\ and\ \bibinfo {author} {\bibfnamefont {J.~J.}\ \bibnamefont
  {de~Swart}},\ }\bibfield  {title} {\bibinfo {title} {{Phase shift analysis of
  0-30 MeV pp scattering data}},\ }\href
  {https://doi.org/10.1103/PhysRevC.38.15} {\bibfield  {journal} {\bibinfo
  {journal} {Phys. Rev. C}\ }\textbf {\bibinfo {volume} {38}},\ \bibinfo
  {pages} {15} (\bibinfo {year} {1988})}\BibitemShut {NoStop}%
\bibitem [{\citenamefont {Machleidt}(1989)}]{Machleidt:1989tm_csb1}%
  \BibitemOpen
  \bibfield  {author} {\bibinfo {author} {\bibfnamefont {R.}~\bibnamefont
  {Machleidt}},\ }\bibfield  {title} {\bibinfo {title} {{The Meson theory of
  nuclear forces and nuclear structure}},\ }\href@noop {} {\bibfield  {journal}
  {\bibinfo  {journal} {Adv. Nucl. Phys.}\ }\textbf {\bibinfo {volume} {19}},\
  \bibinfo {pages} {189} (\bibinfo {year} {1989})}\BibitemShut {NoStop}%
\bibitem [{\citenamefont {Li}\ and\ \citenamefont
  {Machleidt}(1998{\natexlab{a}})}]{Li:1998xh_csb2}%
  \BibitemOpen
  \bibfield  {author} {\bibinfo {author} {\bibfnamefont {G.-Q.}\ \bibnamefont
  {Li}}\ and\ \bibinfo {author} {\bibfnamefont {R.}~\bibnamefont {Machleidt}},\
  }\bibfield  {title} {\bibinfo {title} {{Charge dependence of the
  nucleon-nucleon interaction}},\ }\href
  {https://doi.org/10.1103/PhysRevC.58.3153} {\bibfield  {journal} {\bibinfo
  {journal} {Phys. Rev. C}\ }\textbf {\bibinfo {volume} {58}},\ \bibinfo
  {pages} {3153} (\bibinfo {year} {1998}{\natexlab{a}})},\ \Eprint
  {https://arxiv.org/abs/nucl-th/9807080} {arXiv:nucl-th/9807080} \BibitemShut
  {NoStop}%
\bibitem [{\citenamefont {Miller}\ \emph {et~al.}(1990)\citenamefont {Miller},
  \citenamefont {Nefkens},\ and\ \citenamefont {Slaus}}]{Miller:1990iz_csb3}%
  \BibitemOpen
  \bibfield  {author} {\bibinfo {author} {\bibfnamefont {G.~A.}\ \bibnamefont
  {Miller}}, \bibinfo {author} {\bibfnamefont {B.~M.~K.}\ \bibnamefont
  {Nefkens}},\ and\ \bibinfo {author} {\bibfnamefont {I.}~\bibnamefont
  {Slaus}},\ }\bibfield  {title} {\bibinfo {title} {{Charge symmetry, quarks
  and mesons}},\ }\href {https://doi.org/10.1016/0370-1573(90)90102-8}
  {\bibfield  {journal} {\bibinfo  {journal} {Phys. Rept.}\ }\textbf {\bibinfo
  {volume} {194}},\ \bibinfo {pages} {1} (\bibinfo {year} {1990})}\BibitemShut
  {NoStop}%
\bibitem [{\citenamefont {Li}\ and\ \citenamefont
  {Machleidt}(1998{\natexlab{b}})}]{Li:1998ya_csb4}%
  \BibitemOpen
  \bibfield  {author} {\bibinfo {author} {\bibfnamefont {G.-Q.}\ \bibnamefont
  {Li}}\ and\ \bibinfo {author} {\bibfnamefont {R.}~\bibnamefont {Machleidt}},\
  }\bibfield  {title} {\bibinfo {title} {{Charge asymmetry of the
  nucleon-nucleon interaction}},\ }\href
  {https://doi.org/10.1103/PhysRevC.58.1393} {\bibfield  {journal} {\bibinfo
  {journal} {Phys. Rev. C}\ }\textbf {\bibinfo {volume} {58}},\ \bibinfo
  {pages} {1393} (\bibinfo {year} {1998}{\natexlab{b}})},\ \Eprint
  {https://arxiv.org/abs/nucl-th/9804023} {arXiv:nucl-th/9804023} \BibitemShut
  {NoStop}%
\bibitem [{\citenamefont {Epelbaum}\ and\ \citenamefont
  {Meissner}(1999)}]{Epelbaum:1999zn}%
  \BibitemOpen
  \bibfield  {author} {\bibinfo {author} {\bibfnamefont {E.}~\bibnamefont
  {Epelbaum}}\ and\ \bibinfo {author} {\bibfnamefont {U.-G.}\ \bibnamefont
  {Meissner}},\ }\bibfield  {title} {\bibinfo {title} {{Charge independence
  breaking and charge symmetry breaking in the nucleon-nucleon interaction from
  effective field theory}},\ }\href
  {https://doi.org/10.1016/S0370-2693(99)00776-5} {\bibfield  {journal}
  {\bibinfo  {journal} {Phys. Lett. B}\ }\textbf {\bibinfo {volume} {461}},\
  \bibinfo {pages} {287} (\bibinfo {year} {1999})},\ \bibinfo {note} {[Erratum:
  Phys.Lett.B 467, 308--308 (1999)]},\ \Eprint
  {https://arxiv.org/abs/nucl-th/9902042} {arXiv:nucl-th/9902042} \BibitemShut
  {NoStop}%
\bibitem [{\citenamefont {Walzl}\ \emph {et~al.}(2001)\citenamefont {Walzl},
  \citenamefont {Meissner},\ and\ \citenamefont {Epelbaum}}]{Walzl:2000cx}%
  \BibitemOpen
  \bibfield  {author} {\bibinfo {author} {\bibfnamefont {M.}~\bibnamefont
  {Walzl}}, \bibinfo {author} {\bibfnamefont {U.~G.}\ \bibnamefont
  {Meissner}},\ and\ \bibinfo {author} {\bibfnamefont {E.}~\bibnamefont
  {Epelbaum}},\ }\bibfield  {title} {\bibinfo {title} {{Charge dependent
  nucleon-nucleon potential from chiral effective field theory}},\ }\href
  {https://doi.org/10.1016/S0375-9474(01)00969-1} {\bibfield  {journal}
  {\bibinfo  {journal} {Nucl. Phys. A}\ }\textbf {\bibinfo {volume} {693}},\
  \bibinfo {pages} {663} (\bibinfo {year} {2001})},\ \Eprint
  {https://arxiv.org/abs/nucl-th/0010019} {arXiv:nucl-th/0010019} \BibitemShut
  {NoStop}%
\bibitem [{\citenamefont {Epelbaum}\ and\ \citenamefont
  {Meissner}(2005)}]{Epelbaum:2005fd}%
  \BibitemOpen
  \bibfield  {author} {\bibinfo {author} {\bibfnamefont {E.}~\bibnamefont
  {Epelbaum}}\ and\ \bibinfo {author} {\bibfnamefont {U.-G.}\ \bibnamefont
  {Meissner}},\ }\bibfield  {title} {\bibinfo {title} {{Isospin-violating
  nucleon-nucleon forces using the method of unitary transformation}},\ }\href
  {https://doi.org/10.1103/PhysRevC.72.044001} {\bibfield  {journal} {\bibinfo
  {journal} {Phys. Rev. C}\ }\textbf {\bibinfo {volume} {72}},\ \bibinfo
  {pages} {044001} (\bibinfo {year} {2005})},\ \Eprint
  {https://arxiv.org/abs/nucl-th/0502052} {arXiv:nucl-th/0502052} \BibitemShut
  {NoStop}%
\bibitem [{\citenamefont {Gardestig}(2009)}]{Gardestig:2009ya_csb5}%
  \BibitemOpen
  \bibfield  {author} {\bibinfo {author} {\bibfnamefont {A.}~\bibnamefont
  {Gardestig}},\ }\bibfield  {title} {\bibinfo {title} {{Extracting the
  neutron-neutron scattering length - recent developments}},\ }\href
  {https://doi.org/10.1088/0954-3899/36/5/053001} {\bibfield  {journal}
  {\bibinfo  {journal} {J. Phys. G}\ }\textbf {\bibinfo {volume} {36}},\
  \bibinfo {pages} {053001} (\bibinfo {year} {2009})},\ \Eprint
  {https://arxiv.org/abs/0904.2787} {arXiv:0904.2787 [nucl-th]} \BibitemShut
  {NoStop}%
\bibitem [{\citenamefont {Epelbaum}\ \emph
  {et~al.}(2009{\natexlab{a}})\citenamefont {Epelbaum}, \citenamefont {Krebs},
  \citenamefont {Lee},\ and\ \citenamefont {Meissner}}]{Epelbaum:2009rkz}%
  \BibitemOpen
  \bibfield  {author} {\bibinfo {author} {\bibfnamefont {E.}~\bibnamefont
  {Epelbaum}}, \bibinfo {author} {\bibfnamefont {H.}~\bibnamefont {Krebs}},
  \bibinfo {author} {\bibfnamefont {D.}~\bibnamefont {Lee}},\ and\ \bibinfo
  {author} {\bibfnamefont {U.-G.}\ \bibnamefont {Meissner}},\ }\bibfield
  {title} {\bibinfo {title} {{Ground state energy of dilute neutron matter at
  next-to-leading order in lattice chiral effective field theory}},\ }\href
  {https://doi.org/10.1140/epja/i2009-10755-0} {\bibfield  {journal} {\bibinfo
  {journal} {Eur. Phys. J. A}\ }\textbf {\bibinfo {volume} {40}},\ \bibinfo
  {pages} {199} (\bibinfo {year} {2009}{\natexlab{a}})},\ \Eprint
  {https://arxiv.org/abs/0812.3653} {arXiv:0812.3653 [nucl-th]} \BibitemShut
  {NoStop}%
\bibitem [{\citenamefont {Epelbaum}\ \emph
  {et~al.}(2009{\natexlab{b}})\citenamefont {Epelbaum}, \citenamefont {Krebs},
  \citenamefont {Lee},\ and\ \citenamefont {Meissner}}]{Epelbaum:2009zsa}%
  \BibitemOpen
  \bibfield  {author} {\bibinfo {author} {\bibfnamefont {E.}~\bibnamefont
  {Epelbaum}}, \bibinfo {author} {\bibfnamefont {H.}~\bibnamefont {Krebs}},
  \bibinfo {author} {\bibfnamefont {D.}~\bibnamefont {Lee}},\ and\ \bibinfo
  {author} {\bibfnamefont {U.-G.}\ \bibnamefont {Meissner}},\ }\bibfield
  {title} {\bibinfo {title} {{Lattice chiral effective field theory with
  three-body interactions at next-to-next-to-leading order}},\ }\href
  {https://doi.org/10.1140/epja/i2009-10764-y} {\bibfield  {journal} {\bibinfo
  {journal} {Eur. Phys. J. A}\ }\textbf {\bibinfo {volume} {41}},\ \bibinfo
  {pages} {125} (\bibinfo {year} {2009}{\natexlab{b}})},\ \Eprint
  {https://arxiv.org/abs/0903.1666} {arXiv:0903.1666 [nucl-th]} \BibitemShut
  {NoStop}%
\bibitem [{\citenamefont {Epelbaum}\ \emph
  {et~al.}(2009{\natexlab{c}})\citenamefont {Epelbaum}, \citenamefont
  {Hammer},\ and\ \citenamefont {Meissner}}]{Epelbaum:2008ga}%
  \BibitemOpen
  \bibfield  {author} {\bibinfo {author} {\bibfnamefont {E.}~\bibnamefont
  {Epelbaum}}, \bibinfo {author} {\bibfnamefont {H.-W.}\ \bibnamefont
  {Hammer}},\ and\ \bibinfo {author} {\bibfnamefont {U.-G.}\ \bibnamefont
  {Meissner}},\ }\bibfield  {title} {\bibinfo {title} {{Modern Theory of
  Nuclear Forces}},\ }\href {https://doi.org/10.1103/RevModPhys.81.1773}
  {\bibfield  {journal} {\bibinfo  {journal} {Rev. Mod. Phys.}\ }\textbf
  {\bibinfo {volume} {81}},\ \bibinfo {pages} {1773} (\bibinfo {year}
  {2009}{\natexlab{c}})},\ \Eprint {https://arxiv.org/abs/0811.1338}
  {arXiv:0811.1338 [nucl-th]} \BibitemShut {NoStop}%
\bibitem [{\citenamefont {Machleidt}\ and\ \citenamefont
  {Entem}(2011)}]{Machleidt:2011zz}%
  \BibitemOpen
  \bibfield  {author} {\bibinfo {author} {\bibfnamefont {R.}~\bibnamefont
  {Machleidt}}\ and\ \bibinfo {author} {\bibfnamefont {D.~R.}\ \bibnamefont
  {Entem}},\ }\bibfield  {title} {\bibinfo {title} {{Chiral effective field
  theory and nuclear forces}},\ }\href
  {https://doi.org/10.1016/j.physrep.2011.02.001} {\bibfield  {journal}
  {\bibinfo  {journal} {Phys. Rept.}\ }\textbf {\bibinfo {volume} {503}},\
  \bibinfo {pages} {1} (\bibinfo {year} {2011})},\ \Eprint
  {https://arxiv.org/abs/1105.2919} {arXiv:1105.2919 [nucl-th]} \BibitemShut
  {NoStop}%
\bibitem [{\citenamefont {Weinberg}(1991)}]{Weinberg:1991um_nu}%
  \BibitemOpen
  \bibfield  {author} {\bibinfo {author} {\bibfnamefont {S.}~\bibnamefont
  {Weinberg}},\ }\bibfield  {title} {\bibinfo {title} {{Effective chiral
  Lagrangians for nucleon - pion interactions and nuclear forces}},\ }\href
  {https://doi.org/10.1016/0550-3213(91)90231-L} {\bibfield  {journal}
  {\bibinfo  {journal} {Nucl. Phys. B}\ }\textbf {\bibinfo {volume} {363}},\
  \bibinfo {pages} {3} (\bibinfo {year} {1991})}\BibitemShut {NoStop}%
\bibitem [{\citenamefont {Ordonez}\ \emph {et~al.}(1996)\citenamefont
  {Ordonez}, \citenamefont {Ray},\ and\ \citenamefont {van
  Kolck}}]{Ordonez:1995rz_Q}%
  \BibitemOpen
  \bibfield  {author} {\bibinfo {author} {\bibfnamefont {C.}~\bibnamefont
  {Ordonez}}, \bibinfo {author} {\bibfnamefont {L.}~\bibnamefont {Ray}},\ and\
  \bibinfo {author} {\bibfnamefont {U.}~\bibnamefont {van Kolck}},\ }\bibfield
  {title} {\bibinfo {title} {{The Two nucleon potential from chiral
  Lagrangians}},\ }\href {https://doi.org/10.1103/PhysRevC.53.2086} {\bibfield
  {journal} {\bibinfo  {journal} {Phys. Rev. C}\ }\textbf {\bibinfo {volume}
  {53}},\ \bibinfo {pages} {2086} (\bibinfo {year} {1996})},\ \Eprint
  {https://arxiv.org/abs/hep-ph/9511380} {arXiv:hep-ph/9511380} \BibitemShut
  {NoStop}%
\bibitem [{\citenamefont {Epelbaum}\ \emph {et~al.}(1998)\citenamefont
  {Epelbaum}, \citenamefont {Gloeckle},\ and\ \citenamefont
  {Meissner}}]{Epelbaum:1998ka}%
  \BibitemOpen
  \bibfield  {author} {\bibinfo {author} {\bibfnamefont {E.}~\bibnamefont
  {Epelbaum}}, \bibinfo {author} {\bibfnamefont {W.}~\bibnamefont {Gloeckle}},\
  and\ \bibinfo {author} {\bibfnamefont {U.-G.}\ \bibnamefont {Meissner}},\
  }\bibfield  {title} {\bibinfo {title} {{Nuclear forces from chiral
  Lagrangians using the method of unitary transformation. 1. Formalism}},\
  }\href {https://doi.org/10.1016/S0375-9474(98)00220-6} {\bibfield  {journal}
  {\bibinfo  {journal} {Nucl. Phys. A}\ }\textbf {\bibinfo {volume} {637}},\
  \bibinfo {pages} {107} (\bibinfo {year} {1998})},\ \Eprint
  {https://arxiv.org/abs/nucl-th/9801064} {arXiv:nucl-th/9801064} \BibitemShut
  {NoStop}%
\bibitem [{\citenamefont {Epelbaum}\ \emph {et~al.}(2000)\citenamefont
  {Epelbaum}, \citenamefont {Gloeckle},\ and\ \citenamefont
  {Meissner}}]{Epelbaum:1999dj}%
  \BibitemOpen
  \bibfield  {author} {\bibinfo {author} {\bibfnamefont {E.}~\bibnamefont
  {Epelbaum}}, \bibinfo {author} {\bibfnamefont {W.}~\bibnamefont {Gloeckle}},\
  and\ \bibinfo {author} {\bibfnamefont {U.-G.}\ \bibnamefont {Meissner}},\
  }\bibfield  {title} {\bibinfo {title} {{Nuclear forces from chiral
  Lagrangians using the method of unitary transformation. 2. The two nucleon
  system}},\ }\href {https://doi.org/10.1016/S0375-9474(99)00821-0} {\bibfield
  {journal} {\bibinfo  {journal} {Nucl. Phys. A}\ }\textbf {\bibinfo {volume}
  {671}},\ \bibinfo {pages} {295} (\bibinfo {year} {2000})},\ \Eprint
  {https://arxiv.org/abs/nucl-th/9910064} {arXiv:nucl-th/9910064} \BibitemShut
  {NoStop}%
\bibitem [{\citenamefont {Epelbaum}(2000)}]{Epelbaum:2000kv}%
  \BibitemOpen
  \bibfield  {author} {\bibinfo {author} {\bibfnamefont {E.}~\bibnamefont
  {Epelbaum}},\ }\emph {\bibinfo {title} {{The nucleon nucleon interaction in a
  chiral effective field theory}}},\ \href@noop {} {Master's thesis} (\bibinfo
  {year} {2000})\BibitemShut {NoStop}%
\bibitem [{\citenamefont {Gezerlis}\ \emph {et~al.}(2013)\citenamefont
  {Gezerlis}, \citenamefont {Tews}, \citenamefont {Epelbaum}, \citenamefont
  {Gandolfi}, \citenamefont {Hebeler}, \citenamefont {Nogga},\ and\
  \citenamefont {Schwenk}}]{Gezerlis:2013ipa}%
  \BibitemOpen
  \bibfield  {author} {\bibinfo {author} {\bibfnamefont {A.}~\bibnamefont
  {Gezerlis}}, \bibinfo {author} {\bibfnamefont {I.}~\bibnamefont {Tews}},
  \bibinfo {author} {\bibfnamefont {E.}~\bibnamefont {Epelbaum}}, \bibinfo
  {author} {\bibfnamefont {S.}~\bibnamefont {Gandolfi}}, \bibinfo {author}
  {\bibfnamefont {K.}~\bibnamefont {Hebeler}}, \bibinfo {author} {\bibfnamefont
  {A.}~\bibnamefont {Nogga}},\ and\ \bibinfo {author} {\bibfnamefont
  {A.}~\bibnamefont {Schwenk}},\ }\bibfield  {title} {\bibinfo {title}
  {{Quantum Monte Carlo Calculations with Chiral Effective Field Theory
  Interactions}},\ }\href {https://doi.org/10.1103/PhysRevLett.111.032501}
  {\bibfield  {journal} {\bibinfo  {journal} {Phys. Rev. Lett.}\ }\textbf
  {\bibinfo {volume} {111}},\ \bibinfo {pages} {032501} (\bibinfo {year}
  {2013})},\ \Eprint {https://arxiv.org/abs/1303.6243} {arXiv:1303.6243
  [nucl-th]} \BibitemShut {NoStop}%
\bibitem [{\citenamefont {Lu}\ and\ \citenamefont {Deng}(2023)}]{Lu:2023jyz}%
  \BibitemOpen
  \bibfield  {author} {\bibinfo {author} {\bibfnamefont {B.-N.}\ \bibnamefont
  {Lu}}\ and\ \bibinfo {author} {\bibfnamefont {B.-G.}\ \bibnamefont {Deng}},\
  }\bibfield  {title} {\bibinfo {title} {{Renormalization of many-body
  effective field theory}},\ }\href@noop {} {\  (\bibinfo {year} {2023})},\
  \Eprint {https://arxiv.org/abs/2308.14559} {arXiv:2308.14559 [nucl-th]}
  \BibitemShut {NoStop}%
\bibitem [{\citenamefont {Epelbaum}\ \emph
  {et~al.}(2015{\natexlab{a}})\citenamefont {Epelbaum}, \citenamefont {Krebs},\
  and\ \citenamefont {Mei\ss{}ner}}]{Epelbaum:2014efa_2pe}%
  \BibitemOpen
  \bibfield  {author} {\bibinfo {author} {\bibfnamefont {E.}~\bibnamefont
  {Epelbaum}}, \bibinfo {author} {\bibfnamefont {H.}~\bibnamefont {Krebs}},\
  and\ \bibinfo {author} {\bibfnamefont {U.~G.}\ \bibnamefont {Mei\ss{}ner}},\
  }\bibfield  {title} {\bibinfo {title} {{Improved chiral nucleon-nucleon
  potential up to next-to-next-to-next-to-leading order}},\ }\href
  {https://doi.org/10.1140/epja/i2015-15053-8} {\bibfield  {journal} {\bibinfo
  {journal} {Eur. Phys. J. A}\ }\textbf {\bibinfo {volume} {51}},\ \bibinfo
  {pages} {53} (\bibinfo {year} {2015}{\natexlab{a}})},\ \Eprint
  {https://arxiv.org/abs/1412.0142} {arXiv:1412.0142 [nucl-th]} \BibitemShut
  {NoStop}%
\bibitem [{\citenamefont {Reinert}\ \emph {et~al.}(2018)\citenamefont
  {Reinert}, \citenamefont {Krebs},\ and\ \citenamefont
  {Epelbaum}}]{Reinert:2017usi_4order}%
  \BibitemOpen
  \bibfield  {author} {\bibinfo {author} {\bibfnamefont {P.}~\bibnamefont
  {Reinert}}, \bibinfo {author} {\bibfnamefont {H.}~\bibnamefont {Krebs}},\
  and\ \bibinfo {author} {\bibfnamefont {E.}~\bibnamefont {Epelbaum}},\
  }\bibfield  {title} {\bibinfo {title} {{Semilocal momentum-space regularized
  chiral two-nucleon potentials up to fifth order}},\ }\href
  {https://doi.org/10.1140/epja/i2018-12516-4} {\bibfield  {journal} {\bibinfo
  {journal} {Eur. Phys. J. A}\ }\textbf {\bibinfo {volume} {54}},\ \bibinfo
  {pages} {86} (\bibinfo {year} {2018})},\ \Eprint
  {https://arxiv.org/abs/1711.08821} {arXiv:1711.08821 [nucl-th]} \BibitemShut
  {NoStop}%
\bibitem [{\citenamefont {Kaiser}(2001)}]{Kaiser:2001pc_2pe}%
  \BibitemOpen
  \bibfield  {author} {\bibinfo {author} {\bibfnamefont {N.}~\bibnamefont
  {Kaiser}},\ }\bibfield  {title} {\bibinfo {title} {{Chiral 2 pi exchange N N
  potentials: Two loop contributions}},\ }\href
  {https://doi.org/10.1103/PhysRevC.64.057001} {\bibfield  {journal} {\bibinfo
  {journal} {Phys. Rev. C}\ }\textbf {\bibinfo {volume} {64}},\ \bibinfo
  {pages} {057001} (\bibinfo {year} {2001})},\ \Eprint
  {https://arxiv.org/abs/nucl-th/0107064} {arXiv:nucl-th/0107064} \BibitemShut
  {NoStop}%
\bibitem [{\citenamefont {Epelbaum}\ \emph
  {et~al.}(2015{\natexlab{b}})\citenamefont {Epelbaum}, \citenamefont {Krebs},\
  and\ \citenamefont {Mei\ss{}ner}}]{Epelbaum:2014sza_2pe}%
  \BibitemOpen
  \bibfield  {author} {\bibinfo {author} {\bibfnamefont {E.}~\bibnamefont
  {Epelbaum}}, \bibinfo {author} {\bibfnamefont {H.}~\bibnamefont {Krebs}},\
  and\ \bibinfo {author} {\bibfnamefont {U.~G.}\ \bibnamefont {Mei\ss{}ner}},\
  }\bibfield  {title} {\bibinfo {title} {{Precision nucleon-nucleon potential
  at fifth order in the chiral expansion}},\ }\href
  {https://doi.org/10.1103/PhysRevLett.115.122301} {\bibfield  {journal}
  {\bibinfo  {journal} {Phys. Rev. Lett.}\ }\textbf {\bibinfo {volume} {115}},\
  \bibinfo {pages} {122301} (\bibinfo {year} {2015}{\natexlab{b}})},\ \Eprint
  {https://arxiv.org/abs/1412.4623} {arXiv:1412.4623 [nucl-th]} \BibitemShut
  {NoStop}%
\bibitem [{\citenamefont {Fettes}\ \emph {et~al.}(1998)\citenamefont {Fettes},
  \citenamefont {Meissner},\ and\ \citenamefont
  {Steininger}}]{Fettes:1998ud_2pe}%
  \BibitemOpen
  \bibfield  {author} {\bibinfo {author} {\bibfnamefont {N.}~\bibnamefont
  {Fettes}}, \bibinfo {author} {\bibfnamefont {U.-G.}\ \bibnamefont
  {Meissner}},\ and\ \bibinfo {author} {\bibfnamefont {S.}~\bibnamefont
  {Steininger}},\ }\bibfield  {title} {\bibinfo {title} {{Pion - nucleon
  scattering in chiral perturbation theory. 1. Isospin symmetric case}},\
  }\href {https://doi.org/10.1016/S0375-9474(98)00452-7} {\bibfield  {journal}
  {\bibinfo  {journal} {Nucl. Phys. A}\ }\textbf {\bibinfo {volume} {640}},\
  \bibinfo {pages} {199} (\bibinfo {year} {1998})},\ \Eprint
  {https://arxiv.org/abs/hep-ph/9803266} {arXiv:hep-ph/9803266} \BibitemShut
  {NoStop}%
\bibitem [{\citenamefont {Niskanen}(2002)}]{Niskanen:2001aj_cib1}%
  \BibitemOpen
  \bibfield  {author} {\bibinfo {author} {\bibfnamefont {J.~A.}\ \bibnamefont
  {Niskanen}},\ }\bibfield  {title} {\bibinfo {title} {{Effective field theory
  contribution to charge symmetry breaking N N scattering}},\ }\href
  {https://doi.org/10.1103/PhysRevC.65.037001} {\bibfield  {journal} {\bibinfo
  {journal} {Phys. Rev. C}\ }\textbf {\bibinfo {volume} {65}},\ \bibinfo
  {pages} {037001} (\bibinfo {year} {2002})},\ \Eprint
  {https://arxiv.org/abs/nucl-th/0108015} {arXiv:nucl-th/0108015} \BibitemShut
  {NoStop}%
\bibitem [{\citenamefont {Gonzalez~Trotter}\ \emph {et~al.}(2006)\citenamefont
  {Gonzalez~Trotter} \emph {et~al.}}]{GonzalezTrotter:2006wz_181}%
  \BibitemOpen
  \bibfield  {author} {\bibinfo {author} {\bibfnamefont {D.~E.}\ \bibnamefont
  {Gonzalez~Trotter}} \emph {et~al.},\ }\bibfield  {title} {\bibinfo {title}
  {{Neutron-deuteron breakup experiment at E(n) = 13-MeV: Determination of the
  (1)S(0) neutron-neutron scattering length ann}},\ }\href
  {https://doi.org/10.1103/PhysRevC.73.034001} {\bibfield  {journal} {\bibinfo
  {journal} {Phys. Rev. C}\ }\textbf {\bibinfo {volume} {73}},\ \bibinfo
  {pages} {034001} (\bibinfo {year} {2006})}\BibitemShut {NoStop}%
\bibitem [{\citenamefont {Chen}\ \emph {et~al.}(2008)\citenamefont {Chen} \emph
  {et~al.}}]{Chen:2008zzj_182}%
  \BibitemOpen
  \bibfield  {author} {\bibinfo {author} {\bibfnamefont {Q.}~\bibnamefont
  {Chen}} \emph {et~al.},\ }\bibfield  {title} {\bibinfo {title} {{Measurement
  of the neutron-neutron scattering length using the pi-d capture reaction}},\
  }\href {https://doi.org/10.1103/PhysRevC.77.054002} {\bibfield  {journal}
  {\bibinfo  {journal} {Phys. Rev. C}\ }\textbf {\bibinfo {volume} {77}},\
  \bibinfo {pages} {054002} (\bibinfo {year} {2008})}\BibitemShut {NoStop}%
\bibitem [{\citenamefont {van Kolck}\ \emph {et~al.}(1996)\citenamefont {van
  Kolck}, \citenamefont {Friar},\ and\ \citenamefont
  {Goldman}}]{vanKolck:1996rm_epsilon}%
  \BibitemOpen
  \bibfield  {author} {\bibinfo {author} {\bibfnamefont {U.}~\bibnamefont {van
  Kolck}}, \bibinfo {author} {\bibfnamefont {J.~L.}\ \bibnamefont {Friar}},\
  and\ \bibinfo {author} {\bibfnamefont {J.~T.}\ \bibnamefont {Goldman}},\
  }\bibfield  {title} {\bibinfo {title} {{Phenomenological aspects of isospin
  violation in the nuclear force}},\ }\href
  {https://doi.org/10.1016/0370-2693(96)00009-3} {\bibfield  {journal}
  {\bibinfo  {journal} {Phys. Lett. B}\ }\textbf {\bibinfo {volume} {371}},\
  \bibinfo {pages} {169} (\bibinfo {year} {1996})},\ \Eprint
  {https://arxiv.org/abs/nucl-th/9601009} {arXiv:nucl-th/9601009} \BibitemShut
  {NoStop}%
\bibitem [{\citenamefont {Elhatisari}\ \emph
  {et~al.}(2016{\natexlab{b}})\citenamefont {Elhatisari}, \citenamefont {Lee},
  \citenamefont {Mei\ss{}ner},\ and\ \citenamefont
  {Rupak}}]{Elhatisari:2016hby_phase}%
  \BibitemOpen
  \bibfield  {author} {\bibinfo {author} {\bibfnamefont {S.}~\bibnamefont
  {Elhatisari}}, \bibinfo {author} {\bibfnamefont {D.}~\bibnamefont {Lee}},
  \bibinfo {author} {\bibfnamefont {U.-G.}\ \bibnamefont {Mei\ss{}ner}},\ and\
  \bibinfo {author} {\bibfnamefont {G.}~\bibnamefont {Rupak}},\ }\bibfield
  {title} {\bibinfo {title} {{Nucleon-deuteron scattering using the adiabatic
  projection method}},\ }\href {https://doi.org/10.1140/epja/i2016-16174-2}
  {\bibfield  {journal} {\bibinfo  {journal} {Eur. Phys. J. A}\ }\textbf
  {\bibinfo {volume} {52}},\ \bibinfo {pages} {174} (\bibinfo {year}
  {2016}{\natexlab{b}})},\ \Eprint {https://arxiv.org/abs/1603.02333}
  {arXiv:1603.02333 [nucl-th]} \BibitemShut {NoStop}%
\bibitem [{\citenamefont {Stapp}\ \emph {et~al.}(1957)\citenamefont {Stapp},
  \citenamefont {Ypsilantis},\ and\ \citenamefont
  {Metropolis}}]{Stapp:1956mz_phase}%
  \BibitemOpen
  \bibfield  {author} {\bibinfo {author} {\bibfnamefont {H.~P.}\ \bibnamefont
  {Stapp}}, \bibinfo {author} {\bibfnamefont {T.~J.}\ \bibnamefont
  {Ypsilantis}},\ and\ \bibinfo {author} {\bibfnamefont {N.}~\bibnamefont
  {Metropolis}},\ }\bibfield  {title} {\bibinfo {title} {{Phase shift analysis
  of 310-MeV proton proton scattering experiments}},\ }\href
  {https://doi.org/10.1103/PhysRev.105.302} {\bibfield  {journal} {\bibinfo
  {journal} {Phys. Rev.}\ }\textbf {\bibinfo {volume} {105}},\ \bibinfo {pages}
  {302} (\bibinfo {year} {1957})}\BibitemShut {NoStop}%
\bibitem [{\citenamefont {Vincent}\ and\ \citenamefont
  {Phatak}(1974)}]{Vincent:1974zz_cou}%
  \BibitemOpen
  \bibfield  {author} {\bibinfo {author} {\bibfnamefont {C.~M.}\ \bibnamefont
  {Vincent}}\ and\ \bibinfo {author} {\bibfnamefont {S.~C.}\ \bibnamefont
  {Phatak}},\ }\bibfield  {title} {\bibinfo {title} {{Accurate momentum-space
  method for scattering by nuclear and Coulomb potentials}},\ }\href
  {https://doi.org/10.1103/PhysRevC.10.391} {\bibfield  {journal} {\bibinfo
  {journal} {Phys. Rev. C}\ }\textbf {\bibinfo {volume} {10}},\ \bibinfo
  {pages} {391} (\bibinfo {year} {1974})}\BibitemShut {NoStop}%
\bibitem [{\citenamefont {Bethe}(1949)}]{Bethe:1949yr_r_a}%
  \BibitemOpen
  \bibfield  {author} {\bibinfo {author} {\bibfnamefont {H.~A.}\ \bibnamefont
  {Bethe}},\ }\bibfield  {title} {\bibinfo {title} {{Theory of the Effective
  Range in Nuclear Scattering}},\ }\href
  {https://doi.org/10.1103/PhysRev.76.38} {\bibfield  {journal} {\bibinfo
  {journal} {Phys. Rev.}\ }\textbf {\bibinfo {volume} {76}},\ \bibinfo {pages}
  {38} (\bibinfo {year} {1949})}\BibitemShut {NoStop}%
\bibitem [{\citenamefont {Wilson}(1974)}]{Wilson:1974sk}%
  \BibitemOpen
  \bibfield  {author} {\bibinfo {author} {\bibfnamefont {K.~G.}\ \bibnamefont
  {Wilson}},\ }\bibfield  {title} {\bibinfo {title} {{Confinement of Quarks}},\
  }\href {https://doi.org/10.1103/PhysRevD.10.2445} {\bibfield  {journal}
  {\bibinfo  {journal} {Phys. Rev. D}\ }\textbf {\bibinfo {volume} {10}},\
  \bibinfo {pages} {2445} (\bibinfo {year} {1974})}\BibitemShut {NoStop}%
\bibitem [{\citenamefont {Kogut}\ and\ \citenamefont
  {Susskind}(1975)}]{Kogut:1974ag}%
  \BibitemOpen
  \bibfield  {author} {\bibinfo {author} {\bibfnamefont {J.~B.}\ \bibnamefont
  {Kogut}}\ and\ \bibinfo {author} {\bibfnamefont {L.}~\bibnamefont
  {Susskind}},\ }\bibfield  {title} {\bibinfo {title} {{Hamiltonian Formulation
  of Wilson's Lattice Gauge Theories}},\ }\href
  {https://doi.org/10.1103/PhysRevD.11.395} {\bibfield  {journal} {\bibinfo
  {journal} {Phys. Rev. D}\ }\textbf {\bibinfo {volume} {11}},\ \bibinfo
  {pages} {395} (\bibinfo {year} {1975})}\BibitemShut {NoStop}%
\bibitem [{\citenamefont {Van Der~Leun}\ and\ \citenamefont
  {Alderliesten}(1982)}]{VanDerLeun:1982bhg_D24}%
  \BibitemOpen
  \bibfield  {author} {\bibinfo {author} {\bibfnamefont {C.}~\bibnamefont {Van
  Der~Leun}}\ and\ \bibinfo {author} {\bibfnamefont {C.}~\bibnamefont
  {Alderliesten}},\ }\bibfield  {title} {\bibinfo {title} {{The deuteron
  binding energy}},\ }\href {https://doi.org/10.1016/0375-9474(82)90105-1}
  {\bibfield  {journal} {\bibinfo  {journal} {Nucl. Phys. A}\ }\textbf
  {\bibinfo {volume} {380}},\ \bibinfo {pages} {261} (\bibinfo {year}
  {1982})}\BibitemShut {NoStop}%
\bibitem [{\citenamefont {Ericson}\ and\ \citenamefont
  {Rosa-Clot}(1983)}]{Ericson:1982ei_D25}%
  \BibitemOpen
  \bibfield  {author} {\bibinfo {author} {\bibfnamefont {T.~E.~O.}\
  \bibnamefont {Ericson}}\ and\ \bibinfo {author} {\bibfnamefont
  {M.}~\bibnamefont {Rosa-Clot}},\ }\bibfield  {title} {\bibinfo {title} {{The
  Deuteron Asymptotic $D$ State as a Probe of the Nucleon-nucleon Force}},\
  }\href {https://doi.org/10.1016/0375-9474(83)90516-X} {\bibfield  {journal}
  {\bibinfo  {journal} {Nucl. Phys. A}\ }\textbf {\bibinfo {volume} {405}},\
  \bibinfo {pages} {497} (\bibinfo {year} {1983})}\BibitemShut {NoStop}%
\bibitem [{\citenamefont {Rodning}\ and\ \citenamefont
  {Knutson}(1990)}]{Rodning:1990zz_D26}%
  \BibitemOpen
  \bibfield  {author} {\bibinfo {author} {\bibfnamefont {N.~L.}\ \bibnamefont
  {Rodning}}\ and\ \bibinfo {author} {\bibfnamefont {L.~D.}\ \bibnamefont
  {Knutson}},\ }\bibfield  {title} {\bibinfo {title} {{Asymptotic D-state to
  S-state ratio of the deuteron}},\ }\href
  {https://doi.org/10.1103/PhysRevC.41.898} {\bibfield  {journal} {\bibinfo
  {journal} {Phys. Rev. C}\ }\textbf {\bibinfo {volume} {41}},\ \bibinfo
  {pages} {898} (\bibinfo {year} {1990})}\BibitemShut {NoStop}%
\bibitem [{\citenamefont {Bishop}\ and\ \citenamefont
  {Cheung}(1979)}]{Bishop:1979zz_D27}%
  \BibitemOpen
  \bibfield  {author} {\bibinfo {author} {\bibfnamefont {D.~M.}\ \bibnamefont
  {Bishop}}\ and\ \bibinfo {author} {\bibfnamefont {L.~M.}\ \bibnamefont
  {Cheung}},\ }\bibfield  {title} {\bibinfo {title} {{Quadrupole moment of the
  deuteron from a precise calculation of the electric field gradient in D-2}},\
  }\href {https://doi.org/10.1103/PhysRevA.20.381} {\bibfield  {journal}
  {\bibinfo  {journal} {Phys. Rev. A}\ }\textbf {\bibinfo {volume} {20}},\
  \bibinfo {pages} {381} (\bibinfo {year} {1979})}\BibitemShut {NoStop}%
\bibitem [{\citenamefont {Huber}\ \emph {et~al.}(1998)\citenamefont {Huber},
  \citenamefont {Udem}, \citenamefont {Gross}, \citenamefont {Reichert},
  \citenamefont {Kourogi}, \citenamefont {Pachucki}, \citenamefont {Weitz},\
  and\ \citenamefont {Hansch}}]{Huber:1998zz_D28}%
  \BibitemOpen
  \bibfield  {author} {\bibinfo {author} {\bibfnamefont {A.}~\bibnamefont
  {Huber}}, \bibinfo {author} {\bibfnamefont {T.}~\bibnamefont {Udem}},
  \bibinfo {author} {\bibfnamefont {B.}~\bibnamefont {Gross}}, \bibinfo
  {author} {\bibfnamefont {J.}~\bibnamefont {Reichert}}, \bibinfo {author}
  {\bibfnamefont {M.}~\bibnamefont {Kourogi}}, \bibinfo {author} {\bibfnamefont
  {K.}~\bibnamefont {Pachucki}}, \bibinfo {author} {\bibfnamefont
  {M.}~\bibnamefont {Weitz}},\ and\ \bibinfo {author} {\bibfnamefont {T.~W.}\
  \bibnamefont {Hansch}},\ }\bibfield  {title} {\bibinfo {title}
  {{Hydrogen-Deuterium S-1- S-2 Isotope Shift and the Structure of the
  Deuteron}},\ }\href {https://doi.org/10.1103/PhysRevLett.80.468} {\bibfield
  {journal} {\bibinfo  {journal} {Phys. Rev. Lett.}\ }\textbf {\bibinfo
  {volume} {80}},\ \bibinfo {pages} {468} (\bibinfo {year} {1998})}\BibitemShut
  {NoStop}%
\bibitem [{\citenamefont {Dumbrajs}\ \emph {et~al.}(1983)\citenamefont
  {Dumbrajs}, \citenamefont {Koch}, \citenamefont {Pilkuhn}, \citenamefont
  {Oades}, \citenamefont {Behrens}, \citenamefont {De~Swart},\ and\
  \citenamefont {Kroll}}]{Dumbrajs:1983jd_D29}%
  \BibitemOpen
  \bibfield  {author} {\bibinfo {author} {\bibfnamefont {O.}~\bibnamefont
  {Dumbrajs}}, \bibinfo {author} {\bibfnamefont {R.}~\bibnamefont {Koch}},
  \bibinfo {author} {\bibfnamefont {H.}~\bibnamefont {Pilkuhn}}, \bibinfo
  {author} {\bibfnamefont {G.~c.}\ \bibnamefont {Oades}}, \bibinfo {author}
  {\bibfnamefont {H.}~\bibnamefont {Behrens}}, \bibinfo {author} {\bibfnamefont
  {J.~j.}\ \bibnamefont {De~Swart}},\ and\ \bibinfo {author} {\bibfnamefont
  {P.}~\bibnamefont {Kroll}},\ }\bibfield  {title} {\bibinfo {title}
  {{Compilation of Coupling Constants and Low-Energy Parameters. 1982
  Edition}},\ }\href {https://doi.org/10.1016/0550-3213(83)90288-2} {\bibfield
  {journal} {\bibinfo  {journal} {Nucl. Phys. B}\ }\textbf {\bibinfo {volume}
  {216}},\ \bibinfo {pages} {277} (\bibinfo {year} {1983})}\BibitemShut
  {NoStop}%
\bibitem [{\citenamefont {Wu}\ \emph {et~al.}(2025)\citenamefont {Wu},
  \citenamefont {Wang}, \citenamefont {Lu},\ and\ \citenamefont
  {Li}}]{dataset_citation}%
  \BibitemOpen
  \bibfield  {author} {\bibinfo {author} {\bibfnamefont {C.}~\bibnamefont
  {Wu}}, \bibinfo {author} {\bibfnamefont {T.}~\bibnamefont {Wang}}, \bibinfo
  {author} {\bibfnamefont {B.-N.}\ \bibnamefont {Lu}},\ and\ \bibinfo {author}
  {\bibfnamefont {N.}~\bibnamefont {Li}},\ }\bibfield  {title} {\bibinfo
  {title} {{Dataset of nucleon-nucleon interaction at N3LO in NLEFT[Data set].
  Zenodo.}}\ }\href {https://doi.org/10.5281/zenodo.15468871}
  {10.5281/zenodo.15468871} (\bibinfo {year} {2025})\BibitemShut {NoStop}%
\end{thebibliography}%

\end{document}